\documentclass[graybox, secnum]{svmult}

\usepackage{mathptmx}       
\usepackage{helvet}         
\usepackage{courier}        
\usepackage{makeidx}         
\usepackage{graphicx}        
\graphicspath{{figs/}} 
\usepackage{relsize}
\usepackage{multicol} 
\usepackage[bottom]{footmisc}
\usepackage[bookmarks=true,bookmarksdepth=4]{hyperref} 
\hypersetup{colorlinks=true,urlcolor=blue}

\usepackage[square,numbers]{natbib}
\usepackage{latexsym}
\usepackage{amsmath}
\usepackage{bm}
\usepackage{bbold}
\usepackage{dsfont} 

\newcommand{\GN}{G_\textsc{n}}
\newcommand{\MP}{M_\textsc{p}}
\newcommand{\Hu}{\mathcal{H}}
\newcommand{\Ka}{\mathcal{K}}
\newcommand{\Rez}{\Re\mathrm{e}}
\newcommand{\Imz}{\Im\mathrm{m}}
\newcommand{\MS}{{\textsmaller{2MS}}}

\makeindex             

\begin{document}

\title*{Quantum cosmological gravitational waves?}


\author{Amaury Micheli and Patrick Peter \thanks{corresponding author}}

\institute{Amaury Micheli, Patrick Peter \at ${\cal
    G}\mathbb{R}\varepsilon\mathbb{C}{\cal O}$ -- Institut
  d'Astrophysique de Paris, CNRS \& Sorbonne Universit\'e, UMR 7095 98
  bis boulevard Arago, 75014 Paris, France, \email{peter@iap.fr} \and
  Amaury Micheli \at IJCLab, Laboratoire de Physique des 2 Infinis
  Ir\`ene Joliot-Curie, B\^at. 100 et 200, 15 rue Georges
  Cl\'emenceau, F-91405 Orsay, France
  \email{amaury.micheli@ijclab.in2p3.fr}}

%
%

\maketitle

\abstract{General relativity and its cosmological solution predicts
  the existence of tensor modes of perturbations evolving on top of
  our Friedman-Lemaître-Robertson-Walker expanding Universe.  Being
  gauge invariant and not necessarily coupled to other quantum
  sources, they can be seen as representing pure
  gravity. Unambiguously showing they are indeed to be quantised would
  thus provide an unquestionable proof of the quantum nature of
  gravitation. This review will present a summary of the various
  theoretical issues that could lead to this conclusion.}

\section*{Keywords} 
Cosmological perturbation theory, tensor modes, gravitational waves,
quantum cosmology, perturbative quantum gravity.

\section{Introduction}

Cosmology is a major player when it comes to quantum gravity
effects. Indeed, on top of our Friedman-Lemaître-Robertson-Walker
(FLRW) expanding Universe, one expects various modes of perturbations
to be present, whose classical occurrence is believed to result from
initial quantum vacuum fluctuations. In the usual linear
formalism~\cite{Mukhanov:1990me,PeterUzan2009}, using the FLRW
underlying symmetry group (isotropy and homogeneity), they can be
categorised into three components, namely scalars, vectors and
tensors. At this order, upon which we focus attention below, these
components decouple. In a different situation with a background
endowed with other symmetries, perturbations can still be expanded in
the relevant representations of the associated group; they also
naturally decouple at linear order (see,
e.g. Ref.~\cite{Pereira:2007yy} for Bianchi I).

Scalar modes, detected long ago in the cosmic microwave background,
initiating large-scale structure formation, are distributed in a way
that is compatible with quantum vacuum fluctuations in the very early
times, often during a phase of inflation. This can be seen as
requiring quantisation of gravity, and although many authors consider
it does, others argue that gauge issues and coupling with matter
render the conclusion not as clear as one would wish.

In an ever-expanding FLRW universe with dynamics driven by GR or any
local theory of gravity, with no specific source in the matter fields
to induce them, vector perturbations are expected to have decayed long
ago so as to be mostly undetectable now~\cite{Grishchuk:1993ab}.  One
of the above hypothesis needs to be invalidated to potentially render
them cosmologically relevant. Non local theories are expected to yield
conclusions similar to local ones~\cite{Craps:2014wga}.  A contracting
phase in the universe as implemented in bouncing
models~\cite{Battefeld:2014uga,Brandenberger:2016vhg} can lead to some
increase of vector modes~\cite{Battefeld:2004cd} which are however
limited if produced by means of some coupling with scalar modes
initially set to quantum vacuum
fluctuations~\cite{Pinto-Neto:2020xmb}, leading to the conclusion that
bouncing models are generally stable under vector perturbations.  For
fully quantum cosmological models however, the situation may not be as
clear~\cite{Bojowald:2007hv}. In any case, the question of their
quantum origin would lead to similar doubts regarding the quantumness
of gravity itself; they are conveniently ignored in most studies, and
likewise in the present review.

Finally, one is left with the tensor modes, which are gauge invariant
and with no obvious coupling to other quantum sources. General
relativity (GR) applied to primordial cosmology shows their dynamics
to be that of two time-dependent massive scalar fields; most models
then demand they should be quantised and set in a vacuum state. The
observation of their resulting properties in the absence of quantum
anisotropic pressure, jointly with those of the scalar modes, could
provide an unambiguous and thus indisputable hint that gravitation
itself should acquire the status of a quantum theory.

\section{Tensor modes in general relativistic cosmology}

Before focusing on the quantum features expected from gravitational
waves, let us briefly recap the underlying classical theory.  The
starting point of our discussion is the FLRW background universe,
defined by its scale factor function $a(\eta)$ depending on the
(conformal) time $\eta$ and spatial 3D metric $\gamma_{ij}$, with
tensorial perturbations $h_{ij}$. In that case, ignoring both scalar
and vector modes which are not the subject of this analysis, one sets
the metric as
\begin{equation}
\D s^2 = g_{\mu \nu} \mathrm{d} x^{\mu} \mathrm{d} x^{\nu} =
a^2(\eta)\left[-\D\eta^2+ \left(\gamma_{ij} +h_{ij}\right)\D x^i\D
  x^j\right],
\label{gdg}
\end{equation}
(we use units such that the velocity of light is $c=1$) and with
$h_{ij}$ assumed transverse and traceless, i.e.
$$
D^i h_{ij} = 0 \ \ \ \ \hbox{and} \ \ \ \ h^i_{\ i} = \gamma^{ij}
h_{ij} = 0,
$$
the 3D covariant derivative $D^i$ being derived from the corresponding
metric $\gamma_{ij}$. Noting $\Hu = a^\prime/a$ the conformal Hubble
function\footnote{A prime denotes differentiation with respect to
  conformal time, e.g. $a^\prime := \D a/\D \eta$} and $\Ka$ the
spatial curvature\footnote{In appropriate units for the comoving
  coordinates $x^i$, it can be scaled to $\Ka = 0, \pm 1$. For most of
  the practical applications we shall deal with in this review, we
  shall consider the simplest, and inflation-motivated, flat case with
  $\Ka=0$.}  associated with the background metric $\gamma_{ij}$, the
equation of motion for $h_{ij}$ is found to be
\begin{equation}
h_{ij}^{\prime\prime} + 2 \Hu h_{ij}^{\prime} + \left( 2\Ka -\Delta
\right) h_{ij} = 8\pi\GN a^2 p \pi_{ij},
\label{hiipp}
\end{equation}
where $\Delta = \gamma^{ij} \partial_{i} \partial_{j}$, $p$ is the
background pressure and $\pi_{ij}$ the anisotropic stress. For many of
the known components of matter, it is vanishing (however, see e.g.
\cite{Ganguly:2021pke,Shiraishi:2016yun} and references therein), and
we shall make the assumption that $\pi_{ij}=0$ from now on.

In what follows, we set $\Ka\to 0$ and thus identify the background
spatial metric $\gamma^{ij} \to \delta^{ij}$ as the 3D curvature has
been measured to be vanishingly small. Technically, considering a
non-vanishing curvature merely amounts to changing the spectrum (and
eigenfunctions) of the Laplace-Beltrami operator used below for the
mode decomposition \cite{lifshitzGravitationalStabilityExpanding2017},
so that the calculations and discussions presented below can be
generalised in a straightforward way if applied to epochs in which the
assumption $\Ka=0$ may not be valid.

Let us thus first decompose the tensor perturbations in Fourier modes
through\footnote{The numerical factor $\sqrt{32\pi\GN}$ is included
  here for later convenience.}
\begin{equation}
h_{ij}(\bm{x},\eta) = \sqrt{32\pi\GN}
\int \frac{\D^3 \bm{k}}{(2\pi)^{3/2}a(\eta)}
w_{ij}(\bm{k},\eta)
e^{i\bm{k}\cdot\bm{x}} 
\label{Fourierhij}
\end{equation}
with $w_{ij}^\star(-\bm{k},\eta)=w_{ij}(\bm{k},\eta)$ to ensure
$h_{ij}\in\mathbb{R}$, so that, from Eq.~\eqref{hiipp}, a given mode
satisfies
\begin{equation}
w_{ij}^{\prime\prime} + \omega_{k}^2 w_{ij} = 0 \, ,
\label{mode}
\end{equation}
where we defined the module $k := |\bm{k}| \geq
0$ and the time-varying frequency
\begin{equation}
\omega_{k}^2 = k^2 - \frac{a^{\prime\prime}}{a} \, .
\label{def:frequency}
\end{equation}
At this point, one notes that whenever the scale factor behaves as a
power-law of the conformal time\footnote{We write the absolute value
  of the conformal time in what follows, as it is negative in many
  situations, in particular during inflation.}  $a \left( \eta \right)
\propto \left| \eta \right|^\alpha$, then $a^{\prime \prime}/ a
=\alpha (\alpha-1)/ \eta^{-2} = (\alpha-1) \Hu^2/\alpha$. This in
particular encompasses the cases of cosmological interest where a
single fluid dominates the Friedmann dynamics, as well as the de
Sitter inflationary expansion. The condition $k^2 \gg \left| a^{\prime
  \prime}/ a \right|$ then becomes $k \Hu^{-1} \propto k |\eta| \ll
1$, so that, in terms of the physical wavelength $\lambda \propto
a/k$, one has $\lambda \ll H^{-1}$: such a mode, much smaller than the
Hubble scale $H^{-1}$, is said to be sub-Hubble.  Conversely, modes
with $k^2 \ll \left| a^{\prime \prime} / a \right|$ are called
super-Hubble.

Let us temporarily restrict attention to a sub-Hubble mode $k^2 \gg
\left| a^{\prime \prime}/ a \right|$. Eq.~\eqref{mode} then simplifies
to $w_{ij}^{\prime\prime} + k^2 w_{ij} =0$, whose solution reads
$w_{ij} = \alpha_{ij} \exp (\pm i k\eta)$. For a mode propagating in
the $+\bm{x^3}-$direction, this yields $h_{ij} = \alpha_{ij} \exp [\pm
  i k(x^3 -\eta)] / a(\eta)$.  The first constraint, namely
$\partial^i h_{ij} = 0$, implies $k^i \alpha_{ij} = k \alpha_{zj} =
0$, so that for $k\not= 0$, one is left with $\alpha_{11}$,
$\alpha_{12}$ and $\alpha_{22}$ as the only non vanishing components
(the symmetries of $w_{ij}$ are identical to those of $h_{ij}$). The
second constraint, $h^i_{\ i}=0$, translates into $\alpha_{22} = -
\alpha_{11}$, so the mode has only two independent degrees of
freedom. The matrix $\alpha_{ij}$ can be rewritten explicitly as
\begin{equation}
\label{solModes}
\alpha_{ij} = \left( \begin{array}{ccc}
\alpha_{11} &\alpha_{12} & 0
\\ \alpha_{12} & -\alpha_{11} & 0\\
0&0&0
\end{array}
\right) = \underbrace{
 \left( \begin{array}{ccc}
1 & 0 & 0
\\ 0 & -1 & 0\\
0&0&0
\end{array}
\right)}_{=\sqrt{2}P^+_{ij}}
\alpha_{11} 
+ \underbrace{\left( \begin{array}{ccc}
0 & 1 & 0
\\ 1 & 0 & 0\\
0&0&0
\end{array}
\right)}_{=\sqrt{2}P^\times_{ij}} 
\alpha_{12}.
\end{equation}
The matrices $P^+_{ij}$ and $P^\times_{ij}$ represent the two
polarisations of the gravitational wave, whose associated 
tensor perturbations read
\begin{equation}
    h_{ij} \left( \bm{x} , \eta \right) = h_{\times} \left( t - z
    \right) P^\times_{ij} + h_{+} \left( t - z \right) P^+_{ij},
\label{tmoinsz}
\end{equation}
with $\{t,x,y,z \} = \{a \eta,a x^1,a x^2,a x^3 \}$ the
physical coordinates.

\begin{figure}[h]
\centering
\includegraphics[width=\textwidth]{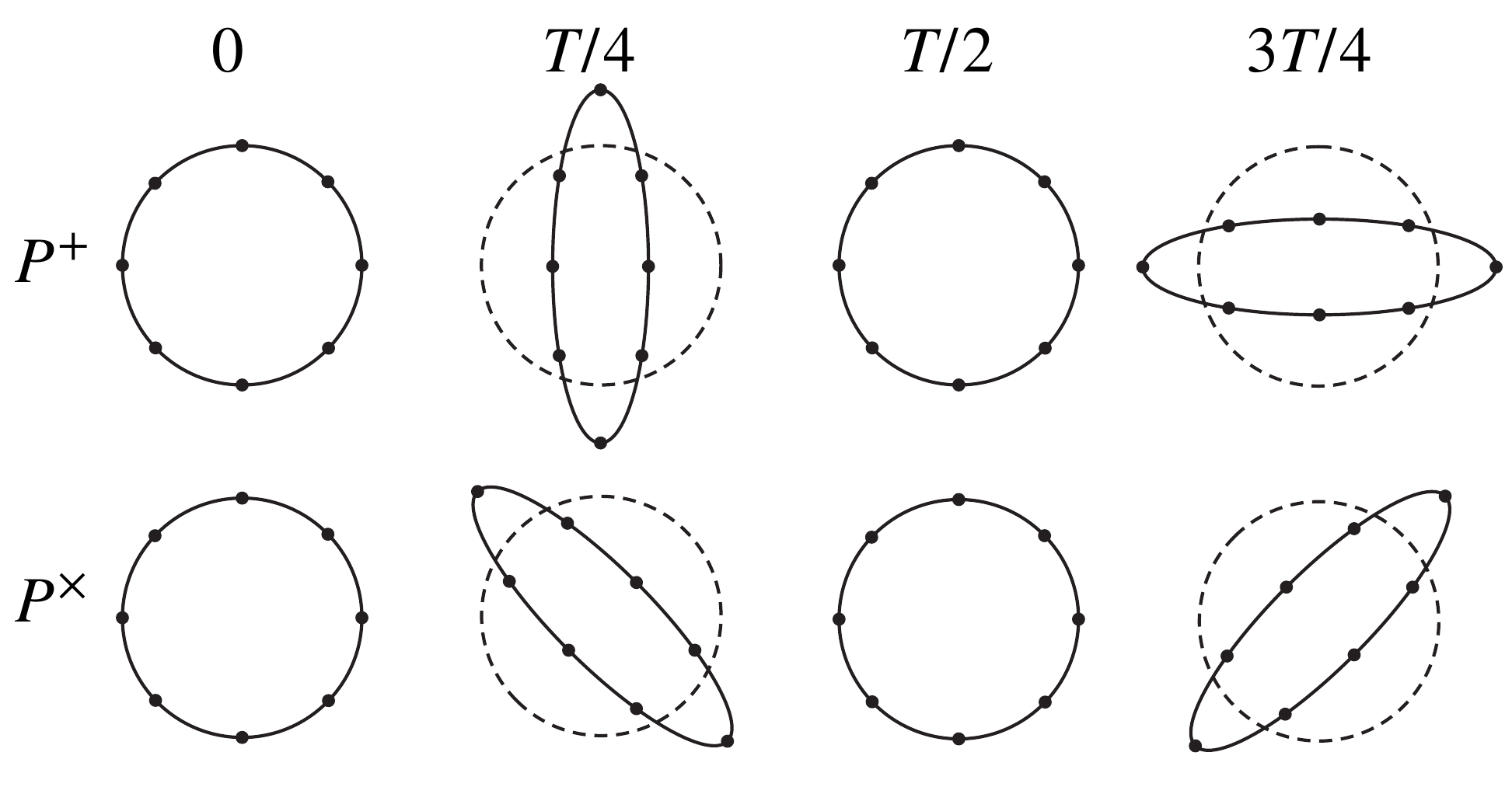}

\caption{Effect of a gravitational wave mode $P^+$ or $P^\times$ as it
  passes through a ring of test particles, producing `$+$' or
  `$\times$' shapes as time goes through a full period of the wave:
  starting with an initially circular ring at $t=0$, its shape is
  modified and shown here for different values of time, namely $T/4$,
  $T/2$ and $3T/4$ for a period $T=2\pi/k$.}

\label{Xplus}
\end{figure}

Consider a test particle following the trajectory of affine parameter
$\lambda$, i.e.  $x^\mu(\lambda)$, and initially at rest in the
TT-frame where the metric has the form \eqref{gdg} with $h_{ij}$ given
by \eqref{tmoinsz}, namely, assuming the scale factor $a$ to be
constant during the passing of the wave,
\begin{equation}
\D s^2 = -\D t^2 + \left[ 1 +h_+ (z-t)\right]
\D x^2 + \left[ 1 - h_+ (z-t)\right] \D y^2 
+ 2 h_\times (z-t) \D x \D y + \D z^2.
\label{MetCrossZ}
\end{equation}
From \eqref{MetCrossZ}, one can evaluate the connections while the
wave passes, and it turns out that $\Gamma^i_{\ \eta\eta} = 0$, so
that the motion of a particle following a geodesic is unaltered as it
moves with the reference frame: it appears at rest at all times. It is
therefore not possible to detect a gravitational wave using a single
particle.

Writing the line element as $\D s^2 = -\D t^2 + \D \ell^2$, we
consider two particles located on the TT-$x$ axis (i,.e. $y=z=0$) with
coordinates $x$ and $x+\Delta x$.  Their proper distance is obtained
from \eqref{MetCrossZ}: the relation $\D \ell_x = \sqrt{1+\alpha_+(t)}
\D x \simeq \left[ 1 +\frac12 h_+(t) \right] \D x$, can be integrated
to yield $\Delta \ell_x = \left[ 1 + \frac12 h_+(t) \right] \Delta
x$. Similarly, considering two particles lying along the $y$ axis, one
obtains $\Delta \ell_y = \left[ 1 - \frac12 h_+(t) \right] \Delta y$,
so that as the separation along one direction is elongated, the other
is compressed, and vice versa.  A similar calculation on particles set
on the $y=\pm x$ lines permits to visualize the effect of the
$\alpha_\times$ polarisation. Setting our test particles along a ring
in the $(x,y)$ plane, such as shown in Fig.~\ref{Xplus}, one gets the
$+$ and $\times$ shapes as the wave propagates in the
$\bm{z}-$direction, hence the names of the polarisation modes.

\begin{figure}[t]
\centering \includegraphics[width=\textwidth]{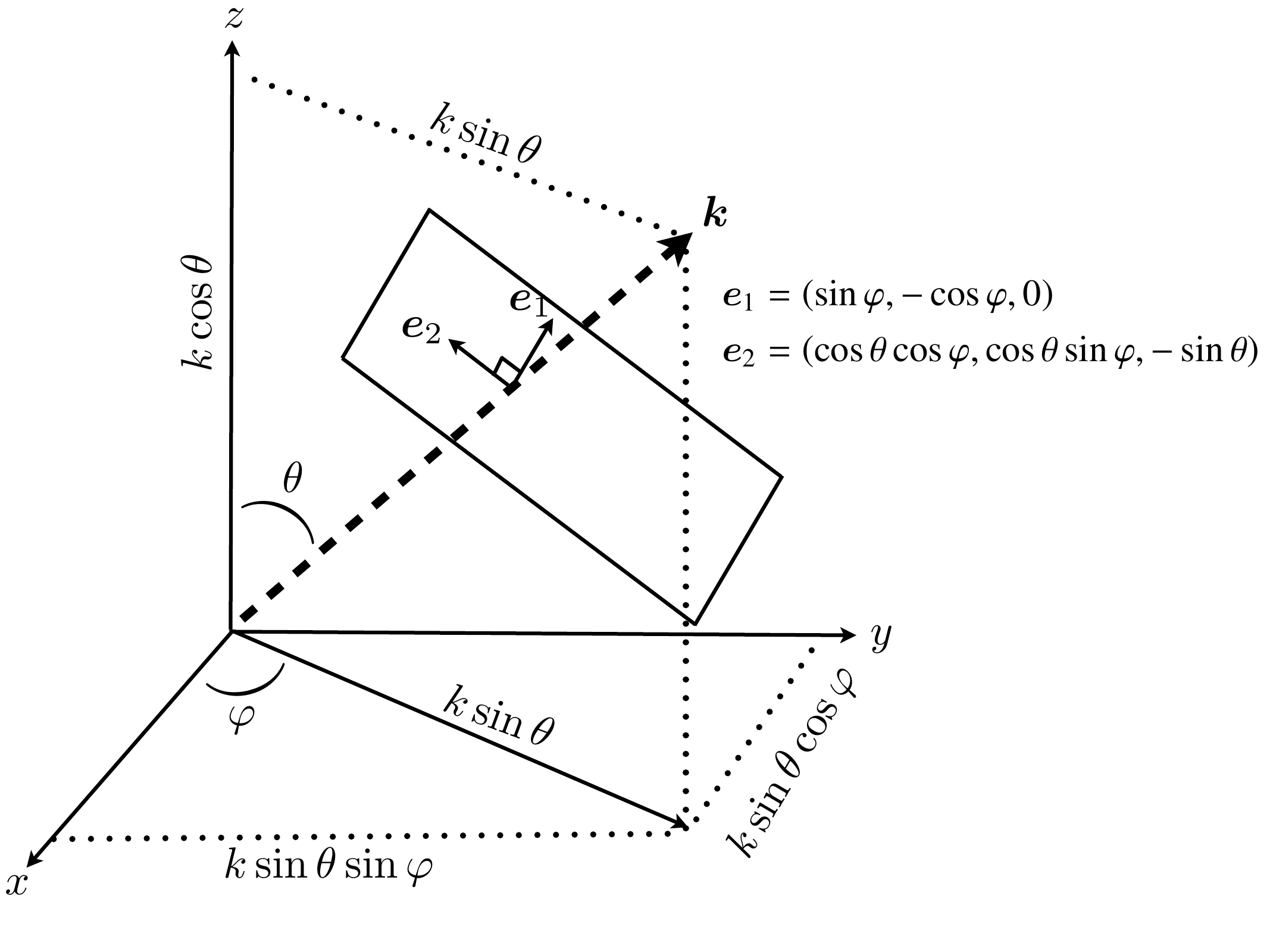}

\caption{Definition of the dyad $\bm{e}_a$ ($a=1,2$) in the plane
  orthogonal to the arbitrary direction $\bm{k}$.
\label{Polarisations}
}

\end{figure}

For a general wave vector $\bm{k} = k \bm{n}$ in the arbitrary
direction parametrised by the angles $\varphi$ and $\theta$ (see
Fig.~\ref{Polarisations}), namely $\bm{n} = \left( \cos\varphi
\sin\theta, \sin\varphi\sin\theta, \cos\theta \right)$, one sets
\begin{equation}
w_{ij} (\bm{k},\eta) = \sum_{\lambda=+,\times}
P^{(\lambda)}_{ij}
(\bm{n}) f_{\lambda} (\bm{k},\eta),
\label{wijP}
\end{equation}
with $P^{(\lambda)}_{ij} (\bm{n})$ the polarisation tensors and
$f_\lambda$ the associated functions solutions of the mode equation
\eqref{mode}.  Fig.~\ref{Polarisations} shows the vectors $\bm{e}_a$
($a=1,2$) generating the plane orthogonal to the direction of
propagation. Defined through
$$ \bm{e}_1 = -\frac{1}{\sin\theta} \frac{\partial \bm{n}}{\partial
  \varphi} = \left( \begin{matrix} \sin\varphi\cr -\cos\varphi \cr
  0 \end{matrix} \right)\ \ \ \ \hbox{and} \ \ \ \ \bm{e}_2 =
\frac{\partial \bm{n}}{\partial \theta} =
\left( \begin{matrix}\cos\theta\cos\varphi\cr \cos\theta\sin\varphi
  \cr -\sin \theta \end{matrix}\right),
$$
so that $\bm{n} = \bm{e}_1\times \bm{e}_2$, they satisfy
$\bm{e}_a\cdot \bm{e}_b =\delta_{ab}$ and
$\bm{n}\cdot\bm{e}_a=0$. Demanding $h_{ij}$ to be transverse and
traceless translates into
\begin{equation}
k^i P^{(\lambda)}_{ij} = 0,\ \ \ \ \hbox{and} \ \ \ \
P^{(\lambda)}_{ij} \delta^{ij} = 0,
\label{tensconst}
\end{equation}
and one can check that the choice
\begin{equation}
P^+_{ij} = \frac{1}{\sqrt{2}} \left[
  \left(\bm{e}_2\right)_i\left(\bm{e}_2\right)_j -
  \left(\bm{e}_1\right)_i\left(\bm{e}_1\right)_j\right]
\ \ \ \ \hbox{and} \ \ \ \ P^{\times}_{ij} = -\frac{1}{\sqrt{2}}
\left[ \left(\bm{e}_1\right)_i \left(\bm{e}_2\right)_j+
  \left(\bm{e}_2\right)_i\left(\bm{e}_1\right)_j \right]
\label{Pij}
\end{equation}
satisfies all the constraints (\ref{tensconst}); they reduce to those
appearing in \eqref{solModes} for $\bm{k}=k\bm{z}$ (choosing
$\varphi\to0$ or $\varphi\to\pi$ as it is then undetermined). One can
check straightforwardly that the relations
\begin{equation}
P^{i+}_{\ j}(\bm{n}) P^{j+}_{\ i}(\bm{n}) = P^{i\times}_{\ j}(\bm{n})
P^{j\times}_{\ i}(\bm{n}) = 1 \ \ \ \ \hbox{and}
\ \ \ \ P^{i+}_{\ j}(\bm{n}) P^{j\times}_{\ i}(\bm{n}) =0
\label{sumepsilon}
\end{equation}
hold.

Let us note at this point that the transformation $\bm{n} \to
-\bm{n}$, which amounts to $(\theta,\varphi)\to
(\pi-\theta,\varphi+\pi)$, implies $\bm{e}_1\to-\bm{e_1}$ and
$\bm{e}_2\to \bm{e_2}$, so that
\begin{equation}
P^+_{ij}(-\bm{n}) = P^+_{ij}(\bm{n}) \ \ \ \ \hbox{and} \ \ \ \ 
P^\times_{ij}(-\bm{n}) = - P^\times_{ij}(\bm{n}).
\label{epsm}
\end{equation}
From \eqref{wijP} and the reality condition below \eqref{Fourierhij},
one then finds that
$$
f_+^\star (-\bm{k}, \eta) = f_+ (\bm{k}, \eta) \ \ \ \ \hbox{and} \ \ \ \ 
f_\times^\star (-\bm{k}, \eta) = -f_\times (\bm{k}, \eta),
$$
the extra minus sign in the cross-polarisation reflecting the fact
that the gravitational wave transforms according to a spin-2
representation and not as a scalar. This sign is however inconvenient
as it requires the functions $f_{\lambda}$ to explicitly depend on the
direction of propagation of the gravitational wave.

This issue is solved by considering another basis instead of the $+$
and $\times$ polarisations:
\begin{equation}
\varepsilon_{ij}^\pm := \frac{1}{\sqrt{2}}
\left( P^+_{ij} \pm i P^\times_{ij}\right),
\label{helicities}
\end{equation}
resulting in the new expansion
\begin{equation}
w_{ij} (\bm{k},\eta) = \sum_{\lambda=\pm}
\varepsilon^{(\lambda)}_{ij}
(\bm{n}) \mu_{\lambda} (\bm{k},\eta),
\label{wij}
\end{equation}
where now one recovers the usual reality conditions in the form
\begin{equation}
\mu_\pm^\star (-\bm{k}, \eta) = \mu_\pm (\bm{k}, \eta),
\label{mustar}
\end{equation}
because $\left[ \varepsilon_{ij}^\pm (-\bm{n}) \right]^\star =
\varepsilon_{ij}^\pm (\bm{n})$. Note also that the orthogonality
relations become $\varepsilon^{i\pm}_{\ j} (\bm{n})
\varepsilon^{j\mp}_{\ i} (\bm{n}) = 1$ and $\varepsilon^{i\pm}_{\ j}
(\bm{n}) \varepsilon^{j\pm }_{\ i} (\bm{n}) = 0$ and that the
coefficients of the expansion are related via
\begin{equation}
\mu_{\pm} (\bm{k}, \eta) = \frac{1}{\sqrt{2}} \left[ f_{+}(\bm{k},
  \eta) \mp i f_{-}(\bm{k}, \eta) \right] \, .
\label{eq:coefficients_helicities_as_a_fn_plus_times}
\end{equation}

Performing a rotation in the plane orthogonal to $\bm{n}$ by an angle
$\alpha$ amounts to rotating $\bm{e}_a$ through
\begin{equation*}
\left\{ \begin{matrix} 
\bm{e}_1 & \to \bm{e}_1 \cos\alpha - \bm{e}_2 \sin\alpha \\
\bm{e}_2 & \to \bm{e}_1 \sin\alpha + \bm{e}_2 \cos\alpha
\end{matrix}\right.
\end{equation*}
and one can check explicitly that the new polarisations transform
according to
\begin{equation}
    \varepsilon_{ij}^\pm \to e^{\pm2i\alpha} \varepsilon_{ij}^\pm \, ,
\end{equation}
i.e. they transform as tensors with helicity $\pm 2$ and are,
therefore referred to as the \emph{helicity} basis. Gathering all the
above, one finds that Eq.~\eqref{wij} permits to show that, in
general, the modes $\mu_+$ and $\mu_-$ both satisfy the same equation
of motion, which is nothing but Eq.~\eqref{mode} with the replacement
$w_{ij} \to \mu_\pm$.

This can be derived directly from Eq.~\eqref{mode} using the expansion
on the helicity basis, or going back to the Einstein-Hilbert action
and performing an expansion in powers of $h_{ij}$
\begin{equation*}
S_\textsc{eh} = \frac{1}{16\pi\GN} \int \D^4 x\,
\sqrt{-g} R =
\frac{1}{16\pi\GN} \int \D^4 x \,
\sqrt{- \left[ g^{(0)} + g^{(2)} \right]  }
\left[ R^{(0)} + R^{(2)}  \right]+ \cdots,
\end{equation*}
where the dots represent higher-order terms and the determinant is
expanded as the exponent of the trace of a logarithm $g = \mathrm{det}
\left( g_{\mu \nu} \right) = \mathrm{det} \left( g_{\mu \nu}^{(0)}
\right) \mathrm{det} \left( \delta^{\mu}_{ \nu} + h^{\mu}_{\nu}
\right) = a^4 \left[ 1 -\frac12 h^{i}_j h^j_i +
  \mathcal{O}\left(h^3\right) \right]$ and the first term vanishes due
to the traceless condition, while the contribution from $R^{(1)}$
vanishes identically if we assume the background to satisfy the
equation of motion. The resulting action at second-order reads
\begin{equation}
\delta^{(2)}S_\textsc{t} = \frac{1}{64\pi \GN}
\int a^2(\eta) \left[ \frac{\partial h^i_{\ j} }{\partial \eta} 
\frac{\partial  h^j_{\ i} }{\partial \eta}
-\left( \partial_k h^i_{\ j}\right)
\partial^k h^j_{\ i} 
\right] \D^4x.
\label{d2Sij}
\end{equation}
Plugging the expansion \eqref{Fourierhij} and the
definition \eqref{wij} into the action \eqref{d2Sij}, leads to
\begin{equation}
\delta^{(2)}S_\textsc{t} = \int\D\eta \sum_{\lambda=\pm}
\frac12\int\D^3\bm{k}\, \left[ \left(
\mu_\lambda^{\prime\star} - \Hu \mu_\lambda^{\star} \right)
\left( \mu_\lambda^\prime - \Hu \mu_\lambda \right)
- \bm{k}^2 \mu_\lambda^\star \mu_\lambda\right].
\label{SModk}
\end{equation}
Upon integrating the $\Hu (\mu_\lambda^\star \mu_\lambda)^\prime$ by
parts, and using Parseval theorem to revert to real space, we get
\begin{equation}
\delta^{(2)}S_\textsc{t} = \int\D\eta \sum_{\lambda=\pm}
\frac12\int\D^3\bm{x}\, \sqrt{\gamma}
\left[ \left(\mu_\lambda'\right)^2 -
\gamma^{ij}\partial_i\mu_\lambda \partial_j\mu_\lambda +
\frac{a''}{a} \mu_\lambda^2 \right],
\label{GravMod2}
\end{equation}
where we wrote $\mu_\lambda = \mu_\lambda(\bm{x},\eta)$ the inverse
Fourier transform of $\mu_\lambda(\bm{k},\eta)$. This is the action
for two independent scalar fields $\mu_+$ and $\mu_-$, with identical
time-varying masses. One can check that the Euler-Lagrange equation
for \eqref{GravMod2} gives back \eqref{mode} for both polarisations.
The form \eqref{SModk} allows for straightforward quantisation of the
gravitational field as a collection of parametric oscillators, which
is the subject of the following section.

\section{Quantisation and time development}

\subsection{Historical perspective}

Parker, in Ref.~\cite{fordQuantizedGravitationalWave1977}, was the
first to use the above separation of the gravitational wave field into
two minimally coupled scalar fields as a simpler route to
quantisation, although previous works on (quantum) fields in curved
spacetime had already identified the crucial prediction of (vacuum)
amplification powered by the expansion of the Universe, including for
gravitational waves.  Particle creation following a change in boundary
conditions of a system was shown in
Ref.~\cite{takahashiGeneralTreatmentExpanding1957}, but creation
powered by an expanding Universe was first demonstrated by Parker in
his seminal articles
\cite{parkerParticleCreationExpanding1968,parkerQuantizedFieldsParticle1969,
  parkerQuantizedFieldsParticle1971}. However,
it was argued that massless non-zero spin fields, including
gravitational waves, had to be conformally coupled to gravity so that
no particle creation could occur. The production of gravitons,
particles associated with gravitational waves, was studied in
anisotropic universes in \cite{bergerQuantumGravitonCreation1974} and
hinted at in \cite{zeldovichParticleProductionVacuum1971} but
Grishchuk \cite{grishchukAmplificationGravitationalWaves1975} was the
first to lift the misunderstanding and to compute the ensuing
gravitational wave amplification in an isotropic expanding
universe. Despite the use of a classical treatment, the corresponding
quantum particle pair creation was noted and the existence of a
primordial gravitational wave background put forward. Several authors
then attempted to compute the spectrum of this background based on
spontaneous pair creation from the vacuum still using a classical
treatment and different initial conditions and renormalisation
procedures as, e.g. in Refs.~\cite{grishchukGRAVITONCREATIONEARLY1977,
  hartleEffectivePotentialApproachGraviton1977,
  starobinskySpectrumRelictGravitational1979}. Finally, in
\cite{starobinskySpectrumRelictGravitational1979}, graviton production
due to an early de Sitter phase of expansion, not yet called
inflation, was considered with the Bunch-Davies vacuum
\cite{bunchQuantumFieldTheory1978} providing the relevant initial
conditions.

Although acknowledged as originating from vacuum fluctuation, the
dynamics of primordial gravitational waves was first analysed
classically as successive stages of parametric amplifications, either
using a classical field and possibly fixing the initial conditions to
match quantum vacuum fluctuations
\cite{grishchukAmplificationGravitationalWaves1975,
  grishchukGRAVITONCREATIONEARLY1977,starobinskySpectrumRelictGravitational1979,
  rubakovGravitonCreationInflationary1982,fabbriEffectPrimordiallyProduced1983},
or using mode functions
\cite{fordQuantizedGravitationalWave1977,abbottConstraintsGeneralizedInflationary1984}.
Another presentation, equivalent to the latter, consists in
understanding the amplification of the waves as successive Bogoliubov
transformations \cite{abbottGravitonProductionInflationary1986} where
the initial state is chosen as the vacuum in an asymptotically
Minkowski region. Finally, it was latter recognised
\cite{grishchukSqueezedQuantumStates1990a}, moving to the Schrödinger
picture, that the evolution puts the gravitational waves in a squeezed
state. A good parallel presentation of the classical and quantum
descriptions can be found in
\cite{grishchukQuantumEffectsCosmology1993}.

In this section, we first proceed to the canonical quantisation of the
field in the Heisenberg picture following
\cite{fordQuantizedGravitationalWave1977}. This is the standard
approach; we refer to Refs.~\cite{birrellQuantumFieldsCurved1982,
  maggioreGravitationalWavesVol2018,
  mukhanovIntroductionQuantumEffects2007} for textbooks dealing with
scalar fields or gravitational waves.  We then review different formal
approaches to the evolution of a quantised gravitational wave field on
an FLRW background.  We begin by using a description in terms of a
Bogoliubov transformation, then make the connection with mode
functions and finally move to the Schrödinger picture, introducing
squeezing parameters and the phase-space representation of the
state. We use these different approaches to discuss the mechanism of
graviton creation in curved spacetime.  This then leads to a
discussion of how these particles back-reacts on the geometry.
Finally, we use these analyses to compute the properties of primordial
gravitational waves produced from the vacuum by the cosmological
expansion and discuss their quantum origin~\footnote{Note that the
  exact same analyses on quantisation and time evolution can be
  repeated for scalar perturbations during inflation with the same
  type of equations
  \cite{martinInflationaryCosmologicalPerturbations2005}.}.

\subsection{Canonical quantisation and Bogoliubov transformation}

Let us consider one of the two fields $\mu_{\lambda}$ in
Eq.~\eqref{GravMod2}.  It so happens that for the study of time
evolution in terms of Bogoliubov transformations and squeezing, it is
useful, and standard \cite{grishchukSqueezedQuantumStates1990a}, to
keep the total derivative that was dropped in the process of
integration by part between eqs.~\eqref{SModk} and
\eqref{GravMod2}. The Lagrangian thus obtained reads
\begin{equation}
L_{\lambda} = \frac{1}{2} \int \mathrm{d}^3 \bm{x} \left[
  \left(\mu_\lambda^{\prime} \right)^2 - 2 \Hu \mu_\lambda^{\prime}
  \mu_\lambda - \partial_i\mu_\lambda \partial^i\mu_\lambda + \Hu^2
  \mu_\lambda^2 \right].
\label{def:tilde_lagrangian_mu}
\end{equation}
The canonically conjugate momentum to $\mu_{\lambda}$ is
\begin{equation}
    \pi_{\lambda} \left( \bm{x} , \eta \right) = \frac{\delta
      L_{\lambda}}{ \delta \mu_{\lambda}^{\prime}} =
    \mu_{\lambda}^{\prime} - \Hu \mu_{\lambda} \, ,
\end{equation}
so the Hamiltonian reads 
\begin{equation}
    H_{\lambda} = \frac{1}{2} \int \mathrm{d}^3 \bm{x} \left[
      \pi_{\lambda}^{2} + \Hu \left( \pi_{\lambda} \mu_{\lambda}
      +\mu_{\lambda} \pi_{\lambda} \right) + \partial_i\mu_\lambda
      \partial^i \mu_\lambda \right],
\label{def:tilde_hamiltonian_mu_real_space}
\end{equation}
the second term being written in a symmetric way, which is classically
irrelevant but prepares for quantisation. We proceed to canonical
quantisation by imposing equal-time canonical commutation relations
(we now drop the $\lambda$ subscripts)
\begin{subequations}
\begin{align}
\left[ \hat{\mu} \left( \bm{x} , \eta \right) , \hat{\pi} \left(
  \bm{x}^{\prime } , \eta \right) \right] & = i \hbar \delta \left(
\bm{x} - \bm{x}^{\prime} \right) \, , \\ \left[ \hat{\mu} \left(
  \bm{x} , \eta \right) , \hat{\mu} \left( \bm{x}^{\prime} , \eta
  \right) \right] & = \left[ \hat{\pi} \left( \bm{x} , \eta \right) ,
  \hat{\pi} \left( \bm{x}^{\prime } , \eta \right) \right] = 0 \, .
\end{align}
\end{subequations}
Going to Fourier-space these relations are equivalent to 
\begin{subequations}
\begin{align}
\left[ \hat{\mu}_{\bm{k}} \left( \eta \right) ,
  \hat{\pi}_{\bm{k}^{\prime}} \left( \eta \right) \right] & = i \hbar
\delta \left( \bm{k} + \bm{k}^{\prime} \right) \, , \\ \left[
  \hat{\mu}_{\bm{k}} \left( \eta \right) , \hat{\mu}_{\bm{k}^{\prime}}
  \left( \eta \right) \right] & = \left[ \hat{\pi}_{\bm{k}} \left(
  \eta \right) , \hat{\pi}_{\bm{k}^{\prime}} \left( \eta \right)
  \right] = 0 \, ,
\end{align}
\end{subequations}
and the Hamiltonian reads
\begin{equation}
 \hat{H} =
 \int_{\mathbb{R}^{3+}} \mathrm{d}^3 \bm{k} \,\hat{H}_{\pm \bm{k}}
 = \int_{\mathbb{R}^{3+}} \mathrm{d}^3 \bm{k}
 \left[ \hat{\pi}_{\bm{k}} \hat{\pi}_{-\bm{k}} + \Hu\left(
   \hat{\pi}_{\bm{k}} \hat{\mu}_{-\bm{k}} + \hat{\mu}_{\bm{k}}
   \hat{\pi}_{-\bm{k}} \right) + k^2
   \hat{\mu}_{\bm{k}} \hat{\mu}_{-\bm{k}} \right],
\label{def:tilde_hamiltonian_mu_Fourier} 
\end{equation}
where $\hat{H}_{\pm \bm{k}}$ is the Hamiltonian for the $\pm \bm{k}$
sector. Observe that, as required by homogeneity, only the modes $\pm
\bm{k}$ are coupled and the coupling only depends on the norm $k$, as
required by isotropy. In order to expand $\hat{H}$ into a sum of
independent Hamiltonians $\hat{H}_{\pm \bm{k}}$ for the bi-modes $\pm
\bm{k}$, we restrict the integration to be over the top-half half of
the Fourier space, denoted by $\mathbb{R}^{3+}$, e.g. by selecting
only the vectors $\bm{k}$ with positive $k_z$ component, and dropping
the original global factor of a half.

Let us first analyse the evolution of one such pair of modes $\pm
\bm{k}$ in a situation where the term in $ \Hu$ can be neglected with
respect to the others, so that $\hat{\mu}$ is just a free scalar field
in Minkowski spacetime. With $\Hu\to 0$, the Hamiltonian $\hat{H}$ is
time-independent and we can introduce the usual creation/annihilation
operators for a real scalar field
\begin{subequations}
\begin{align}
\hat{\mu}_{\bm{k}} \left( \eta \right) & = \sqrt{ \frac{\hbar}{2k} }
\left[ \hat{a}_{\bm{k}} \left( \eta \right) + \hat{a}_{-
    \bm{k}}^{\dagger} \left( \eta \right) \right] \, ,
\\ \hat{\pi}_{\bm{k}} \left( \eta \right) & = - i \sqrt{\frac{\hbar
    k}{2}} \left[ \hat{a}_{\bm{k}} \left( \eta \right) - \hat{a}_{-
    \bm{k}}^{\dagger} \left( \eta \right) \right].
\end{align}
\label{def:minkowski_scalar_creation_operators}
\end{subequations}
The equal-time commutation relations assume the standard form
\begin{equation}
\left[ \hat{a}_{\bm{k}} , \hat{a}_{\bm{k}^{\prime}}^{\dagger}\right] =
\delta \left( \bm{k} - \bm{k}^{\prime} \right) \qquad \hbox{and}
\qquad \left[ \hat{a}_{\bm{k}} , \hat{a}_{\bm{k}^{\prime}}\right] =
\left[ \hat{a}_{\bm{k}}^{\dagger} ,
  \hat{a}_{\bm{k}^{\prime}}^{\dagger}\right] = 0.
\label{eq:commutation_relation_creation_operators}
\end{equation}
The Hamiltonian $\hat{H}_{\pm \bm{k}}$ then separates into two
harmonic oscillators of frequency $k$
\begin{equation}
\hat{H}_{\pm \bm{k}}^{(0)} = \hbar k \left(
\hat{a}^{\dagger}_{\bm{k}} \hat{a}_{\bm{k}} + \frac{1}{2} \right) + \hbar k
\left( \hat{a}^{\dagger}_{- \bm{k}} \hat{a}_{- \bm{k}} + \frac{1}{2}
\right) \, ,
\end{equation}
and the Heisenberg equations of motions 
\begin{equation}
\nonumber
i \hbar \frac{\D\hat{a}^{(\dagger)}_{\pm\bm{k}}}{\D\eta}
= \left[ \hat{a}^{(\dagger)}_{\pm \bm k} , \hat{H}_{\pm
    \bm{k}} \right]
\end{equation}
give $\hat{a}_{\bm{k}}\left( \eta \right) = \hat{a}_{\bm{k}} (0)
e^{-i k \eta }$. Including the friction term proportional to the
Hubble function $\Hu$, the Hamiltonian now reads
\begin{equation}
\hat{H}_{\pm \bm{k}} = \hbar k \left( \hat{a}^{\dagger}_{\bm{k}}
\hat{a}_{\bm{k}} + \frac{1}{2} \right) + \hbar k \left( \hat{a}^{\dagger}_{-
  \bm{k}} \hat{a}_{- \bm{k}} + \frac{1}{2} \right) - i \Hu \hbar
\left(\hat{a}_{- \bm{k}} \hat{a}_{ \bm{k}} - \hat{a}^{\dagger}_{-
  \bm{k}} \hat{a}^{\dagger}_{ \bm{k}} \right) \, .
\label{eq:tilde_hamiltonian_pm_k}
\end{equation}
The additional term corresponds to an interaction with a
time-dependent classical source, the expanding background, acting
through $\Hu$. It couples the $\pm \bm{k}$ modes by
creating/destroying pairs of particles with opposite momentum;
$\hat{a}_{ \bm{k}}$ is paired with $\hat{a}_{- \bm{k}}$ and similarly
for their hermitian conjugate. These terms are the only quadratic
interactions terms that respect homogeneity. The Heisenberg equations
of motions accordingly only mixes $\hat{a}_{ \bm{k}}$ with
$\hat{a}^{\dagger}_{ \bm{k}}$
\begin{align}
\label{eq:tilde_evol_crea}
\frac{\mathrm{d} }{\mathrm{d}  {\eta} } \begin{pmatrix}
\hat{a}_{\bm k} \, \\
\hat{a}^{\dagger}_{- \bm k}
\end{pmatrix}
 = \begin{pmatrix}
-i k & \Hu  \\ \\
\Hu & i k
\end{pmatrix} \begin{pmatrix}
\hat{a}_{\bm k} \, \\
\hat{a}^{\dagger}_{- \bm k}
\end{pmatrix} .
\end{align} 
The operators at any further time $\eta$ can then be expressed as a
linear combination of operators at an earlier time
$\eta_{\mathrm{in}}$
\begin{align}
\label{eq:time_evo_bogo_transfo}
\begin{pmatrix}
\hat{a}_{\bm k} ( \eta ) \, \\
\hat{a}^{\dagger}_{- \bm k} ( \eta )
\end{pmatrix} 
= \begin{pmatrix}
\alpha_{ k}( \eta ) & \beta_{ k} ( \eta ) \\
\beta^{*}_{ k} ( \eta ) & \alpha^{*}_{ k} ( \eta )
\end{pmatrix} \begin{pmatrix}
\hat{a}_{\bm k} \left(  \eta_{\mathrm{in}}\right) \, \\
\hat{a}^{\dagger}_{- \bm k} \left( \eta_{\mathrm{in}}\right)
\end{pmatrix} .
\end{align}
The system \eqref{eq:tilde_evol_crea} is equivalent to
\begin{align}
\label{eq:tilde_evol_bogo}
\frac{\mathrm{d} }{\mathrm{d}  {\eta} } \begin{pmatrix}
\alpha_{ k} \, \\
\beta_{ k}^{\star}
\end{pmatrix}
 = \begin{pmatrix}
-i k & \Hu  \\ \\
\Hu & i k
\end{pmatrix} \begin{pmatrix}
\alpha_{ k} \, \\
\beta_{ k}^{\star}
\end{pmatrix}  ,
\end{align} 
with $\alpha_{k} \left( \eta_{\mathrm{in}}\right) = 1$ and $\beta_{ k}
\left( \eta_{\mathrm{in}}\right) = 0$ as initial conditions.  One can
check that Eq.~\eqref{eq:tilde_evol_bogo} implies the quantity $
\left| \alpha_{ k} \right|^2 - \left| \beta_{ k} \right|^2$ is
conserved, while the commutation relations
\eqref{eq:commutation_relation_creation_operators} impose
\begin{equation}
\left| \alpha_{ k} \right|^2 - \left| \beta_{ k} \right|^2 = 1 \, .
\label{eq:bogo_normalisation}
\end{equation}
At any fixed $\eta$, a transformation like
\eqref{eq:time_evo_bogo_transfo} respecting the condition
\eqref{eq:bogo_normalisation} is called a Bogoliubov transformation
\cite{bogoljubovNewMethodTheory1958}.  Notice that the equations of
motion, and so the Bogoliubov coefficients, only depend on the norm
$k$.  The evolution of the quantum field has thus been reduced to
finding the coefficients of a time-dependent Bogoliubov
transformation. A convenient way to analyse this situation is to
introduce mode functions.

\subsection{Mode functions}

Having observed that the dynamics only mixes $\hat{a}_{\bm{k}}$ and
$\hat{a}_{-\bm{k}}^{\dagger}$, we have a basis on which to expand
$\hat{\mu}$. Inserting \eqref{eq:time_evo_bogo_transfo} in the Fourier
expansion of the field $\hat{\mu}$, we get
\begin{subequations}
\begin{align}
\hat{\mu}_{\bm{k}} (\eta) & = u_{k} \left( \eta \right)
\hat{a}_{\bm{k}} \left( \eta_{\mathrm{in}} \right) + u_{k}^{\star}
\left( \eta \right) \hat{a}_{- \bm{k}}^{\dagger} \left(
\eta_{\mathrm{in}} \right) \, , \\ \hat{\pi}_{\bm{k}} (\eta) & = U_{k}
\left( \eta \right) \hat{a}_{\bm{k}} \left( \eta_{\mathrm{in}} \right)
+ U_{k}^{\star} \left( \eta \right) \hat{a}_{- \bm{k}}^{\dagger}
\left( \eta_{\mathrm{in}} \right) \, ,
\end{align}
\label{eq:field_mode_function_initial_time_operators}
\end{subequations}
where $u_{k}$ and $U_{k}$ are defined by
\begin{subequations}
\begin{align}
 u_{k} \left( \eta \right) & = \frac{\alpha_{k} \left( \eta \right) +
   \beta_{k}^{\star} \left( \eta \right)}{\sqrt{2k}} \, , \\ U_{k}
 \left( \eta \right) & = - i \sqrt{\frac{k}{2}} \left[ \alpha_{k}
   \left( \eta \right) - \beta_{k}^{\star} \left( \eta \right) \right].
\end{align}
\label{def:mode_functions}
\end{subequations}
Using these functions, we get the so-called mode expansion of the
field $\hat{\mu}$
\begin{align}
\hat{\mu} \left( \bm{x} , \eta \right) & = \int \frac{\mathrm{d}^3
  \bm{k}}{(2 \pi)^{3/2}} \left[ e^{i \bm{k}.\bm{x}} u_{k} \left( \eta
  \right) \hat{a}_{\bm{k}}\left( \eta_{\mathrm{in}}\right) + e^{- i
    \bm{k}.\bm{x}} u_{k}^{\star} \left( \eta \right)
  a_{\bm{k}}^{\dagger} \left( \eta_{\mathrm{in}}\right) \right] \, ,
\label{def:mu_Fourier_mode_expansion}
\end{align}
and a similar expression for $\hat{\pi}$ with $U_{k}$ instead of
$u_k$. It can be checked from \eqref{eq:time_evo_bogo_transfo} that
$u_{k}$ simply obeys the same equation of motion \eqref{mode} as the
classical field $\mu \left( \bm{k}, \eta \right)$; the momentum mode
function $U_{k}$ is then determined by
\begin{align}
\label{eq:relation_alpha_plus_beta_alpha_minus_beta}
u_{k}^{\prime}  = \Hu u_{k} + U_{k} \, .
\end{align}
Finally, the conserved quantity $ \left| \alpha_{ k} \right|^2 -
\left| \beta_{ k} \right|^2$ maps to the Wronskian $W \left( u_{k},
u_{k}^{\star} \right)= u_{k}^{\star} u_{k}^{\prime} - u_{k}^{\star
  \prime} u_{k}$, which is a conserved quantity of \eqref{mode}, so
the condition \eqref{eq:bogo_normalisation} translates in the
normalisation
\begin{equation}
    W \left( u_{k}, u_{k}^{\star} \right) = - i \, .
\end{equation}
Any function $u_k$ solution of \eqref{mode} and which satisfies the
normalisation condition of the Wronskian is called a mode function.

We now have a dictionnary between the Bogoliubov and mode function
presentations. Solving the system \eqref{eq:tilde_evol_bogo} with
initial conditions $\alpha_{k} \left( \eta_{\mathrm{in}}\right) = 1$
and $\beta_{ k} \left( \eta_{\mathrm{in}}\right) = 0$ is equivalent to
solving \eqref{mode} for $u_k$ with initial conditions $u_k \left(
\eta _{\rm in} \right)=1/\sqrt{2k}$ and $u_k^{\prime} \left( \eta
_{\rm in} \right)=-i \sqrt{k/2} +\Hu \left( \eta _{\rm in} \right)$,
$U_{k}$ being determined by
Eq.~\eqref{eq:relation_alpha_plus_beta_alpha_minus_beta}.  Using mode
functions the quantum dynamics reduces to the classical one. This
justifies the classical treatment used in works cited in introduction
of this section; it is simply a consequence of working at linear order
and we will encounter other manifestations of this fact when studying
phase-space representation.

\subsection{Squeezed states}
\label{subsec:squeezed_states}

The time evolution was described so far in the Heisenberg picture. We
now show how to move to the Schrödinger picture and introduce the
squeezing formalism.  This formulation was initially proposed in
Ref.~\cite{grishchukSqueezedQuantumStates1990a} and we use conventions
matching those of \cite{martinQuantumDiscordCosmic2016b}. Without loss
of generality, the Bogoliubov coefficients
\eqref{eq:time_evo_bogo_transfo} can be parametrised using three real
coefficients $r_{k}$, $\varphi_k$ and $\theta_k$ through
\begin{subequations}
\begin{align}
    \alpha_{k}(\eta) & = e^{-i \theta_k(\eta)}
    \cosh \left[ r_k(\eta) \right] \, ,
    \\ \beta_{k}(\eta) & = - e^{i \left[ \theta_k(\eta)
    + 2 \varphi_k(\eta) \right]}
    \sinh \left[ r_k(\eta) \right],
\end{align}
\label{def:squeezing_parametrization}
\end{subequations}
where $r_k$ and $\varphi_k$ are respectively called the squeezing
parameter and angle, collectively referred to as the squeezing
parameters.  We define the 2-mode squeezing and the 2-mode rotation
operators by
\begin{subequations}
\begin{align}
\hat{S} \left( r_k , \varphi_k \right) & = \exp \left[
  \int_{\mathbb{R}^{3+} } \mathrm{d}^3 \bm{k} \left( r_k e^{- 2 i
    \varphi_k} \hat{a}_{\bm{k}} \hat{a}_{-\bm{k}} - r_k e^{2 i
    \varphi_k} \hat{a}^{\dagger}_{\bm{k}} \hat{a}^{\dagger}_{- \bm{k}}
  \right) \right], \\ \hat{R} \left( \theta_{k} \right) & = \exp
\left[ - i \int_{\mathbb{R}^{3+} } \mathrm{d}^3 \bm{k} \theta_k \left(
  \hat{a}^{\dagger}_{\bm{k}} \hat{a}_{\bm{k}} + \hat{a}^{\dagger}_{-
    \bm{k}} \hat{a}_{- \bm{k}} \right) \right],
\end{align}
\label{def:squeezing_and_rotation_operators}
\end{subequations}
in which the integrals are again, as in
\eqref{def:tilde_hamiltonian_mu_Fourier}, performed over half the
Fourier space and the creation and annihilation operators are
understood to be evaluated at $\eta_{\mathrm{in}}$.  The operators
$\hat{S}$ and $\hat{R}$ defined through
\eqref{def:squeezing_and_rotation_operators} are unitary and one can
check that
\begin{align}
\hat{a}^{(\dagger)}_{\pm \bm{k}}(\eta) & = \hat{R}^{\dagger} \left( 
\theta_{k} \right)
\hat{S}^{\dagger} \left( r_k , \varphi_k \right) 
\hat{a}^{(\dagger)}_{\pm \bm{k}} \left(
\eta_{\mathrm{in}} \right) \hat{S} \left(  r_k , \varphi_k \right) 
\hat{R} \left( \theta_{k} \right) \, ,
\label{def:evolution_as_queezing_and_rotation}
\end{align}
where the parameters are that of
Eq.~\eqref{def:squeezing_parametrization} and we have made their time
dependence implicit for display convenience.  The time evolution
equation~\eqref{eq:time_evo_bogo_transfo} is seen to correspond to the
application of a rotation of parameter $\theta_k \left( \eta \right)$
followed by a squeezing of parameters $r_k \left( \eta \right)$ and
$\varphi_k \left( \eta \right)$ on the operators.

Any operator $\hat{O} \left( \eta \right)$ in the Heisenberg picture
can be written as a combination of $\hat{a}^{(\dagger)}_{\pm \bm{k}}
\left( \eta \right)$ so we have
\begin{align}
\begin{split}
\left \langle \Psi \left( \eta_{\mathrm{in}} \right) \right| \hat{O}
\left( \eta \right) \left| \Psi \left( \eta_{\mathrm{in}} \right)
\right \rangle & = \left \langle \Psi \left( \eta_{\mathrm{in}}
\right) \right| \hat{R}^{\dagger} \hat{S}^{\dagger} \hat{O} \left(
\eta_{\mathrm{in}} \right) \hat{S} \hat{R} \left| \Psi \left(
\eta_{\mathrm{in}} \right) \right \rangle \, , \\ & = \left \langle
\Psi \left( \eta \right) \right| \hat{O} \left( \eta_{\mathrm{in}}
\right) \left| \Psi \left( \eta \right) \right \rangle \, .
\end{split}
\end{align}
where $\left| \Psi \left( \eta \right) \right \rangle = \hat{S}
\hat{R} \left| \Psi \left( \eta_{\mathrm{in}} \right) \right \rangle $
is the Schrödinger evolved state of the system. Choosing the waves to
be initially in the vacuum of $\hat{a}^{(\dagger)}_{\pm \bm{k}} \left(
\eta_{\mathrm{in}} \right)$ for all modes $\bm{k}$ (we return to this
point later) yields
\begin{equation}
    \left| \Psi \left( \eta \right) \right \rangle = {\displaystyle
      \prod_{\mathbb{R}^{3+}} } \hat{S} \left( r_k , \varphi_k \right)
    \hat{R} \left( \theta_{k} \right) \left| 0_{\bm{k} } , 0_{- \bm{k}
    }\right \rangle = {\displaystyle \prod_{\mathbb{R}^{3+}} }
    \left|\text{\MS}, r_{k} , \varphi_{k} \right \rangle \, ,
\end{equation}
where we have defined the 2-mode squeezed state (\MS) for the modes
$\pm \bm{k}$
\begin{equation}
\label{eq:squeezed_state_ket}
    \left|\text{\MS}, r_{k} , \varphi_{k} \right \rangle = \hat{S}
    \left( r_k , \varphi_k \right) \left| 0_{\bm{k} } , 0_{- \bm{k}
    }\right \rangle = \frac{1}{\cosh \left( 2 r_{\bm{k}} \right) }
    \sum_{n = 0}^{+\infty} \left( - \tanh{2 r_k} e^{2 i \varphi_k}
    \right)^n \left| n_{\bm{k} } , n_{-\bm{k}} \right \rangle \, .
\end{equation}
The last expression can be computed using a Baker-Campbell-Hausdorff
formula on the squeezing operator, now restricted to a single $\pm
\bm{k}$ sector \cite{schumakerNewFormalismTwophoton1985} and $\left|
n_{\bm{k} } , n_{-\bm{k}} \right \rangle$ is the state with $n$
particles in the mode $\bm{k}$ and $-\bm{k}$. Note that the rotation
angle $\theta_k$ has dropped from \eqref{eq:squeezed_state_ket}
because the vacuum is invariant under the rotation operator and the
product involved is over all directions.

Following \cite{polarskiSemiclassicalityDecoherenceCosmological1996b},
one can quickly derive the associated wavefunction of a single pair of
modes by assuming that, at the initial time, the corresponding state
is annihilated by both annihilation operators, i.e.,
\begin{equation}
    \hat{a}_{\pm \bm{k}} \left( \eta_{\mathrm{in}}
\right) \left|  0_{\bm{k} } , 0_{- \bm{k} }\right \rangle = 0 \, .
\end{equation}
Since $\hat{S}$ is unitary ($\hat{S}^\dagger \hat{S}
= \mathbb{1}$), this is also
\begin{align}
\begin{split}
0 & = \hat{S} \left( r_k , \varphi_k \right) \hat{a}_{\pm \bm{k}}
\hat{S}^{\dagger} \left( r_k , \varphi_k \right) \, \hat{S} \left( r_k
, \varphi_k \right) \left| 0_{\bm{k} } , 0_{- \bm{k} }\right \rangle
\, , \\ & = \hat{S} \left( r_k , \varphi_k \right) \hat{a}_{\pm
  \bm{k}} \hat{S}^{\dagger} \left( r_k , \varphi_k \right) \left|
\text{\MS} , r_{k} , \varphi_{k} \right \rangle,
\label{eq:annihilation_schrodinger_state}
\end{split}
\end{align}
where the transformation on the left corresponds to the inverse of
Eq.~\eqref{def:evolution_as_queezing_and_rotation} for $\theta_k =
0$. Inverting the Bogoliubov transformation
\eqref{eq:time_evo_bogo_transfo} and using
\eqref{def:minkowski_scalar_creation_operators}, the relation
\eqref{eq:annihilation_schrodinger_state} becomes
\begin{equation}
\label{eq:eom_wavefunction}
   \left[ \hat{\mu}_{\pm \bm{k}} + \frac{i}{k} \left( \frac{1 - i
       \gamma_{12} }{\gamma_{11} } \right)^{-1} \hat{\pi}_{\pm \bm{k}}
     \right] \left|\text{\MS} , r_{k} , \varphi_{k} \right \rangle = 0
   \,
\end{equation}
where, anticipating the next section, we have introduced
the matrix entries
\begin{subequations}
\begin{align}
\gamma_{11} & = \cosh \left( 2 r_k \right) - \cos \left( 2 \varphi_k \right)
\sinh \left( 2 r_k \right) \, , \\ 
\gamma_{12} & = - \sin \left( 2 \varphi_k \right) \sinh \left( 2 r_k \right).
\end{align}
\end{subequations}

Projecting Eq.\eqref{eq:eom_wavefunction} onto the $\mu_{\pm
  \bm{k}}$-representation of the wavefunction\footnote{Formally, the
  wavefunction is the projection of the relevant state on the basis
  $\{ \mu_{\pm\bm{k}}\}$, i.e.  $\Psi(\mu_{\bm{k}},\mu_{-\bm{k}} ) =
  \langle \mu_{\pm\bm{k}}| \text{\MS} , r_{k} , \varphi_{k}\rangle$.} by
setting $\hat{\mu}_{\pm \bm{k}} \to \mu_{\pm \bm{k}} $ and
$\hat{\pi}_{\pm \bm{k}} \to - i \hbar \partial/\partial \mu_{\mp
  \bm{k}}$. The wavefunction solution of
Eq.~\eqref{eq:eom_wavefunction} reads
\begin{equation}
\label{eq:wavefunction_two_mode_squeezed_state}
    \Psi \left( \mu_{\bm{k}} , \mu_{- \bm{k}} \right) = \sqrt{
      \frac{k}{\pi \hbar \gamma_{11}}} e^{- \frac{k}{\hbar}
      \frac{\left( 1 - i \gamma_{12}\right)}{\gamma_{11}} \mu_{\bm{k}}
      \mu_{-\bm{k}} },
\end{equation}
which we normalised, using \eqref{mustar},
to $\int |\Psi|^2 \D \mu_{\bm{k}} \D \mu_{-\bm{k}} =1$.

When the squeezing parameters are those determined by
Eqs.~\eqref{def:squeezing_parametrization}, this gives the
wavefunction of any $\pm \bm{k}$ mode of the gravitational waves. One
can also provide a description in terms of the squeezed state
parameters only by recasting \eqref{eq:time_evo_bogo_transfo} into a
set of differential equations involving only $r_{k}$, $\varphi_k$ and
$\theta_k$. One finds that the system
\begin{subequations}
\begin{align}
\frac{ \D r_{k} }{\D \eta } & = - \Hu \cos \left( 2
\varphi_{k} \right),
\label{dotrk}\\
\frac{ \D \varphi_{k} }{\D \eta } & = - k + \Hu \coth \left( 2 r_k
\right) \sin \left( 2 \varphi_{k} \right),
\label{dotphik}\\
 \frac{ \mathrm{d} \theta_{k} }{\mathrm{d} \eta} & = k - \Hu \tanh
 \left( r_k \right) \sin \left( 2 \varphi_k \right),
\label{dotthetak}
\end{align}
\label{eq:eom_squeezing_parameters}
\end{subequations}
should indeed hold.  Note that the equations describing the time
evolution of the squeezing parameters $r_{k}$ and $\varphi_{k}$,
namely \eqref{dotrk} and \eqref{dotphik}, are independent of
$\theta_k$.  These equations are however rarely solved directly, as it
is easier to first solve Eq.~\eqref{mode} for the mode function, then
deduce the Bogoliubov coefficients by inverting
\eqref{def:mode_functions} and finally, using
\eqref{def:squeezing_parametrization}, obtain the expression of the
squeezing parameters.  The virtue of the squeezing formalism is rather
to give a clear phase space representation of the system's evolution.
Such representation can be obtained using the Wigner quasi-probability
distribution \cite{caseWignerFunctionsWeyl2008} to which we now turn.

\subsection{Wigner function}
\label{subsec:wigner_function}

Consider a system described by a density matrix $\hat{\rho}$ and
represented by $n$-pairs of canonically conjugate hermitian operators
$\hat{X} = \left\{ \left( \hat{q}_i , \hat{p}_i \right) \right\}_{i
  \in \left[ 1 , n \right] }$ of the same dimension. The Wigner
function is a function of $2n$ phase space variables $X = \left\{
\left( q_i , p_i \right) \right\}_{i \in \left[ 1 , n \right] }$
defined by
\begin{equation}
    W \left( X \right) = \frac{1}{\left( 2 \pi \hbar \right)^n} \int
    \, \mathrm{d}^{n} \vec{x} \, e^{-i \frac{\vec{p}.\vec{x}}{\hbar} }
    \left \langle \vec{q} + \frac{ \vec{x} }{2} \right| \hat{\rho}
    \left| \vec{q} - \frac{ \vec{x} }{2} \right \rangle \, ,
\label{def:wigner_general}
\end{equation}
where the states entering the averaging are product eigenstates of
$\hat{q}_{i}$. The right hand side of \eqref{def:wigner_general} is
the \textit{Weyl transform} of $\hat{\rho}_{\bm{k} } / \left( 2 \pi
\right)^n$. This transform maps any observable $\hat{O}$, which is a
function of operators in $\hat{X}$, to a function $\tilde{O} \left( X
\right)$ of the associated classical variables $X$.  A crucial
property is that the expectation value of any such observable
$\hat{O}$ can be computed by treating the Wigner function as a
probability measure for the Weyl transform
\begin{equation}
\left \langle \hat{O} \right \rangle
= \mathbb{E} 
\left[ \tilde{O}
\left( X \right)
\right]
= \int W \left( X \right)
\tilde{O} \left( X \right) \mathcal{D} X,
\label{def:stochastic_average}
\end{equation}
where the integral is over all the relevant variables in $X$ and we
denoted $\mathbb{E}$ the stochastic average with respect to the Wigner
funcion. Equation \eqref{def:stochastic_average} then allows to
compute averages using the Wigner function as any classical
phase-space probability distribution. Finally, the von-Neumann
equation of motion for the density matrix can be mapped into an
equation of motion for the Wigner function,
namely~\cite{curtrightConciseTreatiseQuantum2014}
\begin{equation}
    i \hbar \dot{W} \left( X \right) =  H \left(
      \vec{q} , \vec{p} \right) \star W - W \star H \left( \vec{q} ,
      \vec{p} \right)  \, ,
\label{eq:eom_wigner_function} 
\end{equation}
where the non-commutative $\star$-product is defined by
\begin{subequations}
\begin{align}
    f \left( \vec{q} , \vec{p} \right) \star g \left( \vec{q} ,
    \vec{p} \right) & = f \left( \vec{q} + \frac{i \hbar}{2}
    \partial_{\vec{p}}, \vec{p} - \frac{i \hbar}{2} \partial_{\vec{q}}
    \right) g \left( \vec{q} , \vec{p} \right) \, , \\ & = f \left(
    \vec{q} , \vec{p} \right) g \left( \vec{q} - \frac{i \hbar}{2}
    \partial_{\vec{p}},\vec{p} + \frac{i \hbar}{2} \partial_{\vec{q}}
    \right).
\end{align}
\end{subequations}
The Wigner function therefore furnishes a complete representation of
the state of the system and its evolution in phase space.

Two remarks are in order here. First, in general, the Wigner function
is \textit{not} everywhere positive making it only a
\textit{quasi}-probability distribution.  It can be shown that, for pure states, it is
everywhere positive only when it takes the form
\cite{hudsonWhenWignerQuasiprobability1974}
\begin{equation}
    W(X)=\frac{1}{\left( \pi \hbar \right)^n \sqrt{\det \gamma}}
    \exp\left(- \frac{X^\mathrm{T}\gamma^{-1} X}{\hbar} \right) \, ,
\label{def:gaussian_wigner_function}    
\end{equation}
which is completely determined by $\gamma$, the covariance matrix,
defined by
\begin{equation}
\gamma_{ab}= \langle \hat{X}_a \hat{X}_b +\hat{X}_a \hat{X}_b \rangle.
\label{def:covariance_matrix}  
\end{equation}
Such states are called Gaussian states and are widely used in quantum
optics, see \cite{adessoContinuousVariableQuantum2014a} for a
review. Second, for evolution under a quadratic Hamiltonian $H \left(
\hat{X} \right)$, the dynamics \eqref{eq:eom_wigner_function} simply
reduces to the classical Liouville equation~\footnote{For a detailed
  derivation in the special case of cosmological perturbations see
  Appendix H of \cite{martinQuantumDiscordCosmic2016b}.}
\begin{equation}
\label{eq:liouville_wigner}
    \dot{W} \left( X \right) = \left\{ H \left( X \right) , W \left( X
    \right) \right\},
\end{equation}
the curly brackets denoting the usual classical Poisson brackets.

Equation~\eqref{eq:liouville_wigner} can be solved by the method of
characteristics i.e. by evolving the initial distribution along the
classical trajectories given by $H$.  This is another manifestation of
the fact that, at quadratic order, the quantum dynamics reduces to the
classical one. In addition, this implies that an initially Gaussian
state will remain Gaussian under a quadratic Hamiltonian and that its
evolution is thus summarised in that of its covariance matrix
$\gamma$.

Both of the above discussed simplifications apply to cosmological
perturbations at linear order, to which we return by considering a
pair of modes $\pm \bm{k}$.  These two degrees of freedom represented
by the four operators $ \hat{\mu}_{\pm \bm{k}} $ and $ \hat{\pi}_{\pm
  \bm{k}} $.  These four operators are \textit{not} hermitian and
related to one another by hermitian conjugation. We can however build
two such pairs of operators by taking the real and imaginary parts of
$ \hat{\mu}_{ \pm \bm{k}} $ and $ \hat{\pi}_{\pm \bm{k}}$ (up to a
factor of $\sqrt{2}$, introduced for further convenience), namely
\begin{subequations}
\begin{align}
\hat{\mu}_{\bm{k}}^{\textsc{r}} & =
\frac{\hat{\mu}_{\bm{k}}+\hat{\mu}_{\bm{k}}^\dagger}{\sqrt{2}}, \qquad
\hat{\mu}_{\bm{k}}^{\textsc{i}} =
\frac{\hat{\mu}_{\bm{k}}-\hat{\mu}_{\bm{k}}^\dagger}{\sqrt{2} i}
\\ \hat{\pi}_{\bm{k}}^{\textsc{r}} & =
\frac{\hat{\pi}_{\bm{k}}+\hat{\pi}_{\bm{k}}^\dagger}{\sqrt{2}}, \qquad
\hat{\pi}_{\bm{k}}^{\textsc{i}} =
\frac{\hat{\pi}_{\bm{k}}-\hat{\pi}_{\bm{k}}^\dagger}{\sqrt{2} i}.
\end{align}
\label{def:vRI}
\end{subequations}
One can straightforwardly check that those are indeed Hermitian and
canonically conjugate
i.e. $[\hat{\mu}_{\bm{k}}^{\textsc{s}},\hat{\pi}_{\bm{k}'}^{\textsc{s}^{\prime}}]
=i\delta(\bm{k}-\bm{k}') \delta_{\textsc{s},\textsc{s}^{\prime}}$ and
$[\hat{\mu}_{\bm{k}}^{\textsc{s}},\hat{\mu}_{\bm{k}'}^{\textsc{s}^{\prime}}]
=[\hat{\pi}_{\bm{k}}^{\textsc{s}},\hat{\pi}_{\bm{k}'}^{\textsc{s}^{\prime}}]=0$
where $\textsc{s}=\textsc{r}, \textsc{i}$. We arrange them in the
vector $\hat{X}_{\textsc{R/I}} = \left( k^{1/2} \hat{\mu}^{\textsc{r}
}_{\pm\bm{k}}, k^{-1/2} \hat{\pi}^{\textsc{r} }_{-\bm{k}} ,k^{1/2}
\hat{\mu}^{\textsc{I} }_{-\bm{k}}, k^{-1/2} \hat{\pi}^{\textsc{r}
}_{-\bm{k}} \right)$, where we have introduced factors of $k$ to give
the same dimension to all entries in the vector, whose associate
vector of classical variables is denoted $X_{\textsc{R/I}}$.  The
Wigner function with respect to these variables is defined by
\begin{equation}
W_{\pm \bm{k}} \left( X_{\pm \bm{k}} \right) = \frac{1}{\left( 2 \pi
  \hbar\right)^2} \int e^{-\frac{i}{\hbar} \left(
  \pi_{\bm{k}}^{\textsc{r} } x + \pi_{\bm{k}}^{\textsc{i}} y \right)}
\left \langle \mu_{\bm{k}}^{ \textsc{r}} + \frac{x}{2} ,
\mu_{\bm{k}}^{\textsc{i}} + \frac{y}{2} \right| \hat{\rho}_{\bm{k} }
\left|\mu_{\bm{k}}^{\textsc{r}} - \frac{x}{2} ,
\mu_{\bm{k}}^{\textsc{i}} - \frac{y}{2} \right \rangle \D x \D y.
\label{def:wigner_RI}
\end{equation}

In terms of the variables \eqref{def:vRI}, the Hamiltonian
$\hat{H}_{\pm \bm{k}}$ separates into two equal Hamiltonian over the
$\textsc{r}/\textsc{i}$ sectors that thus evolve independently
\begin{align}
\label{eq:Hamiltonian:RI}
\hat{H} & =\frac{\hbar}{2} \int_{\mathbb{R}^{3+}} \mathrm{d}^3 \bm
k\sum_{\textsc{s}=\mathrm{R,I}} \left[(\hat{\pi}_{\bm k}^{\textsc{s}})^2 + 2 \Hu \left(
  \hat{\mu}_{\bm{k}}^{\textsc{s}} \hat{\pi}_{\bm k}^{\textsc{s}} + \hat{\pi}_{\bm k}^{\textsc{s}}
  \hat{\mu}_{\bm{k}}^{\textsc{s}} \right) + k^2(\hat{\mu}_{\bm{k}}^{\textsc{s}})^2\right]
=\int_{\mathbb{R}^{3+}} \mathrm{d}^3 \bm k \sum_{\textsc{s}=\mathrm{R,I}}
\hat{H}_{\bm k}^{\textsc{s}} \, .
\end{align}
Similarly, the wavefunction
\eqref{eq:wavefunction_two_mode_squeezed_state} factorises into a
product of two wavefunctions over each sector $\Psi \left(
\mu_{\bm{k}} , \mu_{- \bm{k}} \right) = \Psi \left(
\mu_{\bm{k}}^{\textsc{r}} \right) \Psi \left(
\mu_{\bm{k}}^{\textsc{i}} \right)$ with
\begin{equation}
\label{eq:wavefunction_one_mode_squeezed_state}
\Psi \left( \mu_{\bm{k}}^{\textsc{s}} \right) = \left( \frac{k}{ \pi
  \hbar \gamma_{11}} \right)^{1/4} e^{- \frac{k}{2 \hbar} \frac{\left(
    1 - i \gamma_{12} \right)}{\gamma_{11}} \left(
  \mu_{\bm{k}}^{\textsc{s}} \right)^2},
\end{equation}
and the covariance matrix is block diagonal in the $\textsc{r} /
\textsc{i}$ partition $\gamma = \gamma^{\textsc{r}} \bigoplus
\gamma^{\textsc{i}}$. These separations are in fact imposed by the
homogeneity of the state that requires $\left \langle \hat{a}_{\bm{k}}
\hat{a}^{\dagger}_{\bm{k}} \right \rangle = \left \langle
\hat{a}^2_{\bm{k}} \right \rangle = 0$, which can be recast in the
vanishing of all $\textsc{r} / \textsc{i}$ cross terms
\cite{campoInflationarySpectraPartially2005a}.
Eq.~\eqref{eq:wavefunction_one_mode_squeezed_state} is nothing else
than the wavefunction of a \text{one}-mode squeezed state of parameter
$r_k, \varphi_k$ \cite{albrechtInflationSqueezedQuantum1994}. Going
from the $\pm \bm{k}$ operators to the $\textsc{r}/\textsc{i}$
operators allows to view a 2-mode squeezed state as a product of two
1-mode squeezed states~\footnote{This fact can be directly seen by
  factorizing the 2-mode squeezing operator $\hat{S} \left( r_k ,
  \varphi_k \right)$ into to two 1-mode squeezing operators for the
  $\textsc{r}/\textsc{i}$ creation/annihilation operators defined via
  \eqref{def:minkowski_scalar_creation_operators} where $\pm \bm{k}$
  operators are replaced by $\textsc{r}/\textsc{i}$ operators.}. Such
transformations are studied in details in
\cite{martinDiscordDecoherence2022}.

Since the wavefunction \eqref{eq:wavefunction_one_mode_squeezed_state}
is Gaussian, then so is the associated Wigner function
$W^{\textsc{s}}$; vacuua and squeezed states are indeed Gaussian
states. Note that their gaussianity is preserved by the evolution
because $\hat{H}_{\bm k}^{\textsc{s}}$ is quadratic. The Wigner
function \eqref{def:wigner_RI} also factorises into $W_{\pm \bm{k}} =
W^\textsc{r} \left( \hat{\mu}_{\bm k}^\textsc{r} , \hat{\pi}_{\bm
  k}^\textsc{r} \right) W^\textsc{i} \left( \hat{\mu}_{\bm
  k}^\textsc{i}, \hat{\pi}_{\bm k}^\textsc{i} \right)$. Both sectors
have identical covariance matrix, namely
\begin{equation}
\label{def:gamma}
\gamma^{\textsc{s}} = \begin{pmatrix}
    \gamma_{11} & \gamma _{12}  \\
    \gamma_{21} & \gamma_{22}
  \end{pmatrix}
 \, ,
\end{equation}
with
\begin{subequations}
\label{eq:gammaij}
\begin{align}
\gamma_{11}&=2k\left\langle \left(\hat{\mu}_{\bm
  k}^\textsc{r}\right)^2\right \rangle =2k\left \langle
\left(\hat{\mu}_{\bm k}^\textsc{i}\right)^2 \right \rangle
=k\left\langle \left\lbrace \hat{\mu}_{\bm{k}},
\hat{\mu}_{\bm{k}}^\dagger \right\rbrace\right\rangle,
\label{eq:gamma11}
\\ \gamma_{12} & = \gamma_{21} =\left\langle \hat{\mu}_{\bm
  k}^\textsc{r}\hat{\pi}_{\bm k}^\textsc{r}+\hat{\pi}_{\bm
  k}^\textsc{r}\hat{\mu}_{\bm k}^\textsc{r}\right\rangle =\left\langle
\hat{\mu}_{\bm k}^\textsc{i}\hat{\pi}_{\bm k}^\textsc{i}+
\hat{\pi}_{\bm k}^\textsc{i}\hat{\mu}_{\bm k}^\textsc{i}\right\rangle
=\left\langle \hat{\mu}_{\bm{k}} \hat{\pi}_{\bm{k}}^\dagger +
\hat{\pi}_{\bm{k}} \hat{\mu}_{\bm{k}}^\dagger \right\rangle,
\label{eq:gamma12}
\\ \gamma_{22}& =\frac{2}{k} \left \langle \left(\hat{\pi}_{\bm
  k}^\textsc{r}\right)^2\right\rangle = \frac{2}{k}\left \langle
\left(\hat{\pi}_{\bm k}^\textsc{i}\right)^2\right\rangle
=\frac{1}{k}\left\langle \left\lbrace \hat{\pi}_{\bm{k}},
\hat{\pi}_{\bm{k}}^\dagger \right\rbrace\right\rangle,
\label{eq:gamma22}
\end{align}
\end{subequations} 
where we expressed the entries of the covariance matrix in terms of
two-point function of the original $\hat{\mu}_{\bm{k}}$ and
$\hat{\pi}_{\bm{k}}$ operators (one can also check that $\langle
\hat{\mu}_{\bm{k}} \hat{\pi}_{\bm{k}} + \hat{\pi}_{\bm{k}}^\dagger
\hat{\mu}_{\bm{k}}^\dagger \rangle = 0$). Using \eqref{eq:gammaij} and
the parametrisation \eqref{def:squeezing_parametrization}, the
covariance matrix can be conveniently expressed in terms of the
squeezing parameters
\begin{subequations}
\label{eq:covariance_matrix_squeezed_state}
\begin{align}
\gamma_{11} & = \cosh \left( 2 r_k \right) - \cos \left( 2 \varphi_k
\right) \sinh \left( 2 r_k \right) \, , \\ \gamma_{12} & = \gamma_{21}
= - \sin \left( 2 \varphi_k \right) \sinh \left( 2 r_k\right) \, ,
\\ \gamma_{22} & = \cosh \left( 2 r_k \right) + \cos \left( 2
\varphi_k \right) \sinh \left( 2 r_k \right),
\end{align}
\end{subequations} 
where the expressions for $\gamma_{11}$ and $\gamma_{12}$ correspond
to those defined earlier when computing the wavefunction.  Finally in
order to visualize this probability distribution, we compute its
contour levels. Owing to gaussianity, those are ellipses whose
parameters can be computed through diagonalizing the quadratic form
appearing in the argument of the exponential in
Eq.~\eqref{def:gaussian_wigner_function}. It is readily done by
performing a rotation in phase space $\widetilde{X}^{\textsc{s}} =
R(-\varphi_k) X^{\textsc{s}}$ so that the covariance matrix of
$X^{\textsc{s}}$ reads
\begin{align}
\label{eq:gamma_diagonalized}
(\widetilde{\gamma^{\textsc{s}}})^{-1}=
\begin{pmatrix}
e^{2 r_k} & 0\\
0 & e^{-2 r_k}
\end{pmatrix} .
\end{align}

\begin{figure}
\centering
\includegraphics[width=0.7\textwidth]{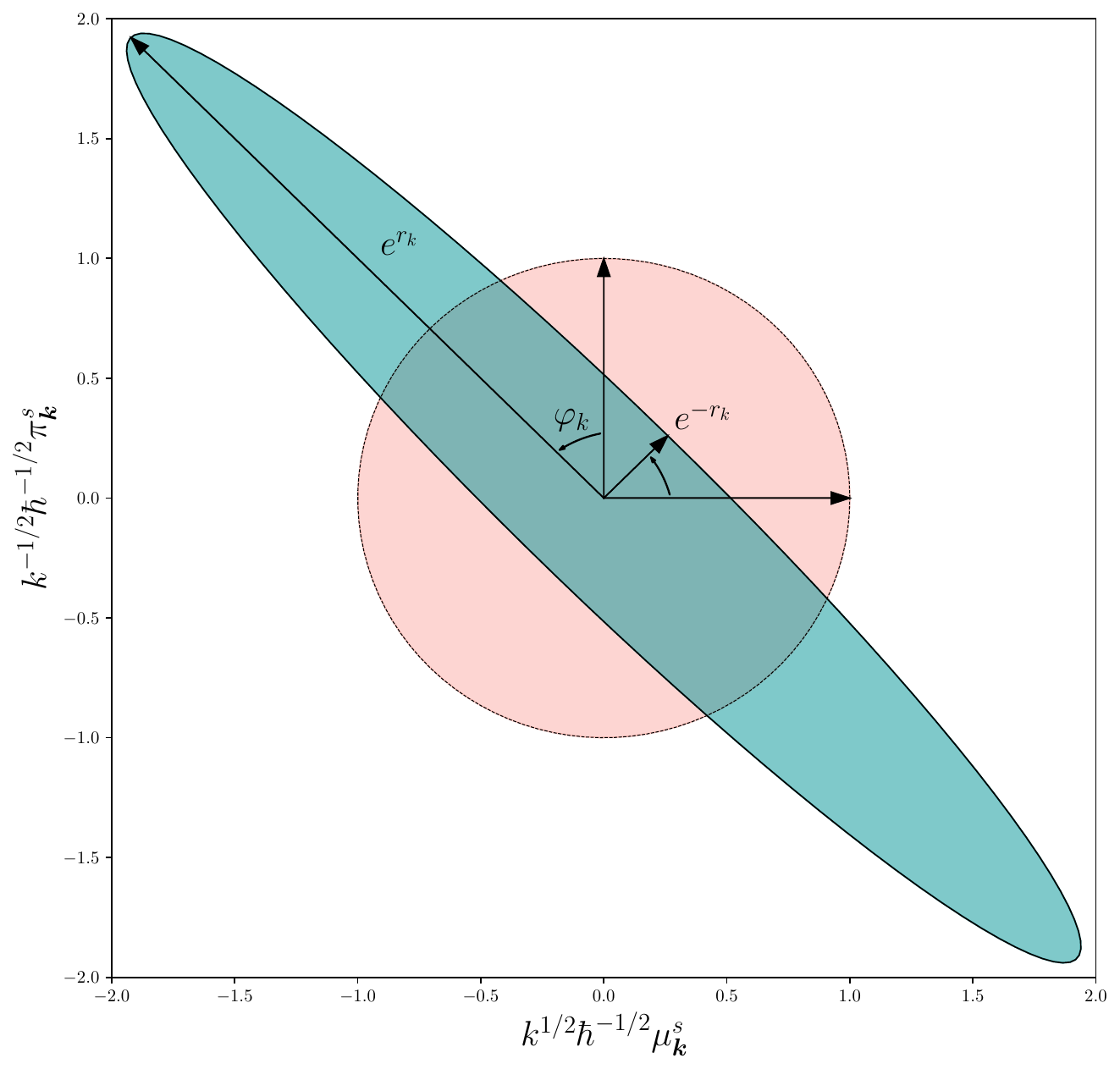}
\caption{$\sqrt{2}$-$\sigma$ contour level of the Wigner function
  $W^{\textsc{s}}$ for $\varphi_k = \pi / 4 $, $r_k = 1$ (green
  ellipse) and the vacuum state $r_k = 0$ (pink circle). This figure
  is adapted from \cite{martinDiscordDecoherence2022}.}
\label{fig:wigner_function_squeezing_parameters}
\end{figure}

Some contour levels of $W^{\textsc{s}}$ are plotted in
Fig.~\ref{fig:wigner_function_squeezing_parameters}; they provide a
geometrical representation of the state of the tensor perturbations in
phase space and illustrate the meaning of the squeezing parameters:
the ellipse representing the $\sqrt{2}$-$\sigma$ contour has
semi-minor and semi-major axes of length $A_k =\sqrt{\hbar} e^{r_k}$
and $B_k =\sqrt{\hbar} e^{-r_k}$, which are tilted by the angle
$\varphi_k$ in phase space. The fluctuations of the operator in the
direction of the semi-major axis are exponentially amplified with
respect to the vacuum; this is called a super-fluctuant mode. On the
other hand, the fluctuations of the operator related to the semi-major
axis are exponentially suppressed, thus defining a sub-fluctuant mode.

The presence of amplification and suppression is a manifestation of
the existence of a growing and a decaying solution in Eq.~\eqref{mode}
\cite{albrechtInflationSqueezedQuantum1994}.  Their complementary can
be traced back to the purity of the state which, for a Gaussian state,
can be computed directly in terms of the covariance matrix
via~\cite{adessoContinuousVariableQuantum2014a}
\begin{align}
\label{eq:purity_Gaussian_States}
p_k = \mathrm{tr} \left( \hat{\rho}^2 \right) = \frac{1}{\sqrt{\det
    \left(\gamma \right)}} = \frac{1}{\gamma_{11} \gamma_{22} -
  \gamma_{12}^2 } = \frac{\hbar^2}{A_k^2 B_k^2} = \frac{\hbar^2 \pi^2}{S_k^2}   \, ,
\end{align}
where $S_k$ is the area of the $\sqrt{2}$-$\sigma$ contour defined by
the points where the argument of the exponential in
Eq~\eqref{def:gaussian_wigner_function} is unity.  Since the purity of
the state is preserved under Hamiltonian evolution, so is
$S_k$. Therefore, the amplification in a given direction has to be
balanced out with squeezing in another. Conversely, if the
fluctuations in one direction are reduced, they increase in
another. For any quantum state, $p_k \leq 1$ and so the area is
minimal for a pure state $p_k =1$, like the one we consider here,
where $S_k = \pi \hbar$; this is a geometrical translation of the
Heisenberg uncertainty principle forbidding to localise the system too
precisely. Note that in general, due to the rotation $\varphi_k$, the
product uncertainty of the original pair $\left(\hat{\mu}_{\bm{k}} ,
\hat{\pi}_{\bm{k}} \right)$ does not saturate the inequality anymore.

In addition to granting an elegant geometrical representation of the
state, the presentation in terms of 2-mode squeezed states is often
used in the literature to discuss the quantumness of primordial
gravitational waves and scalar perturbations alike. These aspects are
discussed in Sec.~\ref{sec:quantumness_primordial_GW}.

\subsection{Particle production}
\label{subsec:graviton_production}

Having laid out the formalisms to follow the evolution of
gravitational waves in cosmology, we want to give more physical
insights into the evolution and show that, under certain conditions,
it can be understood as a process of particle creation. Bogoliubov
transformations and mode functions are the appropriate way to describe
this process in curved spacetime. We start by analysing their relation
to particle content.
 
Consider two pairs of operators $\left( \hat{a} ,
\hat{a}^{\dagger}\right)$ and $\left( \hat{b} ,
\hat{b}^{\dagger}\right)$ related by a constant Bogoliubov
transformation
\begin{equation}
 \hat{b} =  \alpha \hat{a} + \beta \hat{a}^{\dagger} \, ,
 \label{eq:bogo_transfo_operators}
\end{equation}
with $\left( \alpha , \beta \right) \in \mathbb{C}^2$ such that $
\left| \alpha \right|^2 - \left| \beta \right|^2 = 1$. We define two
vacua: $\left| 0 \right \rangle_{a}$ with respect to the $\hat{a}$
operators and $\left| 0 \right \rangle_{b}$ with respect to the
$\hat{b}$ operators. The crucial observation is that these vacua do
\textit{not} coincide. The number of $b$-particles in the $a$-vacuum
is always non-vanishing when the Bogoliubov transformation is
non-trivial
\begin{equation}
{}_{a} \left \langle 0 \right| \hat{b}^{\dagger} \hat{b} \left| 0
\right \rangle_{a} = \left| \beta \right|^2 > 0 \, .
\end{equation}
The analysis carries over to the study of $\pm \bm{k}$
modes. Equation~\eqref{eq:squeezed_state_ket} shows that the vacuum of
the operators $\hat{a}_{\pm \bm{k}}(\eta)$ is filled with particles
associated to $\hat{a}_{\pm \bm{k}}(\eta_{\rm in})$. We thus already
see that the number of particles will be different for the same state
using the operators $\hat{a}_{\pm \bm{k}} \left( \eta \right)$ of
Eq.~\eqref{def:minkowski_scalar_creation_operators} at two different
times.

What are then the appropriate operators to describe the particle
content of the field $\hat{\mu}$ and define a vacuum as we have in
Sec.~\ref{subsec:squeezed_states}?  We have so far considered
operators defined by \eqref{def:minkowski_scalar_creation_operators}.
In Minkowski spacetime ($a^{\prime}=0$), this form is uniquely
selected (up to a phase) by requiring that the Hamiltonian is diagonal
and that the vacuum thus defined is invariant under the Poincaré group
so that it is shared by all inertial observers, or, equivalently, the
vacuum is the ground state of the Hamiltonian
\cite{mukhanovIntroductionQuantumEffects2007}.  In this situation
there \textit{is} a preferred set of operators selected by physical
symmetries, which subsequently define preferred notions of vacuum and
particle.

The procedure described above breaks down in an expanding Universe,
$a^{\prime} \neq 0$, as the Poincaré group is no longer a symmetry of
spacetime, $\omega_k^2$ [see
  Eq.~\eqref{def:tilde_hamiltonian_mu_Fourier}] is time-dependent and
can even become negative so that the existence of an energy minimum is
not guaranteed anymore.  We are left with no physically preferred
vacuum in which no inertial detector would record the presence of
particles.  In this context, the choice of $\hat{a}_{\bm{k}} \left(
\eta_{\mathrm{in}} \right)$ to perform the expansion
\eqref{def:mu_Fourier_mode_expansion} appears arbitrary.

A choice of operators in fact corresponds to a choice of mode
functions, the latter being more convenient to work with. Consider the
operators $ \hat{b}_{\pm \bm{k}}$ related to $\hat{a}_{\pm \bm{k}}
\left( \eta_{\mathrm{in}} \right)$ by the following
\textit{time-independent} Bogoliubov transformation
\begin{equation}
	\hat{b}_{\bm{k}} = \rho_{k}^{\star} \hat{a}_{\bm{k}} \left(
        \eta_{\mathrm{in}} \right) + \chi_{k} \hat{a}_{-
          \bm{k}}^{\dagger} \left( \eta_{\mathrm{in}} \right) \, ,
	\label{eq:bogo_operators_fixed_time}
\end{equation}
with $\left( \rho_{k} , \chi_{k} \right) \in \mathbb{C}^2$ such that $
\left| \rho_{k} \right|^2 - \left| \chi_{k} \right|^2 = 1$. Inverting
this transformation and inserting in
\eqref{def:mu_Fourier_mode_expansion}, we get
\begin{equation*}
	\hat{\mu} \left( \bm{x} , \eta \right) = \int
        \frac{\mathrm{d}^3 \bm{k}}{(2 \pi)^{3/2}} \left[ e^{i
            \bm{k}.\bm{x}} v_{k} \left( \eta \right) \hat{b}_{\bm{k}}
          + e^{- i \bm{k}.\bm{x}} v_{k}^{\star} \left( \eta \right)
          \hat{b}_{ \bm{k}}^{\dagger} \right],
\end{equation*}
where
\begin{equation}
 v_{k} = \rho_{k}^{\star} u_{k} - \chi_{k} u_{k}^{\star}
\label{eq:bogo_mode_functions}
\end{equation}
can be checked to be a mode function, i.e. a solution of \eqref{mode}
with a Wronskian normalised to $W(v_{k}, v_k^{\star})=\left( \left|
\rho_{k} \right|^2 - \left| \chi_{k} \right|^2\right) W(u_{k},
u_{k}^{\star}) = -i$. A similar expansion is found for $\hat{\pi}$,
with $v_k$ replaced by new functions $V_{k}$ defined as the $U_k$s
through the replacement $u_k\to v_k$.

We then have an alternative expansion of $\hat{\mu}$ and $\hat{\pi}$
over another set of mode functions and operators. The meaning of the
operators in the expansion is set once the associated mode functions
are fixed~\footnote{This can be seen by expressing $\hat{b}_{\bm{k}}$
  in terms of the mode function and the fields $\hat{b}_{\bm{k}} =
  -i\left[ V_{k}^{\star} \hat{\mu} \left(\bm{k},\eta \right) -
    v_{k}^{\star} \hat{\pi} \left( \bm{k},\eta \right)\right]$.}, and
a choice of mode functions corresponds to a choice of initial
conditions for the solutions of \eqref{mode}. The normalisation of the
Wronskian fixes one condition, and one is left to choose.  For
instance, the Minkowski operators
\eqref{def:minkowski_scalar_creation_operators} are associated to the
mode function
\begin{equation}
	u_{k}^{\textsc{(m)}} \left( \eta \right) = \frac{e^{- i k
            \eta}}{\sqrt{2k}} ,
	\label{def:Minkowski_mode_function}
\end{equation}
corresponding to the initial conditions
\begin{equation}
  u_{k} \left( \eta_0 \right) = e^{- i k \eta_0} \quad \text{and}
  \quad u_{k}^{\prime} \left( \eta_0 \right) = i k \eta_0 e^{- i k
    \eta_0} .
\label{eq:minkowski_initial_conditions_mode_function}
\end{equation}

For non-vanishing $\Hu$, $u_{k}^{\textsc{(m)}}$ is no longer a
solution of \eqref{mode}.  Yet, when analysing the evolution of the
two helicities of the gravitational field in this context, we have
used the associated operators
\eqref{def:minkowski_scalar_creation_operators}.  Their
time-dependence then does not simply factorise in the running phase of
$u_{k}^{\textsc{(m)}}$ and we have to deal with a continuous change of
reference operators parametrised by a \textit{time-dependent}
Bogoliubov transformation.  These operators correspond at any time
$\eta$ to what would be the Minkowskian definition of particle and
vacuum if the modulation were to stop at this instant. Alternatively,
we can work with the operators defined at some fixed time $\eta_{\rm
  in}$, as we did in
Eq.~\eqref{eq:field_mode_function_initial_time_operators}, in which
case the time-dependence is that of a mode function satisfying
Eq.~\eqref{mode} which differs from that of $u_{k}^{\textsc{(m)}}$.
As just discussed, when the background is time-dependent neither of
these two sets of operators can be favoured to discuss the particle
content of the field.

There are some situations where one can unambiguously define particles
and their properties.  One such
case~\cite{parkerQuantizedFieldsParticle1969} is that of a spacetime
which is asymptotically Minkowski at both very early and very late
times, i.e. one for which the scale factor varies in-between two
asymptotic constant values
\begin{equation*}
 a \left( \eta \right) \xrightarrow[\eta \to - \infty]{} a_{\rm in}
 \quad \text{and} \quad a \left( \eta \right) \xrightarrow[\eta \to +
   \infty]{} a_{\rm out} \, .
\end{equation*}
We can therefore define asymptotically Minkowski ``in" and ``out" mode
functions $u^{(\mathrm{ \text{in} / \text{out}})}_{k}$ and associated
operators $\hat{a}^{(\mathrm{ \text{in} / \text{out}})}_{\bm{k}}$ by
requiring as initial condition that they match the Minkowski solution
\begin{equation*}
	u^{(\mathrm{in})}_{k} \xrightarrow[\eta \to -\infty]{} \frac{
          e^{- i k \eta}}{ \sqrt{2k} } \quad \text{and} \quad
        u^{(\mathrm{out})}_{k} \xrightarrow[\eta \to +\infty]{} \frac{
          e^{- i k \eta}}{ \sqrt{2k} } \, .
\end{equation*}
These mode functions are both solution of \eqref{mode} for any time
$\eta$ and are therefore related by a \textit{time-independent}
Bogoliubov transformation
\begin{equation}
	u^{(\mathrm{in})}_{k} = \rho_{k} u^{(\mathrm{out})}_{k} +
	\chi_{k} u^{\star \, (\mathrm{out})}_{k} \, ,
\end{equation}
and, via \eqref{eq:bogo_operators_fixed_time}, so are
$\hat{a}^{(\mathrm{ \text{in} / \text{out}})}_{\bm{k}}$, and it is
straightforward to evaluate the number of particles produced by the
non-trivial evolution of the background. We assume that the field is
initially (in the ``in" region) in the vacuum defined by the ``in"
operators where there exists a preferred notion of vacuum; we denote
$\left| 0 \right \rangle_{ \text{in} }$ this ``in" vacuum.  In order
to read the particle content at the end of evolution (in the ``out"
region) we need to use the ``out" operators that define the Minkowski
notion of particle there. The number of particles in the ``out" region
is given by
\begin{equation}
    n^{\rm out}_{\pm \bm{k}} = {}_{\text{in}} \left \langle 0 \right|
    \hat{a}^{\dagger \, (\mathrm{out})}_{\pm \bm k}
    \hat{a}^{(\mathrm{out})}_{\pm \bm k} \left| 0 \right \rangle_{
      \text{in} } = \left| \chi_{k} \right|^2 \, .
\end{equation}
This number is strictly positive and the same in the modes $\pm
\bm{k}$; this is the well-known phenomenon of pair production out of
the vacuum, here powered by the background expansion. To evaluate the
extent of this production quantitatively we have to compute the mode
equation for both ``in" and ``out" conditions and match them. This
computation can, for example, be done exactly in a 2d model where the
scale factor evolves as a hyperbolic tangent between its asymptotic
values \cite{bernardRegularizationRenormalizationQuantum1977}.

Let us make the connection in this idealised case with the
time-dependent Bogoliubov coefficients solving the dynamics of
\eqref{eq:time_evo_bogo_transfo}. First, note that the operators
\eqref{def:minkowski_scalar_creation_operators} coincide with those
defined with respect to a mode function $u_{k}$ at times $\eta_0$
where it satisfies the Minkowski
conditions~\eqref{eq:minkowski_initial_conditions_mode_function}.
This can be checked directly upon inserting
\eqref{def:minkowski_scalar_creation_operators} in the expression of
the operator in terms of the mode function and the fields at time
$\eta_0$. This applies in both the ``in" and ``out" regions
\begin{align*}
\hat{a}_{\pm \bm k} \left( \eta \right) & \xrightarrow[\eta \to -
  \infty]{} \hat{a}^{(\mathrm{in})}_{\pm \bm k} \, , \\ \hat{a}_{\pm
  \bm k} \left( \eta \right) & \xrightarrow[\eta \to + \infty]{}
\hat{a}^{(\mathrm{out})}_{\pm \bm k} \, .
\end{align*}
The time-independent Bogoliubov coefficients between the ``in" and the
``out" states, therefore, correspond to the late time limit of the
time-dependent Bogoliubov coefficients of
Eq.~\eqref{eq:tilde_evol_bogo}
\begin{equation}
\rho_{k} = \alpha_{k} \left( \eta \to + \infty  \right) \quad
\mathrm{and} \quad \chi_{k} = \beta_{k} \left(\eta \to + \infty \right)
\, ,
\end{equation}
where the associated number of particles $\left \langle \hat{a}_{\pm
  \bm k}^{\dagger} \left( \eta \right) \hat{a}_{\pm \bm k} \left( \eta
\right) \right \rangle$ and correlations are now meaningful.  While it
is not \textit{a priori} the case at any intermediate times, since the
scale factor is varying, we discuss in
Sec.~\ref{sec:connection_observations} how it is often possible to
identify ``in" and ``out" regions for certain ranges of modes $k$ in
the cosmological evolution.

Anticipating these considerations, we conclude by making a connection
with Sec.~\ref{subsec:squeezed_states} and studying the particle
content of a 2-mode squeezed state. Those can also be fully
characterised by the following three non-vanishing expectation values
(two of them being equal)
\begin{subequations}
\begin{align}
    n_{k} & = \left \langle \hat{a}^{\dagger}_{\bm{k}}
    \hat{a}_{\bm{k}} \right \rangle = \left \langle
    \hat{a}^{\dagger}_{-\bm{k}} \hat{a}_{-\bm{k}} \right \rangle =
    \frac{\gamma_{11}+\gamma_{22} - 2}{4} = \sinh^2 \left( r_k\right)
    \, , \\ c_{k} & = \left \langle \hat{a}_{ \bm{k}} \hat{a}_{-
      \bm{k}} \right \rangle = \frac{\gamma_{11}-\gamma_{22}}{4} + i
    \frac{\gamma_{12}}{2} = - \frac12 \sinh \left( 2 r_k \right)
    e^{2i\varphi_k} \, .
\label{eq:nk_ck}
\end{align}
\end{subequations}
These expressions are obtained by inverting
Eq.~\eqref{def:minkowski_scalar_creation_operators} and making use of
Eqs.~\eqref{eq:gammaij} and
\eqref{eq:covariance_matrix_squeezed_state}.  The first expectation
value $n_k$ gives the number of particles in the modes $\bm{k}$ and
$-\bm{k}$, which must be identical because of isotropy, while $c_k$
encodes the 2-mode coherence of the pairs.  Imposing the purity to be
less than unity, $p_k = \gamma_{11} \gamma_{22} - \gamma_{12}^2 \leq
1$, yields the following bound on the magnitude of this coherence:
\begin{equation}
\left| c_k \right| \leq \sqrt{n_k \left( n_k +1 \right)}.
\label{eq:bound_ck}
\end{equation}
For a pure state like that of the gravitons, the bound is saturated
$\left| c_k \right| = \sqrt{n_k \left( n_k +1 \right)}$, while for a
thermal state $c_k = 0$.  In this sense, the modes are uncorrelated in
the thermal state and maximally correlated in a 2-mode squeezed state;
they are even \textit{entangled}
\cite{campoInflationarySpectraPartially2005a}. We come back to this
important point in Sec.~\ref{sec:quantumness_primordial_GW}.

\subsection{Anomaly-induced semiclassical theory}

The concept of particle associated with a quantum field is a global
one in the sense that it is defined through modes; somehow, it can be
understood, as described above, as the effect of geometry on matter,
even when ``matter'' consists of tensor-like perturbations of the
gravitational field itself. When coupled to classical GR in a
semiclassical way, the quantum nature of gravitational waves, just
like any other particle, may also manifest itself in another way,
namely in the back reaction of their quantum fields on geometry; (see,
e.g., the historical papers by M.~J.~Duff
\cite{Duff:1977ay,Deser:1976yx,Duff:1993wm} who proposed it for the
first time, and Refs.~\cite{Birrell:1982ix,Buchbinder:1992rb} as well
as the more recent Ref~\cite{Shapiro:2008sf}).  This approach is,
therefore, the opposite of the above, making extensive use of the
stress-energy tensor $T_{\mu\nu}(x)$, which is a local quantity.

In this section, for the sake of notational simplicity,
we set $\hbar \to 1$ as all the effects are quantum by
nature.

\subsubsection{Gravity with quantum fields}

When quantum fields are described in a geometric background, it is
customary to write the corresponding Einstein's equations in the
semiclassical form
\begin{equation}
R_{\mu\nu} -\frac12 Rg_{\mu\nu} + \Lambda g_{\mu\nu}
=8\pi\GN \langle
T_{\mu\nu}\rangle_\mathrm{ren}, \label{semi}
\end{equation}
so that geometry is now sourced by the renormalised
stress-energy tensor $\langle T_{\mu\nu}\rangle_\mathrm{ren}$.

As the classical Einstein equations are derived from a variation of
the vacuum Einstein-Hilbert term\footnote{We do not consider the
  Gibbons–Hawking–York boundary term in these discussions; it can be
  set to zero by assuming a compact manifold.}  (possibly including a
cosmological constant contribution),
\begin{equation}
S_{\textsc{eh}\Lambda}
=\frac{1}{16\pi \GN}\int \D^4 x\sqrt{-g}\left( R+2\Lambda \right),
\label{EH}
\end{equation}
the stress-energy tensor being derived from the classical matter
action $\mathcal{S}_\mathrm{m}$ through
\begin{equation}
T_{\mu\nu}^\mathrm{class} =- \frac{2}{\sqrt{-g}} \frac{\delta
  \mathcal{S}_\mathrm{m}}{\delta g^{\mu\nu}},
\label{deltaSm}
\end{equation}
one can recover the semiclassical case \eqref{semi} by similarly
defining an effective action $\Gamma[g_{\mu\nu}]$ such that
\begin{equation}
\langle T_{\mu\nu}\rangle = -\frac{2}{\sqrt{-g}} \frac{\delta
  \Gamma}{\delta g^{\mu\nu}}.
\label{deltaGamma}
\end{equation}
It can be shown that for a set of matter fields denoted generically by
$\phi$, and which can include scalar, gauge and fermion fields, whose
dynamics is driven by the action $S[\phi;g_{\mu\nu}]$, one finds
\begin{equation}
e^{i \Gamma [g_{\mu\nu}]} = \int \mathcal{D}\phi e^{i
  S[\phi;g_{\mu\nu}]},
\label{EffAct}
\end{equation}
and the expectation value in \eqref{deltaGamma} is then understandable
in terms of "in" and "out" vacuum states:
\begin{equation}
\frac{2}{\sqrt{-g}} \frac{\delta \Gamma}{\delta g^{\mu\nu}} =
\frac{_\mathrm{out}\langle 0| T_{\mu\nu} | 0
  \rangle_\mathrm{in}}{_\mathrm{out}\langle 0 |0\rangle_\mathrm{in}},
\label{inout0}
\end{equation}
thereby automatically providing the required normalisation.

In order to integrate explicitly \eqref{inout0} and obtain the
relevant effective action, one needs to know the matter content and
its corresponding action. Compared to their flat space counterparts,
fermionic and vectorial contributions are merely obtained by the
minimal coupling, namely making the replacements $\partial\to\nabla$
and using the metric $g_{\mu\nu}$ to integrate.  The scalar field case
can also include an extra term, not present in the flat Minkowski
situation, and one gets
\begin{equation}
S_\varphi = -\frac12\int\D^4x \sqrt{-g} \left[ \left( \partial
  \bm{\varphi} \right)^2 + \xi_{ij}
  \varphi^i \varphi^j R \right],
\label{GenVarphi}
\end{equation}
where we considered a set of scalars $\{ \varphi^i\} = \bm{\varphi}$;
a possible extra potential term $V\left(\bm{\varphi}\right)$ can be
added to this action.  Eq.~\eqref{GenVarphi} involves a set of new
dimensionless numbers $\{\xi_{ij}\}$ which are called non-minimal
parameters. For a single scalar field, this reduces to a single
parameter; its special value $\xi=\frac16$ yields conformal
invariance.

It turns out that the action derived from this procedure contains
ultraviolet divergences that thus need to be renormalised. These lead
to contributions that are purely geometrical, involving only scalars
made out of the Riemann tensor $R_{\mu\nu\alpha\beta}(x)$ and its
contractions. This is understandable as short wavelengths are only
sensitive to local features of spacetime. Regularising and
renormalising forces to introduce counterterms involving higher-order
derivatives, and one is naturally led to the conclusion that in order
to obtain a renormalisable theory of quantum matter on a classical
curved spacetime, one must demand a geometrical framework that goes
beyond general relativity.

Applying the procedure described above, the relevant vacuum classical
action
\begin{equation}
S_\mathrm{vac}=S_{\textsc{eh}\Lambda}+S_\textsc{hd}
\label{vacuum}
\end{equation}
is found to include the usual Einstein-Hilbert term \eqref{EH} in
which both $\GN$ and $\Lambda$ are renormalised quantities, but
another contribution, containing higher derivatives (HD) terms, needs
be included, namely
\begin{equation}
S_\textsc{hd} = \int \D^4x \sqrt{-g} \left( a_1C^2+a_2E+a_3
{\Box}R+a_4R^2 \right),
\label{HD}
\end{equation}
where 
$$
C^2=R_{\mu\nu\alpha\beta}^2 -
2 R_{\alpha\beta}^2 + \frac13 R^2
$$
is the square of the Weyl tensor and 
$$
E = R_{\mu\nu\alpha\beta}^2 - 4
R_{\alpha\beta}^2 + R^2
$$
represents the Gauss-Bonnet topological term. The action
\eqref{vacuum} has been shown~\cite{Stelle:1976gc} to lead to a
renormalisable (albeit containing unphysical ghosts or having
non-unitarity issues) theory of quantum gravity.  Details can be found
in particular in \cite{Ohta:2022zqs} in the present volume. The
parameter $a_3$ is irrelevant for the equations of motion since $\Box
R$ is a surface term, while the $R^2$ term is at the origin of the
most serious inflation model proposed by
Starobinsky~\cite{Starobinsky:1980te}.

\subsubsection{Conformal anomalies}

Let us consider a conformally invariant theory, i.e. for which the
transformations
\begin{equation}
g_{\mu\nu} \to \bar{g}_{\mu\nu} = \Omega^2(x)
g_{\mu\nu}\ \ \ \ \ \hbox{and}\ \ \ \ \ \varphi \to \varphi /\Omega(x)
\label{Conformal}
\end{equation}
(vector fields being left unchanged and spinors transforming with
$\Omega^{-3/2}$) leaves the action $S$ unchanged. From this
requirement, one finds that the trace of the energy-momentum
tensor~\cite{Birrell:1982ix}
\begin{equation}
T^\mu_{\ \ \mu}[g_{\alpha\beta}(x)] = - \frac{\Omega(x)}{\sqrt{-g(x)}}
\frac{\delta S[\bar{g}_{\mu\nu}]}{\delta\Omega(x)}\Big|_{\Omega\to 1},
\label{TraceConf}
\end{equation}
should vanish if \eqref{Conformal} is a symmetry of $S$. This implies
that the scalar fields are massless and $\xi\to\frac16$.  The identity
\eqref{TraceConf} is true at the classical level, and indeed the
conserved Noether current in this case reads
\begin{equation}
\left(2 g_{\mu\nu} \frac{\delta}{\delta g_{\mu\nu}} + \sum_i k_i
\phi_i \frac{\delta}{\delta \phi_i} \right)
S[g_{\alpha\beta}(x),\phi(x)] =0,
\label{NoetherConformal}
\end{equation}
in which the weights $k_i$ correspond to the various fields involved,
with $k_\mathrm{s}=-1$ for scalar fields, $k_\mathrm{f}=-3/2$ for the
fermions and $k_\mathrm{v}=0$ for the gauge fields.

At the quantum level, however, the trace $\langle T^\mu_{\ \ \mu}
\rangle$ is no longer vanishing, as explicitly calculating it with the
given matter content (scalar, vector and spinor fields) yields a
renormalised expectation value~\cite{Birrell:1982ix}
\begin{equation}
\langle T^\mu_{\ \ \mu} \rangle = -\left( \omega C^2 + bE + c\Box R
\right),
\label{tracean}
\end{equation}
where the $\beta-$functions $\omega$, $b$ and $c$ depend on the
numbers of real scalar degrees of freedom $N_0$, four-component spinor
fermions $N_{1/2}$ and vector fields $N_1$ in the underlying particle
physics model. In practice, they are found to be
\begin{equation}
\left(
\begin{array}{c}
\omega  \\
b       \\
c        \\
\end{array}
\right)
=
\frac{1}{360 (4\pi)^2}
\left(
\begin{array}{ccc}
3 N_0 + 18 N_{1/2} + 36 N_1
\\
-  N_0 - 11 N_{1/2} - 62 N_1
\\
2 N_0 + 12 N_{1/2} - 36 N_1    \\
\end{array}
\right).
\label{wbc}
\end{equation}
In the standard model (SM) of particle physics, where the
SU(3)$\times$SU(2)$\times$U(1) is broken to SU(3)$\times$U(1) through
a Higgs doublet, the relevant numbers are $N^\textsc{sm}_0=4$,
$N^\textsc{sm}_1=12$ (eight gluons, the intermediate $W^\pm$ and $Z^0$
and the photon) and $N^\textsc{sm}_{1/2} =24$ (leptons and quarks,
assuming a massive neutrino), one finds
$$
\omega^\textsc{sm}=\frac{73}{480\pi^2}, \ \ \ \ \ b^\textsc{sm} =
-\frac{253}{1440\pi^2} \ \ \ \ \ \hbox{and} \ \ \ \ \ c^\textsc{sm} =
-\frac{17}{720\pi^2}.
$$
Note that although $b$ is negative definite, the sign of $c$ depends
on the exact matter content: measuring this sign somehow, e.g. through
that of the primordial gravitational wave spectrum, could be an
indirect way of getting information about the physics that should
apply at high energies such as the grand unification (if any)
scale. Note for instance that in the case of the minimal
supersymmetric extension of the standard model (MSSM), the number of
vector modes is unchanged ($N^\textsc{mssm}_1=12$), while the number
of fermions is increased to $N^\textsc{mssm}_{1/2}=32$ and the
proliferation of new scalar modes then yields $N^\textsc{mssm}_0=104$,
leading to $c^\textsc{mssm}=1/(36\pi^2) >0$.

Integrating the trace of \eqref{deltaGamma} using \eqref{tracean}
is a non-trivial task that has been achieved in
Refs.~\cite{Riegert:1984kt,Fradkin:1983tg}.
Ref~\cite{Shapiro:1994ww} suggested to rewrite the
action in terms of two auxiliary scalar fields $\sigma$ 
and $\rho$ (see also \cite{Mazur:2001aa} for an independent
but equivalent formulation) which happens to be particularly
useful for the gravitation wave discussion. It reads
\begin{eqnarray}
\Gamma = S_\text{c}[g_{\mu\nu}] & + & \displaystyle \int \D^4x
\sqrt{-g} \left( \frac12 \sigma\Delta_4\sigma -
\frac12\rho\Delta_4\rho + \ell_1 C^2 \rho\right) \nonumber \\ & + &
\displaystyle \int \D^4x \sqrt{-g} \left\{ \sigma\left[ k_1 C^2 +
  k_2\left( E - \frac{2}{3}\Box R\right) \right] - \frac{1}{12} k_3
R^2 \right\},
\label{tota}
\end{eqnarray}
where the integration constant $S_\text{c}[g_{\mu\nu}]$ is conformally
invariant, the covariant conformal fourth-order operator is (see
Refs.~\cite{Riegert:1984kt,Fradkin:1983tg})
$$ \Delta_4=\Box^2+2R^{\mu\nu}\nabla_\mu \nabla_\nu
-\frac23\,R\Box+\frac13\,R^{;\mu}\nabla_\mu
$$
and the coefficients are given in terms of those of \eqref{tracean}
through
\begin{equation}
k_1 = - \frac{\omega}{2\sqrt{|b|}}
 ,\quad
k_2=\frac{\sqrt{|b|}}{2}
 ,\quad
k_3 = c + \frac{2}{3}b
 \quad \hbox{and} \quad
\ell_1=\frac{\omega}{2\sqrt{|b|}}
\end{equation}
(recall $b<0$). This effective action stemming from the
conformal anomaly (the Noether current is not conserved at
the quantum level) should be added to the vacuum term
$S_\text{vac}$ of Eq.~\eqref{vacuum}.

\subsubsection{Anomaly-induced cosmology and gravitational waves}

Let us apply the above discussion to the specific case of a
cosmological framework which is our main subject, first by considering
a background FLRW (conformally flat) solution and its tensorial
perturbations.

The FLRW metric can be written as a conformal transformation of the
Minkowski metric $\eta_{\mu\nu}$ by setting $g_{\mu\nu} = a^2(\eta)
\eta_{\mu\nu}$.  In this very simple case, variations of \eqref{tota}
with respect to the auxiliary fields $\sigma$ and $\rho$ yields
\begin{equation}
\left( \partial_t^2 - \bm{\nabla}^2\right) \left( \sigma
+ 8\pi\sqrt{|b|} \ln a \right) = 0\ \ \ \hbox{and} \ \ \
\left( \partial_t^2 - \bm{\nabla}^2\right) \rho = 0,
\label{sigmarho}
\end{equation}
with solutions
\begin{equation}
\sigma = \sigma_\text{h} - 8\pi\sqrt{|b|} \ln a 
\qquad \hbox{and} \qquad
\rho =  \rho_\text{h},
\label{sigmarho0}
\end{equation}
in which $\sigma_\text{h}$ and $\rho_\text{h}$ are solutions of the
homogeneous equation, $\left( \partial_t^2 - \bm{\nabla}^2\right)
f_\text{h} = 0$; they can be set to zero in the cosmological context.
In this case, one finds the relation
$$
\frac{\D^n \sigma}{\D t^n} = -8\pi \sqrt{|b|}
\frac{\D^{n-1} H}{\D t^{n-1}}
$$
where $H=\dot{a}/a$.

The above solution \eqref{sigmarho0} with the FLRW metric can now be
inserted into the full theory containing both \eqref{tota} and the
original vacuum \eqref{vacuum}. It leads to the modified Friedmann
equation
\begin{equation}
\frac{\ddot{a}}{a}+ H^2 -\frac23\Lambda
= \frac{c}{\MP^2} \left[
\frac{\ddddot{a}}{a}
+3 H \frac{\dddot{a}}{a}
+\left(\frac{\ddot{a}}{a}\right)^2
-\left( 5+\frac{4b}{c}\right) H^2
\frac{\ddot{a}}{a}
\right],
\label{AnomalousFriedmann}
\end{equation}
in which we defined the Planck mass $\MP^{-2}=8\pi\GN$.  As could have
been anticipated, this solution depends on $b$ and $c$, but neither on
$\omega$ and $a_1$ since the Weyl tensor is conformally invariant, nor
on $a_2$ and $a_3$ (surface terms), and we have set $a_4\to0$ to
ensure the original theory is conformally invariant.

Inflationary solutions for \eqref{AnomalousFriedmann} can be found in
Refs.~\cite{Starobinsky:1980te,Antoniadis:1986tu,Starobinsky:1981vz,
  Shapiro:2001rh,Pelinson:2002ef}. A simple case consists of a de
Sitter solution $a \propto \exp (H t)$ with $H$ constant, which
transforms \eqref{AnomalousFriedmann} into a quadratic algebraic
equation for $H$ whose solutions
\begin{equation}
H^2 = \frac{\MP^2}{2|b|} \left( 1\pm\sqrt{1+\frac{4|b|
    \Lambda}{3\MP^2}} \right) \xrightarrow[|b|\Lambda \ll \MP^2]{}
\begin{cases}
    H^2_\text{inf} = \displaystyle\MP^2/|b| & (+) \\
    H^2_\Lambda = \displaystyle 2\Lambda/3 & (-)
  \end{cases}
\label{infLambda}
\end{equation}
produce the two relevant extreme cases of present-day cosmological
constant domination and initial inflation, with $H_\text{inf} \gg
H_\Lambda$.

Tensor perturbations of the kind \eqref{gdg} in this context are
slightly different from those of ordinary GR discussed in the previous
sections. In particular, the mode equation \eqref{mode} is now
replaced by the slightly more involved fourth order equation (see
Ref.~\cite{Fabris:2011qq} for details)
\begin{eqnarray}
& & \left( 2 f_1 + \frac{f_2}{2}\right) \ddddot{h} + \left[ 3H(4f_1 +
    f_2) + 4 \dot{f}_1 + \dot{f}_2\right] \dddot{h} + \left[
    3H^{2}\left(6f_1 + \frac{f_2}{2} - 4f_3\right)\right.  \nonumber
    \\ \nonumber
& &\hskip5mm
+\left. H\left(16 \dot{f}_1 + \frac{9}{2}\dot{f}_2\right)
+ 6\dot{H}(f_1 - f_3) -\frac{16 \pi^2}{3}|b|
\left( H^2-\dot{H} \right) \right] \ddot{h}
\\
\nonumber
& &\hskip5mm
- \left( 4 f_1 + f_2 \right) \frac{\nabla^2 \ddot{h}}{a^2}
+ \Biggl[ 2\dot{H}(2\dot{f}_1 - 3 \dot{f}_3)
- \frac{21}{2}\, H \dot{H} \left(f_2 + 4 f_3\right)
- \frac32\,\ddot{H}\left(f_2 + 4 f_3\right)
\\
\nonumber
& &\hskip5mm
+ 3H^{2}\left( 4 \dot{f}_1 + \frac{1}{2} \dot{f}_2
- 4 \dot{f}_3\right)
-9H^{3}\left(f_2 + 4f_3\right)
+ H \left( 4\ddot{f}_1 + \frac{3}{2}\ddot{f}_2
 + \frac{3\MP^2}{4} \right)
\\
\nonumber
& &\hskip5mm
+ \frac{16 \pi^2}{3} |b| \left( \ddot{H}
+ H\dot{H}-3H^3 \right)
\Biggr]  \dot{h}
- \left[ H(4 f_1 + f_2) + 4 \dot{f}_1
+ \dot{f}_2\right] \frac{\bm{\nabla}^{2} \dot{h}}{a^{2}}
\\
\nonumber
& &\hskip5mm
+ \Biggl[ \frac{16\pi^2}{3} |b| \left( 2
\dddot{H} + 12 H \ddot{H}
+9\dot{H}^2 - 6 H^2 \dot{H} -15 H^4 \right)
+\frac{\MP^2}{2}\left( 2 \dot{H} + 3 H^{2}\right)
\nonumber
\\
\nonumber
& &\hskip5mm
- 4H \dot{H}\left( 8 \dot{f}_1 + 9 \dot{f}_2 +
30 \dot{f}_3 \right)
- 8\ddot{H}\left( \dot{f}_1 + \dot{f}_2 + 3 \dot{f}_3)
- H^{2}(4 \ddot{f}_1 + 6 \ddot{f}_2 + 24 \ddot{f}_3\right)
\\
\nonumber
& &\hskip5mm
- 4\dot{H}\left(\ddot{f}_1 + \ddot{f}_2 + 3 \ddot{f}_3\right)
- H^{3}\left(8 \dot{f}_1 + 12 \dot{f}_2 + 48 \dot{f}_3\right)
\\
\nonumber
& &\hskip5mm
- \left(36 \dot{H} H^{2}
+ 18 \dot{H}^{2}
+ 24 H \ddot{H}
+ 4 \dddot{H} \right) (f_1 + f_2 + 3 f_3)
 \Biggl] h
\\
\nonumber
& &\hskip5mm
+ \Bigl[
2\left(2H^2 + \dot{H}\right)(f_1 + f_2 + 3 f_3)
+ \frac{1}{2} H \left(4 \dot{f}_1 +  \dot{f}_2\right)
+\frac{\MP^2}{2} -\frac12 \ddot{f}_2
\\
& &\hskip5mm
- \frac{16\pi^2}{3} |b| \left( \dot{H}+5H^2 \right)
\Bigr] \frac{\bm{\nabla}^{2} h}{a^{2}}
+ \Bigl(2 f_1
+ \frac{1}{2} f_2\Bigl) \frac{\bm{\nabla}^{4} h}{a^{4}} = 0,
\label{diff}
\end{eqnarray}
stemming from the variation of the second-order Lagrangian function
\begin{equation}
\mathcal{L}
= \frac{\MP^2}{2} R
+ f_1 R^2_{\alpha\beta\mu\nu}
+ f_2 R^2_{\alpha\beta}
+ f_3 R^{2}
-\frac{4\pi}{3}\,\sqrt{|b|} \sigma\Box R
+ \frac12 \sigma \Delta \sigma,
\label{LagTensor}
\end{equation}
and we have set $\rho = \rho_\text{h} \to 0$ and $\sigma_\text{h} \to
0$ as the background depends only on time; the perturbation
$h({\bm{x},t})$ is the amplitude of the tensor mode $h_{ij}$ for a
given polarisation. In Eqs.~\eqref{diff} and \eqref{LagTensor}, the
coefficients $f_1$, $f_2$ and $f_3$ are time-dependent functions that
take the values
\begin{eqnarray*}
f_1&=&
a_{1} + a_{2} + \frac{|b| - \omega}{2 \sqrt{|b|}}
\sigma, 
\\
f_2&=&
-2a _{1} - 4a_{2}
+ \frac{\omega  - 2|b|}{\sqrt{|b|}}
\sigma,
\\
f_3
&=&
\frac{a _{1}}{3} + a_{2}  - \frac{3c - 2|b|}{36}
+ \frac{3|b| - \omega}{6\sqrt{|b|}}
\sigma.
\end{eqnarray*}
By inspection of the combinations of $f$'s entering Eq.~\eqref{diff},
one notes that the equation of motion does not depend on $a_2$, as
expected from the fact that this comes from a surface term.

Eq.~\eqref{diff} was obtained by assuming the value \eqref{sigmarho0}
for the auxiliary field $\sigma(t)$ in terms of the background Hubble
variable, and so can be used for any admissible solution for the scale
factor, including the inflating case of \eqref{infLambda}.  Expanding
in Fourier modes, i.e. replacing $\bm{\nabla}$ by $-\bm{k}^2$, in
principle permits to evaluate the gravitational wave stochastic
spectrum in such a theory, with a catch: contrary to GR, the mode
equation is no longer that of a parametric oscillator, so that its
quantisation, and consequently the vacuum initial conditions, are not
that well defined.

This issue, still under discussion, can be handled by assuming that
our semiclassical framework provides a perturbation to GR, so that the
extra (higher derivative) terms may be neglected while quantising in a
regime in which one can manage to construct a consistent Hilbert space
of state. Setting quantum vacuum fluctuation initial conditions
exactly then allows setting initial values for the gravitational wave
amplitude and its first three time derivatives.

Moreover, the presence of the higher derivative terms potentially
implies instabilities. Setting initial conditions as discussed above,
one finds~\cite{Salles:2014rua,Peter:2017xxf} that the time development,
and hence
the resulting predictions, is very sensitive to the properties of the
background. Assuming, for instance, a de Sitter inflation phase with
constant Hubble rate $H=H_\text{inf}$, initial trans-Planckian runaway
solutions can be redshifted to become sub-Planckian and then rapidly
damped by the expansion: the instabilities indeed present in the
theory can end up harmless in a cosmological setup. We assume in what
follows that this is indeed the case.

\subsection{Primordial gravitational-wave background}

Independently of the underlying quantum theory leading to the
production of primordial tensor modes, one must now evolve them
through the expanding universe to evaluate their current
contribution. As we know GR to be valid for the most part of the FLRW
evolution, we consider from now on that the higher derivative terms
discussed above are either not present at all, or contribute only
negligibly. In order to clearly distinguish classical from quantum
effects, we include again the relevant factors of $\hbar$ when
necessary.

In Sec.~\ref{subsec:graviton_production}, we have laid out three
equivalent ways to describe the evolution of perturbations for a
general time-dependent background $a(\eta)$: the use of Bogoliubov
transformations, mode functions and squeezing parameters. We now solve
the dynamics of the gravitational wave field in a simplified model of
the cosmological evolution to discuss the properties of the primordial
gravitational waves generated and make a connection with observations.

\subsubsection{Cosmological evolution}
\label{subsubsec:cosmo_evo}

In FLRW the curvature of spacetime is contained in the scale factor
$a$, whose dynamics is related to the matter content of the Universe
through the Friedmann equations. In what follows, we first solve them
in the standard approximation that there is always a single fluid
dominating the energy budget of the Universe and that transitions
between two phases are instantaneous. One can thus model the
cosmological evolution as a succession of three eras: first an
accelerated expansion phase for $- \infty \leq \eta \leq
\eta_{\text{r}}$, whose dynamics is that of a slow-roll inflation
phase \cite{stewartMoreAccurateAnalytic1993}, then a radiation
dominated phase for $\eta_{\text{r}} \leq \eta \leq \eta_{\text{m}}$
and finally a matter domination for $\eta \geq \eta_{\text{m}}$. For
the sake of simplicity, we ignore the late-time accelerated expansion.

The evolution of the gravitational waves contained in the universe is
controlled by Eq.~\eqref{mode} where the expansion enters through the
scale factor $a \left( \eta \right)$ and its second derivative.
Connecting the scale factor and its derivative continuously across the
transitions, we have
\begin{align}
\frac{a \left( \eta \right)}{a_\text{r} } = 
\begin{cases}
\displaystyle \frac{\eta_{\text{r}}^{1+ \epsilon}}{\left( 2
  \eta_{\text{r}} - \eta \right)^{1+\epsilon}} \approx
\frac{\eta_{\text{r}}}{ 2 \eta_{\text{r}} - \eta } +
\mathcal{O}(\varepsilon) \quad & \text{for} \quad - \infty \leq \eta
\leq \eta_{\text{r}}, \\[10pt] \displaystyle
\frac{\eta}{\eta_{\text{r}} } \quad & \text{for} \quad \eta_{\text{r}}
\leq \eta \leq \eta_{\mathrm{m}}, \\ \displaystyle
\frac{\eta_{\text{m}}}{2 \eta_{\text{r}}} \left(
\frac{\eta^2}{\eta_{\text{m}}^2} +1 \right) \quad & \text{for} \quad
\eta_{\text{m}} \leq \eta,
\end{cases}
\label{eq:cosmo_evo_scale_factor}  
\end{align}
where $\eta_{\text{r}} >0$. The first expression in inflation is at
first order in $\epsilon = 1 - \Hu^{\prime} / \Hu^2$ the first
slow-roll parameter considered time-independent and we have also given
the de Sitter limit $\epsilon=0$.  From this, one computes the
time-dependent part of the frequency $\omega_{k}^2$ defined in
Eq.~\eqref{def:frequency}
\begin{align}
\frac{a^{\prime \prime}}{a} =
\begin{cases}
\displaystyle \frac{2+3\epsilon}{\left( 2 \eta_{\text{r}} - \eta
  \right)^2} \approx \frac{2}{\left( \eta_{\text{r}} - \eta \right)^2}
+ \mathcal{O}(\varepsilon) \quad & \mathrm{for} \quad - \infty \leq
\eta \leq \eta_{\text{r}}, \\ 0 \quad & \mathrm{for} \quad
\eta_{\text{r}} \leq \eta \leq \eta_{\mathrm{m}}, \\ \displaystyle
\frac{2}{\eta^2} \quad & \mathrm{for} \quad \eta_{\mathrm{m}} \leq
\eta.
\end{cases}
\label{eq:cosmo_evo_frequency}  
\end{align}
Solving Eq.~\eqref{mode} with \eqref{eq:cosmo_evo_frequency} yields
reference mode functions in each era, namely
\begin{subequations}
\begin{align}
u_{k}^{\text{(infl.)}} \left( \eta \right) & = \sqrt{\frac{- \left(
    \eta - 2 \eta_{\text{r}} \right) \pi}{4}} H^{(1)}_{\frac{3}{2} +
  \epsilon } \left[ - k \left( \eta - 2 \eta_\text{r} \right)
  \right] \enspace & \mathrm{for} & \enspace - \infty \leq \eta \leq
\eta_{\text{r}}, \\ & \approx \nonumber \frac{e^{- i k \left(\eta
    - 2 \eta_{\text{r}} \right)}}{\sqrt{2k}} \left[ 1 - \frac{i}{k
    \left(\eta - 2 \eta_{\text{r}} \right)}\right] \quad & &
\\ u_{k}^\text{(r)} \left( \eta \right) & = \frac{e^{- i k
    \eta}}{\sqrt{2k}} = u_{k}^{\textsc{(m)}} \left( \eta \right) \quad
& \mathrm{for} & \quad \eta_{\text{r}} \leq \eta \leq
\eta_{\mathrm{m}}, \\ u_{k}^\text{(m)} \left( \eta \right) & =
\frac{e^{- i k \eta}}{\sqrt{2k}} \left( 1 - \frac{i}{k \eta}\right)
\quad & \mathrm{for} & \quad \eta_{\mathrm{m}} \leq \eta,
\label{eq:cosmo_evo_ref_mode_functions}  
\end{align}
\end{subequations}
where in the first line, $H^{(1)}_{\kappa}$ is the Hankel function of
the first kind of index $\kappa$ and the approximation corresponds to
the de Sitter limit. We refer to \cite{baumannCosmology2022} for a
recent textbook in which all details of the computations of the
inflationary mode function can be found. Note that during radiation
domination, the solution is given by the Minkowski mode function
because $a^{\prime \prime} = 0$.  Since two solutions of \eqref{mode}
are related by a Bogoliubov transformation, a mode function solution
of \eqref{mode} for the whole cosmological evolution is related by a
Bogoliubov transformation to the associate reference mode function
\eqref{eq:cosmo_evo_ref_mode_functions} in each era.

One can construct a global solution $u_{k} \left( \eta \right)$
starting in the inflationary period. The reference mode function there
was chosen to match the Minkowski mode function $u_{k}^{\textsc{(m)}}$
in the asymptotic past $\eta \to - \infty$. This gives us an ``in"
region in which we can set the initial condition for the state of the
system in terms of a well-defined particle content. We therefore pick
$u_{k} \left( \eta \right) = u_{k}^{\textsc{(m)}} \left( \eta \right)$
during inflation. The expressions for the radiation and matter
domination are then
\begin{align}
u_{k}\left( \eta \right) =
\begin{cases}
 \alpha^\text{(r)}_{k} u_{k}^\text{(r)} \left( \eta \right) +
 \beta^\text{(r)}_{k} u_{k}^{\star \, \text{(r)}} \left( \eta \right)
 \quad & \mathrm{for} \quad \eta_{\text{r}} \leq \eta \leq
 \eta_{\mathrm{m}}, \\[5pt] \alpha^\text{(m)}_{k} u_{k}^\text{(m)}
 \left( \eta \right) + \beta^\text{(m)}_{k} u_{k}^{\star \,
   \text{(m)}} \left( \eta \right) \quad & \mathrm{for} \quad
 \eta_{\mathrm{m}} \leq \eta \, ,
\end{cases}
\label{eq:cosmo_evo_global_mode_functions}    
\end{align}
where the Bogoliubov coefficients are found by requiring that the mode
function and its first time-derivative are continuous across the
transition. Their expressions are worked-out in full in
Ref.~\cite{albrechtInflationSqueezedQuantum1994}. The mode $u_k\left(
\eta \right)$ is then completely determined for both polarisations
and, using \eqref{def:mu_Fourier_mode_expansion}, one achieves a fully
quantum description of the evolution of the gravitational wave field.

The analysis is completed once one specifies the initial state of the
gravitational waves as $k \eta \to - \infty$. The standard choice is
to assume that, in the far past, the inflation phase somehow wiped out
any initial perturbation, leaving no graviton to start with: this is
the motivation behind choosing the vacuum state for every mode. This
vacuum initial state is often referred to the Bunch-Davies
vacuum~\cite{bunchQuantumFieldTheory1978}, although it should be more
appropriately be called Minkowski vacuum.  This choice implies that
the state of the perturbation consists of a collection of independent
2-mode squeezed states as discussed in
Sec.~\ref{subsec:squeezed_states}.

For scalar perturbations, the above vacuum choice turns out to be in
excellent agreement with the observations of the Cosmic Microwave
Background \cite{planckcollaborationPlanck2018Results2020}.  For
gravitational waves, we are so far short of equivalent observations so
that other states could be chosen as initial condition
\cite{martinNonVacuumInitialStates2000}. Although such alternative
choices do not modify our description of the subsequent evolution,
they change the values of the Bogoliubov coefficients and therefore
the prediction on the amplitude of gravitational waves or,
equivalently, the number of gravitons produced.

\begin{figure}[!t]
\centering
\includegraphics[width=\textwidth]{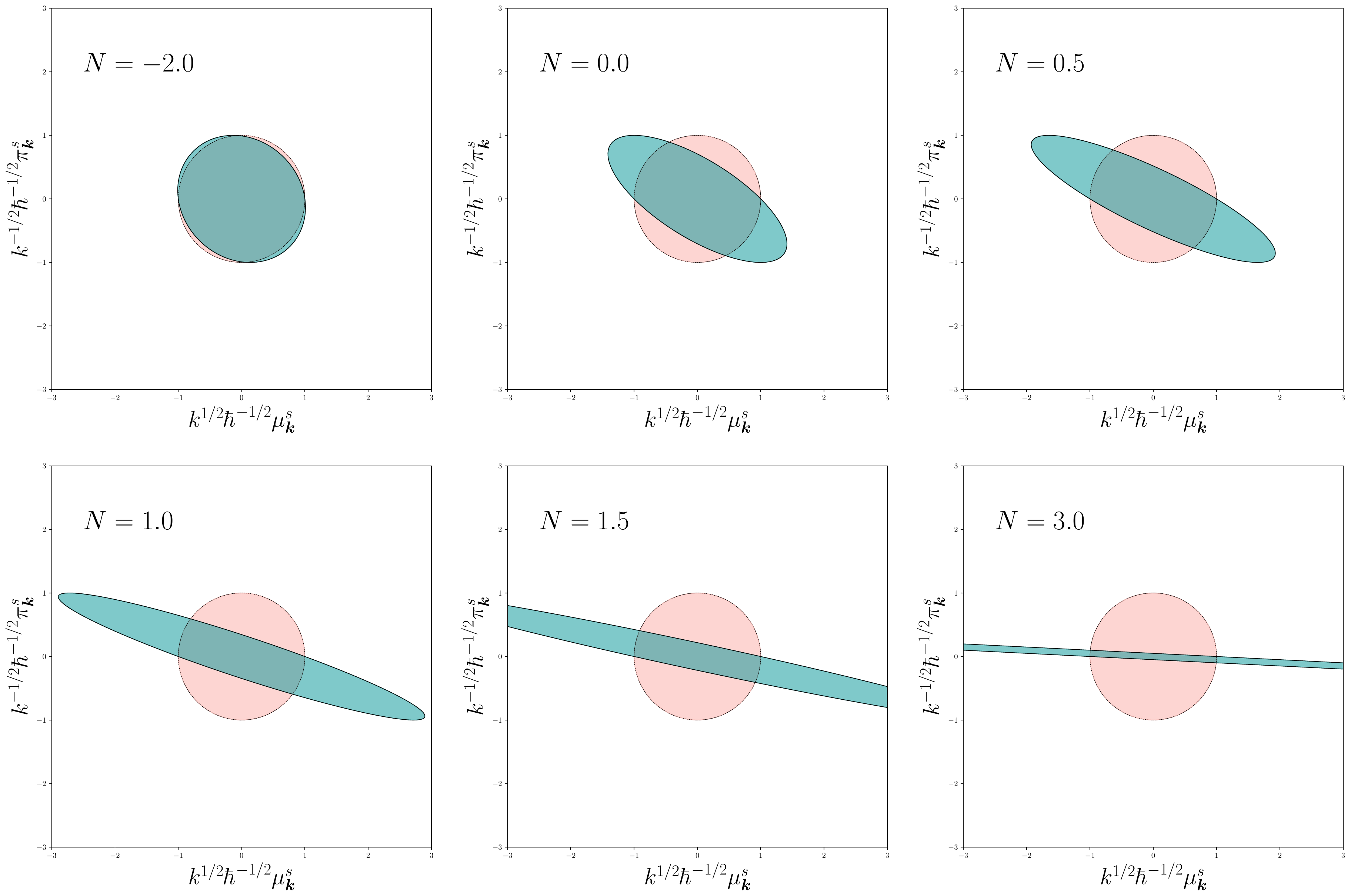}
\caption{Phase space ellipse in the plane $(k^{1/2}
  \mu_{\bm{k}}^{\textsc{s}},k^{-1/2} \pi_{\bm{k}}^{\textsc{s}})$ at
  different instants during inflation, labelled by $N =\ln \left[ a/a
    \left( \eta_k \right) \right]$, i.e. the number of $e$-folds
  measured from the Hubble-crossing time of the mode under
  consideration. On sub-Hubble scales, the ellipse remains roughly a
  circle, while it gets squeezed and rotates in the super-Hubble
  regime.}
\label{fig:wigner_de_sitter_no_deco}
\end{figure}

We have explained in Sec.~\ref{subsec:graviton_production} that, most
of the time, this number is ambiguous due to the time-dependent part
of $\omega_k^2$. Let us explain how to make sense of it for primordial
gravitational waves.  First, in the sub-Hubble regime $k^2 \gg
a^{\prime \prime}/a$ the frequency reduces to $\omega_{k} \sim k$ i.e.
the mode $\bm{k}$ does not feel the expansion of space and effectively
oscillates as in flat spacetime. In this sub-Hubble limit, the
reference mode functions~\eqref{eq:cosmo_evo_ref_mode_functions}
reduce to the Minkowski one, and we can treat the mode as if evolving
in Minkowski. On the other hand, in the super-Hubble regime $k^2 \ll
a^{\prime \prime}/a$, the mode behaves as an inverted harmonic
oscillator $\omega_{k} \sim - a^{\prime \prime}/a < 0$.  One therefore
expects its amplitude to be amplified, and it is indeed where most of
the squeezing happens, as illustrated in the first two panels of
Fig.~\ref{fig:wigner_de_sitter_no_deco}.

\begin{figure}
\centering
\includegraphics[width=0.7\textwidth]{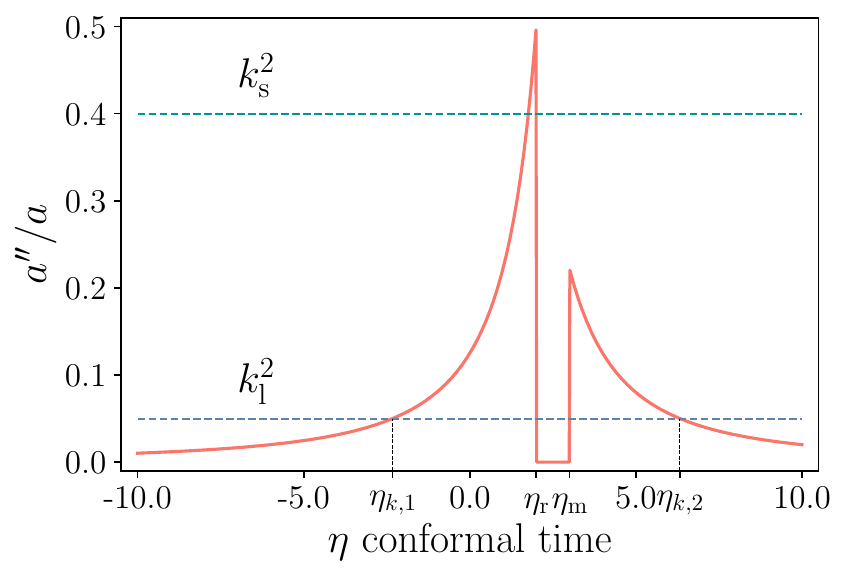}
\caption{Sketch of the potential for the tensor mode within the toy
  model \eqref{eq:cosmo_evo_frequency}. The full red line represents
  the evolution of $a^{\prime \prime}/a$ in arbitrary units in our
  simplified cosmological evolution~\eqref{eq:cosmo_evo_scale_factor}.
  The blue and green dotted lines represent two comoving frequencies
  $k_\text{s}^2$ and $k_\text{l}^2$ in arbitrary units which are
  constant during the evolution.  }
\label{fig:frequency_evolution}
\end{figure}

The evolution~\eqref{eq:cosmo_evo_frequency} of the time-dependent
piece $a^{\prime \prime}/a$ is plotted in
Fig.~\ref{fig:frequency_evolution} and is compared to the square of
the comoving frequencies of two different modes $k_{\text{s}}^2$ and
$k_{\text{l}}^2$.  Note that at the beginning of inflation and during
the radiation era, since $a^{\prime \prime} = 0$, all modes are
sub-Hubble and effectively living in Minkowski there\footnote{Recall
  that in this limiting case, the relation between the dominant term
  in the frequency and the wavelength size compared to the Hubble
  radius does not hold.  One cannot, strictly speaking, employ the
  terminology sub or super-hubble here.}. This second aspect is due to
our simplistic modelling of the transition in
Eq.~\eqref{eq:cosmo_evo_scale_factor}.  In a realistic cosmological
model, $a^{\prime \prime}$ is continuous and part of the modes
progressively reach the sub-Hubble regime.  In
Fig.~\ref{fig:frequency_evolution}, the mode $\bm{k}_\text{s}$ has a
short wavelength and is always sub-Hubble. It is not affected by the
amplification process.  The mode $\bm{k}_\text{l}$ has a larger
wavelength and becomes super-Hubble during inflation after $\eta_{k ,
  1}$, is insensitive to the expansion during radiation domination,
becomes super-Hubble again during matter domination, until $\eta_{k ,
  2}$ where it settles in the sub-Hubble regime.  The modes of
interest for cosmological observations are of the second type (or
become and stay super-Hubble during radiation domination).

The picture that we have just sketched for these modes, putting aside
radiation domination, is reminiscent of the idealized situation
described in Sec.~\ref{subsec:graviton_production} where the ``in"
region corresponds to $\eta \ll \eta_{k , 1}$ and the ``out" region to
$\eta \gg \eta_{k , 2}$.  For such modes, we are thus justified in
talking about graviton production.

Two remarks are in order here. First, modes progressively reenter the
Hubble radius during the neglected current accelerated expansion. For
the modes of interest here, this is of no consequence.  Second, one
should be careful when discussing modes responsible for the B modes of
polarisation in the CMB since some of them had not yet reached the
sub-Hubble regime when generating polarisation.

To close this discussion, we compute the relevant quantities
describing the gravitons in the different formalisms. For simplicity
we only consider an inflationary period where most of the
amplification occurs. For this estimate, we neglect slow-roll
corrections and model inflation by a period of de Sitter expansion
ending at $\eta_{\text{r}}$.  After the transition to radiation
domination, the mode does not feel the expansion anny more, so that
its particle content can be computed. In de Sitter, the covariance
matrix elements can be computed exactly using the mode function in
Eqs.~\eqref{eq:cosmo_evo_ref_mode_functions}. We evaluate them at
$\eta_{\text{r}}$ \cite{martinQuantumDiscordCosmic2016b}
\begin{equation}
    \gamma_{11} = 1 + \frac{1}{k^2 \eta_{\text{r}}^2} \approx e^{2N}
    \, , \quad \gamma_{12} = - \frac{1}{k \eta_{\text{r}}} \approx
    e^{N} \, , \quad \gamma_{22} = 1 \, .
\label{eq:dS_covariance}
\end{equation}
where, since we are considering a mode which is in the super-Hubble
regime during inflation, we have taken the limit $k \eta_{\text{r}}
\ll 1$. These last expressions are given in terms of the number of
$e$-folds $N$ defined by $N = \ln \left[ a \left( \eta \right) / a
  \left( \eta_k \right) \right] = \ln \left[ k \left( 2
  \eta_{\text{r}} - \eta \right) \right] $ where $a \left( \eta_k
\right)$ is the scale factor evaluated at Hubble crossing time $k
\left( 2 \eta_{\text{r}} - \eta_k \right) = 1$.

At this point, one notes that $ \left \langle
\left(\hat{h}^{\prime}_{\lambda , \bm k} \hat{h}^{\prime}_{\lambda , -
  \bm k} \right)^2\right \rangle \propto \gamma_{22} / a^2 (\eta)$, so
that, for a super-Hubble mode, it decays exponentially during
inflation.  The fact that the matrix element $\gamma_{11} =
2k\left\langle \left(\hat{\mu}_{\bm k}^\textsc{r}\right)^2\right
\rangle$ grows faster than $\gamma_{12}$ and $\gamma_{22}$ leads to
squeezing in a direction close to that of the $\mu_{\bm k}^\textsc{r}$
axis. This can be verified by computing explicitly the squeezing
parameters: inverting Eq.~\eqref{eq:covariance_matrix_squeezed_state},
we deduce the squeezing parameters as
\begin{equation}
    r_k = \mathrm{arcsinh} \left( \frac{1}{2 k \eta_{\text{r}}}\right)
    \quad , \quad \varphi_k = \frac{\pi}{2} - \frac{1}{2} \arctan
    \left(2 k \eta_{\text{r}} \right),
\label{eq:dS_squeezing}
\end{equation}
which, in the super-Hubble limit, yields
\begin{equation}
\label{eq:dS_superH_squeezing}
    r_k \approx  \ln \left(  \frac{1}{2 k \eta_{\text{r}}}  \right) \approx N
    \, , \quad \varphi_k \approx \frac{\pi}{2} - k \eta_{\text{r}}
    \approx \frac{\pi}{2} - e^{-N} \, .
\end{equation}
$\varphi_k \to \pi / 2$ so that indeed the ellipse will be squeezed in
a direction close to $\mu_{\bm k}^\textsc{r}$.

For scales of cosmological interest, one typically expects $N = \ln
\left( k \eta_{\text{r}} \right) \sim 50$ at the end of inflation so
that $r \approx 50$. This is to be compared with the best quantum
optics experiments where one can hardly achieve $r \approx 2$; the
squeezing is extreme \cite{martinCosmicInflationQuantum2019}. The
resulting evolution of the Wigner function is plotted for a few
$e$-folds after Hubble exit in Fig.~\ref{fig:wigner_de_sitter_no_deco}
where the very large squeezing is manifest.

Finally, we compute the number of particles created and their pair
correlation: Eq.~\eqref{eq:nk_ck} in the de Sitter and super-Hubble
limits gives
\begin{align}
n_{k} & = \frac{1}{4 k^2 \eta^2} = \frac{e^{2N}}{4} \, , \\ c_k & =
\frac{1}{4 k^2 \eta^2} - \frac{i}{2 k \eta} \approx \frac{e^{2N}}{4}
\, .
\label{eq:dS_nk_ck}
\end{align}
The number of pairs and their correlation grow at the same rate;
squeezing necessarily creates entangled pairs. After $50$ $e$-folds of
inflation, one finds $n_k \sim 10^{43}$. This number might appear very
large, but the physical field $h_{ij}$ is diluted by the inverse of
the scale factor that will keep acting even when the creation process
stops, following Eq.~\eqref{Fourierhij}.  In addition, the number of
gravitons is not directly observable; we observe gravitational waves
or their imprint on other fields, e.g. the electromagnetic field in
the CMB, but not individual gravitons. One therefore needs to compute
the physical quantities that are more directly relevant in forecasting
future observations.

\subsubsection{Connection to observations \label{sec:connection_observations}}

There is hope that observable signatures of these primordial
gravitational waves will be found either in the $\bm{B}$-modes of the
CMB or directly in future gravitational wave interferometers. We refer
to Ref.~\cite{Caprini:2018mtu} or Chapter 19, 20 and 23 in
Ref.~\cite{maggioreGravitationalWavesVol2018} for a detailed
account. The waves we have described are stochastic in nature owing to
their quantum origin. They account for part of the stochastic
gravitational-wave background (SGWB), the rest being produced by
unresolved astrophysical sources or possibly other high-energy
phenomena such as topological defects. The SGWB is usually assumed to
be statistically homogeneous and isotropic, as the FLRW background
metric, Gaussian, either due to the sum of a large number of
independent sources or because it is sourced by a Gaussian state as
considered here, and unpolarised (same content in both polarisations
and polarisations are uncorrelated) \footnote{It can be checked using
  \eqref{eq:coefficients_helicities_as_a_fn_plus_times} that the
  assumptions that the waves are both unpolarised $\left \langle
  \mu_{+} \mu_{+}^{\star} \right \rangle = \left \langle \mu_{\times}
  \mu_{\times}^{\star} \right \rangle$ and that polarisations
  uncorrelated $\left \langle \mu_{+} \mu_{\times}^{\star} \right
  \rangle = 0$ in terms of the $+ , -$ helicity basis is equivalent to
  the same two assumptions on the $+ , \times$ basis.} because there
is no significant source of parity violation in the Universe
\cite{Caprini:2018mtu}. All these assumptions only have to be made on
the initial state as the dynamics is the same for both fields
$\hat{\mu}_{\lambda}$ and preserves isotropy and homogeneity. They are
in particular satisfied for primordial gravitational waves produced
from the Bunch-Davies vacuum. A typical quantity used to characterize
a stochastic ensemble of waves is their power spectrum which, within
the gaussianity assumption, contains all the information. The power
spectrum $\mathcal{P}_\textsc{t}$ of gravitational waves $h_{ij}$ at
time $\eta$ is then defined (working classically for the moment) by
\begin{equation}
\left \langle \mu_{\lambda} \left( \bm{k} , \eta \right)
\mu_{\lambda^{\prime}}^{\star} \left( \bm{k^{\prime}} , \eta \right)
\right \rangle = \frac{\pi a^2 \left( \eta \right)}{16 \GN k^3}
\delta^{(3)} \left( \bm{k} - \bm{k^{\prime}} \right) \delta_{\lambda ,
  \lambda^{\prime} } \mathcal{P}_\textsc{t} \left( k , \eta \right),
\label{def:power_spectrum_GW}
\end{equation}
where the Dirac delta comes from homogeneity, and
$\mathcal{P}_\textsc{t}$ only depends on $k$ since the background is
isotropic and unpolarised.  The index ``$\textsc{t}$'' stands for
``tensor'', to differentiate the latter from the scalar power spectrum
$\mathcal{P}_\textsc{s}$.  Using \eqref{Fourierhij}, \eqref{wij} and
the orthogonality relations of the tensors below \eqref{mustar}, one
finds the two-point correlation function of the Fourier coefficients
of $h_{ij}$
\begin{equation*}
    \left \langle h^{ij} \left( \bm{k} , \eta \right) h^{\star}_{ij}
    \left( \bm{k} , \eta \right) \right \rangle = \delta^{(3)} \left(
    \bm{k} - \bm{k^{\prime}} \right) \frac{4 \pi^2}{k^3}
    \mathcal{P}_\textsc{t} \left( k , \eta \right),
\end{equation*}
as well as the two-point correlation of $h_{ij}$ in real space, namely
\begin{equation*}
    \left \langle h^{ij} \left( \bm{x} , \eta \right) h_{ij} \left(
    \bm{x} , \eta \right) \right \rangle = 2 \int \mathrm{d} \ln
    \left( k \right) \mathcal{P}_\textsc{t} \left( k , \eta \right).
\end{equation*}
The power spectrum $\mathcal{P}_\textsc{t} \left( k , \eta \right)$
corresponds to the typical squared amplitude of the wave, per
logarithm of $k$, and per polarisation, at the time $\eta$. For
perturbations made of sub-Hubble modes $k \gg \left| \Hu \right|$, the
time dependent term of \eqref{mode} can be neglected and the energy
density of gravitational waves reads \footnote{Averaging is necessary
  even for a deterministic source of gravitational waves to make sense
  of their energy. The averaging can either be performed over a
  certain volume or a certain duration, see
  \cite{maggioreGravitationalWavesVol2018}. In the context of this
  review, the averaging in \eqref{def:energy_density_GW} refers to an
  ensemble average.}
\begin{equation}
    \rho_\textsc{gw} = \frac{1}{32 \pi \GN} \left \langle \dot{h}^{ij}
    \left( \bm{x} , \eta \right) \dot{h}_{ij} \left( \bm{x} , \eta
    \right) \right \rangle.
\label{def:energy_density_GW}
\end{equation}
For sub-Hubble modes, $h_{+,\times} \left( \bm{k} , \eta \right)
\propto e^{ i \left( \bm{k}.\bm{x} - k \eta \right)} / a\left(\eta
\right) $ so that neglecting terms in $\Hu$ with respect to $k$ we get
\begin{equation}
\left \langle \dot{h}^{ij} \left( \bm{k} , \eta \right) \dot{h}_{ij}
\left( \bm{k} , \eta \right) \right \rangle \approx k^2 \frac{ \left
  \langle h^{ij} \left( \bm{k} , \eta \right) h_{ij} \left( \bm{k} ,
  \eta \right) \right \rangle }{a^2 \left( \eta \right)}.
\end{equation}
Note that, since $h_{ij} $ dilutes as $a^{-1}$, $\rho_\textsc{gw}$
dilutes as $a^{-4}$, i.e. sub-Hubble modes dilute as standard
radiation. Expanding the energy density in Fourier space and
normalising by the critical energy density $\rho_\text{c} = 3 H^2 / 8
\pi \GN$, we get the energy fraction per logarithm of $k$ that is
directly expressed as a function of the power spectrum
\begin{equation}
    \Omega_\textsc{gw} \left( k , \eta \right) = \frac{1}{\rho_c}
    \frac{\mathrm{d} \rho_\textsc{gw}}{\mathrm{d} \ln{k}} =
    \frac{k^2}{6 H^2 a^2 \left( \eta \right)} \mathcal{P}_\textsc{t}
    \left( k , \eta \right).
\label{def:energy_fraction_GW}
\end{equation}
The power spectrum \eqref{def:power_spectrum_GW} and the energy
density fraction \eqref{def:energy_fraction_GW} are the two quantities
customarily used to assess the observability and constrain the models
of primordial gravitational waves. More precisely, we often estimate
the \textit{primordial} power spectrum, i.e. the power spectrum at the
beginning of radiation domination. The rest of the evolution is
encoded in so-called transfer functions; these can be estimated using
the previous computations. For actual comparison with observations,
they have to be computed numerically by solving Boltzmann-like
equations.

Let us then evaluate the primordial power spectrum by considering only
the initial phase of single field slow-roll inflation in the
cosmological evolution Eq.~\eqref{eq:cosmo_evo_scale_factor} and
assuming Bunch-Davies vacuum for both polarisations $\pm$. Using
\eqref{def:mu_Fourier_mode_expansion}, the power spectrum is
straightforwardly expressed in terms of the mode function for
$\hat{\mu}_{\bm{k}}$
\begin{equation}
    \mathcal{P}_\textsc{t} \left( k , \eta \right) = \frac{32 \GN
      k^3}{\pi a^2 \left( \eta \right)} \left| u_{k}^{\text{(infl.)}}
    \left( \eta \right) \right|^2.
\label{def:power_spectrum_mode_function}
\end{equation}
Making use of \eqref{eq:cosmo_evo_ref_mode_functions} and expanding
all the quantities at first order in the slow-roll parameter
$\epsilon$, we get
\begin{equation}
    \mathcal{P}_\textsc{t} \left( k , \eta \right) = H_k^2 \left( 1 -
    2 \epsilon \right) \left[ - k \left( \eta - 2 \eta_{\text{r}}
      \right) \right]^{3 + 2 \epsilon} \left| H^{(1)}_{\frac{3}{2} +
      \epsilon } \left[ - k \left( \eta - 2 \eta_{\text{r}} \right)
      \right] \right|^2,
\label{def:primordial_power_spectrum_slow_roll}
\end{equation}
where $H_k = H \left( \eta_{k} \right)$ and $\eta_{k}$ is the Hubble
crossing time $k / a \left( \eta_{k} \right) = H \left( \eta_{k}
\right)$.  This time is often taken in the super-Hubble limit $|k
\left( \eta_k - 2 \eta_{\text{r}} \right)| \ll 1$ relevant for
cosmological scales. We get
\begin{equation}
    \mathcal{P}_\textsc{t} \left( k , \eta \right) = \frac{2
      H_k^2}{\pi} \left[ 1 + 2 \left( 1 + \log 2 + \gamma_\textsc{e}
      \right) \epsilon \left( \eta_{k} \right) \right],
\label{eq:primordial_power_spectrum_superH}
\end{equation}
where $\gamma_\textsc{e}$ is the Euler-Mascheroni constant. The exact
magnitude of \eqref{eq:primordial_power_spectrum_superH} depends on
the values of $\epsilon$ and $H$ at Hubble crossing, which are
model-dependent quantities.

Current experiments have not been able to detect the primordial
gravitational-wave background but the combined (non)-observations of
Planck and \textsmaller{BICEP} experiments allow us to put bounds on
the tensor-to-scalar ratio in single-field slow-roll inflation
\cite{planckcollaborationPlanck2018Results2020a}. A discussion of its
potential observability in future gravitational wave interferometers
and with future CMB experiments can be found in \cite{Caprini:2018mtu,
  abazajianCMBS4ForecastingConstraints2022,
  campetiMeasuringSpectrumPrimordial2021}. Notice that different
models of the very early universe would change the prediction
\eqref{def:primordial_power_spectrum_slow_roll}: initially excited
states \cite{martinNonVacuumInitialStates2000}, temporary departures
from the single field slow-roll scenario
\cite{fumagalliPrimordialGravitationalWaves2022} or coupling with
extra fields \cite{dimastrogiovanniPrimordialGravitationalWaves2017}
might for instance be able to generate larger signatures than
single-field slow-roll inflation, while modifications of gravity in
the high energy regime could also lead to changes in the spectrum at
high frequencies, e.g. through introdution of a cut-off in theories of
lower dimensionality in the ultraviolet
\cite{mureikaDetectingVanishingDimensions2011}.  Finally, we want to
emphasise that the toy model of cosmological evolution of
Eq.~\eqref{eq:cosmo_evo_scale_factor} makes the unrealistic assumption
of an instantaneous reheating.  Adding a period of reheating is known
to significantly modify the resulting spectrum, e.g. the frequency at
which it starts to decay, thereby modifying observational perspectives
\cite{nakayamaProbingReheatingTemperature2008}.

\subsubsection{Quantum origin of the primordial gravitational waves}

To close this part, we want to comment on how quantumness enters the
prediction~\eqref{def:primordial_power_spectrum_slow_roll}.

First, a subtle point hidden in \eqref{def:power_spectrum_GW} is the
meaning of the averaging $\langle \rangle$. In the discussion of the
dynamics of perturbations, we have been computing averages in the
sense of expectation value for observables in a given quantum
state. It is a basic assumption of quantum mechanics that this would
be the expected average value of the physical quantity after repeated
measurements of it when the system is prepared in the same
state. Unfortunately, we only have one realisation of the history of
the Universe. Yet, using statistical isotropy, we can treat each
(sufficiently large) patch of sky as an independent realisation of the
same underlying random process and compute average values over this
ensemble of patches. Under an \textit{ergodicity} assumption, the
resulting correlation functions can then be compared to
\eqref{def:primordial_power_spectrum_slow_roll}, a procedure applied
to CMB data analysis
\cite{grishchukBestUnbiasedEstimates1997}. Additional arguments to
justify trading quantum averages for classical ones will be discussed
in Sec. \ref{sec:quantumness_primordial_GW}.

Second, as we repeatedly emphasised, since the linear evolution is the
same in the classical and quantum settings, the quantum aspect has to
be confined to the choice of initial state. The result
\eqref{def:primordial_power_spectrum_slow_roll} reflects the choice
that the waves emerged from initial \textit{vacuum fluctuations}. For
primordial gravitational waves, we are short of observational data to
test this prediction. Still, if we were to insist on having a purely
classical treatment, then a classical vacuum of gravitational waves,
i.e. $a_{\pm \bm{k}} \left( \eta_{\mathrm{in}} \right) = 0$ would
persist throughout the evolution. There would simply be \textit{no}
primordial gravitational waves. On the contrary, initial gravitational
waves would be classically amplified by cosmological expansion, but we
then have to motivate a specific choice for the initial distribution
of perturbations. For scalar perturbations, the CMB observations
already demonstrated a tremendously good agreement with the predicted
power spectrum $\mathcal{P}_\textsc{s}$ of initial vacuum fluctuations
for the modes observed
\cite{planckcollaborationPlanck2018Results2020a,
  dodelsonCoherentPhaseArgument2003}. Giving up on a quantum treatment
in the inflationary paradigm would then require providing an
\textit{ad hoc} classical theory that yields the same initial
conditions as the quantum vacuum. We could therefore argue that
observations of the scalar sector give indirect proof that
gravitational degrees of freedom should be quantised.

Yet, third, it is sometimes argued see
e.g. \cite{maggioreGravitationalWavesVol2018,
  hsiangNoIntrinsicDecoherence2022}, that the verification of the
prediction \eqref{def:primordial_power_spectrum_slow_roll} would
provide additional insights on the quantum aspect of gravity with
respect to the observation of the scalar perturbations. In the
treatment of scalar perturbations in single-field slow-roll inflation,
the appropriate gauge-invariant variable is the Mukhanov-Sasaki (MS)
field related to the perturbations of the inflaton $\delta \phi$ and
the gravitational potential $\Psi$ through
\begin{equation}
    v_{\textsc{ms}} = \frac{z}{\kappa} \left( \Psi + \Hu \frac{\delta
      \phi}{\phi_0^{\prime}} \right),
\label{def:mukhanov_sasaki}
\end{equation}
where $z = a \sqrt{2 \epsilon}$, $\epsilon$ being the first slow-roll
parameter, $\phi_0$ the homogeneous background inflaton field and
$\kappa$ the reduced Planck mass. This is a scalar field whose
Lagrangian is the same as \eqref{def:tilde_lagrangian_mu} upon
substituting $a \to z$, and up to normalisation. The MS field is then
quantised in exactly the same manner as the two polarisations of the
graviton and the power spectrum evaluated by initially choosing the
Bunch-Davies vacuum. However, in the absence of perturbations $\delta
\phi$ for the scalar field, the equation of motion of the scalar part
of the metric perturbations show that they can be set to zero. This is
the so-called synchronous gauge
\cite{maggioreGravitationalWaveExperiments2000}. The existence of
scalar perturbations then requires the presence of the scalar field
$\delta \phi$ and is not intrinsic to the gravitational degrees of
freedom. Even when $\delta \phi \neq 0$, in the synchronous gauge
$\Psi =0$ and only the scalar field contributes to the
perturbations. In this gauge, the whole quantification process and
evaluation of the power spectrum only deals with the physics of a
quantum scalar field that is not related to gravitational degrees of
freedom. This could therefore cast some doubt on whether some
gravitational degrees of freedom were even quantised in the first
place. Such ambiguity does not exist when dealing with gravitational
waves. In the absence of any anisotropic stress, the gravitational
waves persist, and the quantisation procedure is undeniably a
perturbative quantisation of the gravitational field. Verification of
\eqref{def:primordial_power_spectrum_slow_roll} would then be an
indirect observational proof that the gravitational field must be
quantised.

Finally, to mitigate the above discussion, let us also mention an
argument against our line of reasoning, as discussed, e.g. in
Ref.~\cite{martinInflationaryCosmologicalPerturbations2005}.  The
argument made above can be reversed, as one can also find a gauge in
which the perturbation of the field vanishes altogether, with $\delta
\phi = 0$, while the metric part is $\Psi \neq 0$; in this gauge, the
quantisation is then over an element of the metric only. In addition,
since perturbations of matter and geometry appear on each side of
Einstein equations (however, note that their quantum counterpart is
unknown), it is inconsistent to quantise only one degree of
freedom. The observational verification of the prediction for the
scalar power spectrum thus can be argued to be an indirect proof that
the gravitational field should be quantised.

\section{Quantum features in primordial gravitational waves ?}
\label{sec:quantumness_primordial_GW}

Given the quantum origin of primordial gravitational waves, it may
seem natural to wonder about their state's quantum or classical
character at present. While it is expected that we will never be able
to detect the signal produced by a single graviton
\cite{dysonGravitonDetectable2013}, a discrete spectrum of excitations
is not the only specific feature of a quantum theory. For instance,
entanglement is a \textit{statistical} quantum feature that can be
experimentally verified using Bell inequalities
\cite{bellEinsteinPodolskyRosen1964,allenSqueezingRelicGravitational1999}. The
exciting possibility that primordial gravitational waves exhibit such
features has been investigated since this idea was first put forward
by Grishchuk and Sidorov in
\cite{grishchukSqueezedQuantumStates1990a}. These discussions have
gradually introduced many concepts borrowed from low-energy quantum
physics, particularly quantum optics: squeezing, quasi-probability
distribution, decoherence, quantum discord. In this section, we will
review this line of research following a historical approach and
trying to show the progress brought by each contribution.

This section is structured as follows. First, arguments based on the
very squeezed character of the state are used to justify a classical
treatment to compute cosmological observables
\cite{albrechtInflationSqueezedQuantum1994}. This approach, sometimes
called ``decoherence without decoherence’'
\cite{polarskiSemiclassicalityDecoherenceCosmological1996b}, and its
critics are reviewed in Sec.~\ref{subsec:wkb_classicality}. It
turns out, however, that the classicality identified by these
works does not do away with all the quantum features of the state;
the state of the perturbations could
for instance violate a Bell inequality
\cite{campoInflationarySpectraViolations2006a}.  We review these
``quantum information" approaches in
Sec.~\ref{subsec:quantum_information_approach}. Lastly, taking into
account the weak interactions of the perturbations is necessary as
they would induce \textit{decoherence} which might erase the quantum
features exhibited at the linear level. This aspect is reviewed
in Sec.~\ref{subsec:decoherence_perturbations}.

For the most part, these works are based on analysing the state of a
quantum scalar field, which can either represent the MS field of
scalar perturbations or one of the polarisations of the tensor
perturbations. The mechanisms and arguments being the same for both,
we do not distinguish when citing works which refers to which
and are only specific when necessary.

\subsection{Classicalisation of perturbations without decoherence}
\label{subsec:wkb_classicality}

In \cite{grishchukSqueezedQuantumStates1990a}, the authors argue that
the perturbations exhibit non-classical features due to the fact that
the relevant quantum state is strongly squeezed. In order to make the
discussion precise yet simple, we focus again on the inflationary
period modeled by a de Sitter phase of expansion and assume initial
Bunch-Davies vacuum\footnote{The reasoning can also be extended to
  certain non-vacuum initial states
  \cite{lesgourguesQuantumtoclassicalTransitionCosmological1997}.};
the relevant equations were derived at the end of
Sec.~\ref{subsubsec:cosmo_evo}, and the squeezing is shown in
Fig~\ref{fig:wigner_de_sitter_no_deco}.

One of the arguments developed in
\cite{grishchukSqueezedQuantumStates1990a} is that the trajectory in
phase space of a classical system with given initial conditions is
represented by a point moving on a single curve.  The situation is
different for a quantum system.  Due to the intrinsic uncertainty
stemming from the Heisenberg principle, the trajectory is represented
by a moving surface. The quantum state that comes closest to mimicking
a classical trajectory would then be a coherent state. Indeed, its
trajectory in phase space is represented by a circle moving along a
single curve: the system is located within a tube of minimal
uncertainty around the classical trajectory. On the contrary, the
surface representing an increasingly squeezed state is stretched
around its centre delocalising the position of the system away from
any single curve. Therefore, they argue, a very squeezed state like
that of the cosmological perturbations is a very \textit{quantum}
state.

In a couple of works written in response
\cite{albrechtInflationSqueezedQuantum1994,
  polarskiSemiclassicalityDecoherenceCosmological1996b}, the authors
reproduce and complement the computations made in
\cite{grishchukSqueezedQuantumStates1990a}, but give a different
interpretation of the result. The gist of their arguments, which we
reproduce below, is that the properties of a system in an extremely
squeezed state are indistinguishable from that of a classical system
whose state is represented by a \textit{classic stochastic
  distribution}; an argument borrowed
from~\cite{guthQuantumMechanicsScalar1985a}. In other words, although
the intrinsic quantum uncertainty on the outcome of a measurement
dramatically spreads due to the evolution of the system, this
uncertainty cannot be distinguished from a purely classical one.

To demonstrate this, let us consider the wavefunction of the
perturbations in the modes $\pm \bm{k}$ decomposed in the
$\textsc{r}/\textsc{i}$ sector and given by
Eq.~\eqref{eq:wavefunction_one_mode_squeezed_state}. Discarding the
indices ``$k$'', we recall that this is the wavefunction of a 1-mode
squeezed state. We can show that for large $r$, it satisfies very well
the conditions of the WKB approximation. For a general wavefunction
$\Psi \left( \mu \right) = C \left( \mu \right) \exp \left[ i S \left(
  \mu \right) / \hbar \right]$ the WKB approximation is valid when the
amplitude $C$ varies slowly compared to the phase $S$: $\left|
\partial S/\partial\mu \right| \gg \left| C^{-1} \partial
C/\partial\mu \right|$. Since the WKB approximation is generally
understood as a semi-classical limit, this property is sometimes
referred to as the ''WKB-classicality" of the state. Using the
wavefunction~\eqref{eq:wavefunction_one_mode_squeezed_state}, we have
\begin{subequations}
\begin{align}
\label{eq:amplitude_phase_wavefunction}
C \left( \mu \right) & = \left( \frac{1}{ \pi \gamma_{11} }
\right)^{1/4} e^{- \frac{k \mu^2}{2 \hbar \gamma_{11}}} \, , \\ S
\left( \mu \right) & = k \mu^2 \frac{ \gamma_{12} }{ 2 \hbar
  \gamma_{11} },
\end{align}
\end{subequations}
where we have dropped the exponent $\textsc{s}$ and the index $\bm{k}$
for simplicity. We get
\begin{equation}
\label{eq:WKB_squeezed_state}
\left| \frac{C}{\partial C/\partial\mu}
\frac{\partial S/\partial\mu}{\hbar}
\right| = \left| \sin \left( 2 \varphi_k \right) \sinh \left( 2 r_k
\right) \right| \, .
\end{equation}
In the de Sitter case, using Eq.~\eqref{eq:dS_squeezing}, one has
$\sin \left( 2 \varphi_k \right) \sinh \left( 2 r_k \right)
\approx e^N$; the condition is perfectly
satisfied. We then compute the action of $\hat{\mu}$ and $\hat{\pi}$
on such a state
\begin{subequations}
\begin{align}
\hat{\mu} \Psi \left( \mu \right) & = \mu \Psi \left( \mu \right),
\\ \hat{ \pi } \Psi \left( \mu \right) & = - i \hbar \frac{\partial
  \Psi}{\partial \mu} = \frac{\partial S}{\partial\mu}
  \left( 1 - i \hbar
\frac{\partial C/\partial\mu}{C \partial S/\partial\mu}
\right) \Psi \left( \mu
\right) \approx\frac{\partial S}{\partial\mu}
\left( \mu \right) \Psi \left( \mu
\right),
\end{align}
\end{subequations}
where in the last line we have used
Eq.~\eqref{eq:WKB_squeezed_state}. This last equality suggests that,
neglecting sub-dominant contributions, we could attribute an
unambiguous value to the ``momentum" $\pi$ through the relation $\pi
\approx \partial S /\partial
\mu$~\cite{albrechtInflationSqueezedQuantum1994} while the value of
the position $\mu$ would be controlled by the probability distribution
given by the $\mu$-representation of the wavefunction, namely
\begin{equation}
\label{eq:proba_distribution_wavefunction}
P \left( \mu \right) = C \left( \mu \right)^2 = \left( \frac{k}{\pi
  \hbar \gamma_{11} } \right)^{1/2} e^{- \frac{k \mu^2}{\hbar
    \gamma_{11}}} \, .
\end{equation}
To make this intuition rigorous, which is not always possible as we
explain at the end of the section, we have to use a phase space
representation of the state. The Wigner function $W^{\textsc{s}}
\left( \mu , \pi \right)$ can be factorised
\begin{align}
\begin{split}
    W^{\textsc{s}} \left( \mu , \pi \right) & = \sqrt{\frac{k}{\pi
        \hbar \gamma_{11}}} e^{- \frac{ k \mu^2}{\hbar \gamma_{11}}}
    \sqrt{ \frac{\gamma_{11}}{k \pi \hbar } } e^{-
      \frac{\gamma_11}{\hbar k} \left( \pi -
      \frac{\gamma_{12}}{\gamma_{11}} k \mu \right)^2 } \, , \\ & = P
    \left( \mu \right) \sqrt{ \frac{\gamma_{11}}{k \pi \hbar } } e^{-
      \frac{\gamma_{11}}{\hbar k} \left( \pi -
      \frac{\gamma_{12}}{\gamma_{11}} k \mu \right)^2 } \, ,
\end{split}    
\end{align}
where the relation $\det \left( \gamma \right) = \gamma_{11}
\gamma_{22} - \gamma_{12}^2 = 1$ (we are using
a pure state) was
used. The first piece is the probability distribution
\eqref{eq:proba_distribution_wavefunction}. The second piece controls
the value of $\pi - \gamma_{12} k \mu/ \gamma_{11} = \pi -
\partial S/\partial\mu $, i.e. the difference between the actual value of
$\pi$ and that attributed to it following the WKB-classicality
approach. It can be read out from the above, or shown by a
straightforward computation using covariance matrix elements, that
\begin{equation}
\label{eq:wkb_error}
    \left \langle \left( \hat{\pi} - \frac{\gamma_{12}}{\gamma_{11}} k
    \hat{\mu} \right)^2 \right \rangle = \frac{k}{2}
    \frac{1}{\gamma_{11}} \approx \frac{k}{2} e^{-2 N} \, ,
\end{equation}
where we have taken the super-Hubble limit in the last equality. Since
the state is Gaussian, and $\hat{\pi}$, $\hat{\mu}$ are centred, this
is the only quantity that controls the error induced by replacing
$\hat{\pi}$ by its WKB counterpart $\gamma_{12} k
\hat{\mu}/\gamma_{11}$ in the expectation values. As inflation
proceeds, this error becomes exponentially small while the
fluctuations of $\hat{\mu}$ get exponentially large, and that of
$\gamma_{12} k \hat{\mu}/\gamma_{11}$ tends to a constant.  Therefore,
to compute the expectation value of any operator which is a polynomial
in $\hat{\mu}$ and $\hat{\pi}$, one can safely make the WKB
replacement. We emphasise that, to have meaningful operators, the
coefficients of these polynomials must not depend on the state of the
system. In such polynomials, when expanding $\hat{\pi}$ as $(\hat{\pi}
- \gamma_{12} k \hat{\mu} / \gamma_{11} ) + \gamma_{12} k \hat{\mu} /
\gamma_{11}$, the coefficients of $\hat{\mu}$ and $\hat{\pi}$ cannot
conspire to yield an expression depending only on the subdominant
combination $\hat{\pi} - \gamma_{12} k \hat{\mu} / \gamma_{11}$ since
it explicitly depends on the squeezing parameters. The translation of
this approximation in terms of the Wigner function is to take the
limit of infinite $r_k$, with $\gamma_{11} \to \infty$, and to replace
the Gaussian over $\pi - \partial S/\partial\mu$ by a Dirac delta
\cite{grishchukAmplificationGravitationalWaves1975,
  albrechtInflationSqueezedQuantum1994,
  polarskiSemiclassicalityDecoherenceCosmological1996b}
\begin{equation}
\label{eq:wigner_delta}
W^{\textsc{s}} \left( \mu , \pi \right) \approx P \left( \mu \right)
\delta \left( \pi - \frac{\gamma_{12}}{\gamma_{11}} k \mu \right) \, .
\end{equation}
The interpretation of this equation is straightforward: when computing
expectation values using the Wigner function and
Eq.~\eqref{def:stochastic_average}, up to very sub-dominant
contributions, we can replace $\pi$ by $\partial S /\partial \mu$ in
the Weyl transform and take the average on $\mu$ using the classical
stochastic variable of distribution
Eq.~\eqref{eq:proba_distribution_wavefunction}.  In the limit of
Eq.~\eqref{eq:wigner_delta}, the contour levels of the Wigner function
are squashed from ellipses to lines, and this implies that the size of
the sub-fluctuant mode has been neglected. This line-like limit of the
Wigner function is visible in the last panels of
Fig.~\ref{fig:wigner_de_sitter_no_deco}.

We conclude with a series of remarks on this result.  First, it is
clear that the replacement $\hat{\pi} \to \partial S/\partial\mu$
cannot be exact as it implies $\left[ \hat{\mu} , \hat{\pi} \right] =
0 \neq i \hbar$, thus violating the canonical commutation relations,
although those must be verified irrespective of the state of the
system. Yet, the contribution of this non-vanishing commutator to the
expectation value of operators $O \left( \hat{\mu} , \hat{\pi}
\right)$ which are polynomial in $\hat{\mu}$ and $\hat{\pi}$ is
negligible.

The second remark is that, as explained in
\cite{martinCosmicInflationQuantum2019}, we want to emphasise that the
Wigner function of a WKB state does \textit{not} in general give rise
to a Dirac delta; in fact, it needs not even be positive
everywhere. The naive intuition is only verified here because the
state is also Gaussian. In addition, the fact that the Wigner function
can be negative suggests taking with a grain of salt the idea that any
WKB state is understandable as an approximate classical state.

Thirdly, as stressed in \cite{hsiangNoIntrinsicDecoherence2022}, the
distribution \eqref{eq:wigner_delta} has some undesirable features for
a Wigner function. For instance, computing the purity using the
function \eqref{eq:wigner_delta} and
Eq.~\eqref{def:stochastic_average} yields an infinite result. This is
obviously incorrect since for any quantum state $p_k \leq 1$, and, in
this pure case, we had derive earlier $p_k =1$. Geometrically, by
squeezing the ellipse to a line, one looses the information on the
area that encodes the purity and the non-commutation of the variables
through the Heisenberg uncertainty principle. This additionally
informs us that there exist quantities of interest that crucially
depend on the sub-leading contributions that were neglected, and so on
the sub-fluctuant mode.

The fourth point we want to stress concerns classicality.  The part of
the argument based on analysing the phase space distribution does not
actually require large squeezing to be formulated. Indeed, even before
taking any limit, the Wigner function of the state is everywhere
positive and obeys the classical equations of motion
\eqref{eq:liouville_wigner}, so that using
Eq.~\eqref{def:stochastic_average}, any observable can be computed
using a classical stochastic distribution.

As a fifth point, let us note that the above statement has to be made
more precise because it hides several subtle points. To start with, as
pointed out in \cite{martinQuantumDiscordCosmic2016b}, beyond
quadratic order, the Weyl transform of an observable $O \left(
\hat{\mu} , \hat{\pi} \right)$ is, in general, not obtained by
replacing the operators $\hat{\mu} $ and $ \hat{\pi}$ by the
corresponding phase space variables i.e. $O \left( \mu , \pi \right)
\neq \tilde{O} \left( \mu , \pi \right)$. For instance
\begin{equation}
    \widetilde{ \hat{\mu}^2_{ \bm{k}} \hat{\pi}^2_{\bm{k}}} +
    \widetilde{\hat{\pi}^2_{\bm{k}} \hat{\mu}^2_{\bm{k}} } = 2
    \mu_{\bm{k}}^2 \pi_{\bm{k}}^2 - \hbar \, ,
\end{equation}
so that, using Eq.~\eqref{def:stochastic_average}
\begin{equation}
    \left \langle \hat{\mu}^2_{ \bm{k}} \hat{\pi}^2_{\bm{k}} +
    \hat{\pi}^2_{\bm{k}} \hat{\mu}^2_{\bm{k}} \right \rangle = 2
    \mathbb{E} \left( \mu^2_{\bm{k}} \pi^2_{\bm{k}} \right) - \hbar.
\end{equation}
This extra $\hbar$ is a contribution of the commutator that the
Wigner-Weyl formalism takes into account. Therefore, despite the
Wigner function being everywhere positive and acting as a measure in
Eq.~\eqref{def:stochastic_average}, these terms introduce a slight
difference with classical stochastic distributions. The culprit is the
Weyl transform of the operators rather than the Wigner function. As
argued above, in the large squeezing limit, these extra contributions
to the Weyl transform of $\hat{\mu}$ and $\hat{\pi}$ are expected to
become negligible. The second subtle point is precisely that these
distortions will \textit{not} become negligible for all observables so
that the classicality argument does not apply to these.  The fact that
certain quantum features persist should not be a surprise since we
have shown that the gravitons produced by the evolution remain in
entangled pairs in the absence of other
interactions~\cite{hsiangNoIntrinsicDecoherence2022}.

The findings of this section can be summarised as follows: as long as
we measure only $\hat{\mu}$ and $\hat{\pi}$, or observables which are
polynomials of it, super-Hubble modes behave classically since their
expectation values can be completely reproduced by a classical
stochastic distribution
\cite{campoInflationarySpectraViolations2006a,martinQuantumDiscordCosmic2016b}.

\subsection{Quantum information approaches \label{subsec:quantum_information_approach}}

It has to be mentioned that the authors of
Ref.~\cite{albrechtInflationSqueezedQuantum1994} do recognise the
possibility that other operators would exhibit quantum features since
squeezed states are known to possess such features in quantum optics
experiments.  However, they dismiss this possibility by arguing that,
contrary to quantum optics, one can only perform measurements of the
values of the fields $\hat{\mu}_{\bm{k}}$ and $ \hat{\pi}_{\bm{k}}$
and not, say, of the number of particles $\hat{n}_k$. Therefore the
`decoherence without decoherence' argument is sufficient to claim that
the perturbations are \textit{practically} classical.  Setting
temporarily aside the question of their observability, we now derive
examples of operators revealing non-classicality features in the state
of primordial gravitational waves.

We have already mentioned that the purity of the state cannot be
computed if the sub-dominant contributions of the non-vanishing
commutators are dropped. In
\cite{lesgourguesPhaseSpaceVolumePrimordial1997}, the authors showed
that in order to correctly compute the entropy of the state using the
von Neumann entropy $ S \left( \hat{\rho} \right) = - \mathrm{Tr}
\left[ \hat{\rho} \log \left( \hat{\rho} \right) \right]$ the
sub-dominant contributions have to be restored.  For a 2-mode mode
squeezed state , the von Neumann entropy reads
\cite{demariePedagogicalIntroductionEntropy2018}
\begin{align}
\begin{split}
      S \left( \hat{\rho} \right) = 2 f \left[ \mathrm{det} \left(
        \gamma \right) \right] \, ,
\label{def:entropy_neumann}  
\end{split}
\end{align}
where the growing function $f$ is defined for $x \geq 1$ by
\begin{align}
\label{def:function_f}
 f(x)=\left(\frac{x+1}{2}\right) \log_2\left(\frac{x+1}{2}\right)
 -\left(\frac{x-1}{2}\right) \log_2\left(\frac{x-1}{2}\right) \, .
\end{align}
The entropy and purity are both controlled by the determinant of the
covariance matrix, which requires the inclusion of sub-dominant
contributions to be correctly evaluated. For the pure 2-mode squeezed
state of perturbations, one gets $\det \left( \gamma \right) = 1$ and
the definition of $f$ gives $f(1)=0$, so we recover that the entropy
vanishes.

We have so far only shown that, for certain operators, it is not
appropriate to neglect the sub-fluctuant mode. We now go further and
exhibit quantities whose values \textit{cannot} be accounted for if
the system is described by a classical stochastic distribution. The
prime example of such quantities is the combinations of expectation
values of spin operators entering the famous Bell inequalities
\cite{bellEinsteinPodolskyRosen1964}. To design a Bell inequality, one
has to exhibit a combination of operators $C \left( \hat{O}_1 , ... ,
\hat{O}_n \right)$ such that, if the expectation values of the
$\hat{O}_i$s are described by a stochastic probability
distribution\footnote{The precise assumption is that their values are
  described by a local realistic theory. For a discussion of this
  subtle and important point we refer to
  \cite{maudlinWhatBellDid2014}.}, then $C$ is bounded by a real
number $c$
\begin{equation}
    C \left( O_1 , ... , O_n\right) \leq c \, .
\end{equation}
As a consequence, if a quantum state is such that $\left \langle C
\left( \hat{O}_1 , ... , \hat{O}_n\right) \right \rangle > c$, then we
have proven that not all expectation values of this state can be
accounted for by a classical probabilistic theory.

A necessary condition for a state to violate a Bell inequality is that
it is not \textit{separable}
\cite{wernerQuantumStatesEinsteinPodolskyRosen1989}. A state
$\hat{\rho}$ of a system that can be partitioned in two subsystems $A$
and $B$ is said to be separable in this partition if its density
matrix can be written as \begin{equation} \hat{\rho} = \sum_i p_i
  \hat{\rho}_{\mathrm{A}}^i \bigotimes \hat{\rho}_{\mathrm{B}}^i \, ,
\label{def:separability}
\end{equation}
where $p_i \geq 0$ and $\sum_i p_i \geq 0$. Such a state can be
constructed using a classical protocol
\cite{wernerQuantumStatesEinsteinPodolskyRosen1989}. The
interpretation of Eq.~\eqref{def:separability} is that $p_i$ is the
probability of finding the system in the sector
$\hat{\rho}_{\mathrm{A}}^i \bigotimes \hat{\rho}_{\mathrm{B}}^i$ where
the subsystems $A$ and $B$ are independent since the density matrix is
factorised. The correlations between the subsystems are thus
controlled only by the probabilities $\left\{ p_i \right\}$ and deemed
classical. Non-separable states are generally called
\textit{entangled} states. In general, it is very difficult to
determine whether a state is separable. Fortunately, for Gaussian
states, the Peres-Horodecki criterion allows us to check separability
using the covariance matrix elements only
\cite{simonPeresHorodeckiSeparabilityCriterion2000a}. This method was
first applied to cosmological perturbations by Campo and Parentani in
\cite{campoInflationarySpectraPartially2005a}. We explain their result
in the terms used in this review.

We first need to choose a partition of the system. The separable
character of the state or not depends on the subsystems considered;
for a general discussion of the notion of partition, see
\cite{martinDiscordDecoherence2022}. Using the vectors of conjugate
operators introduced in Sec.~\ref{subsec:wigner_function}, we define a
(bi)partition of the system by sorting the operators into two vectors
of smaller dimensions
\begin{equation}
    \hat{X} = \hat{X}_{A} \bigoplus \hat{X}_{B} \, .
\end{equation}
To represent the state of the perturbations we have used the
$\textsc{r}/\textsc{i}$ partition defined by
$\hat{X}_{\textsc{r}/\textsc{i}} = \left( k^{1/2}
\hat{\mu}_{\bm{k}}^{\textsc{r}}, k^{-1/2}
\hat{\pi}_{\bm{k}}^{\textsc{r}}, k^{1/2}
\hat{\mu}_{\bm{k}}^{\textsc{i}}, k^{-1/2}
\hat{\pi}_{\bm{k}}^{\textsc{i}}\right)$ where the two subsystems
decouple. These operators will, however, mix the creation/annihilation
operators \eqref{def:minkowski_scalar_creation_operators} defining the
modes $\pm \bm{k}$. If we are interested in the correlations between
these modes we have to build separate hermitian operators describing
the mode $\bm{k}$ and $- \bm{k}$. This is readily done by considering
\begin{equation}
    \hat{q}_{\pm \bm{k}} = \sqrt{ \frac{\hbar}{2k} } \left(
    \hat{a}_{\pm \bm{k}} + \hat{a}^{\dagger}_{\pm \bm{k}} \right)
    \quad \text{and} \quad \hat{p}_{\pm \bm{k}} = - i
    \sqrt{\frac{\hbar k}{2}} \left( \hat{a}_{\pm \bm{k}} +
    \hat{a}^{\dagger}_{\pm \bm{k}} \right) \, .
\end{equation}
These operators define the $\pm \bm{k}$ partition $\hat{X}_{\pm
  \bm{k}} = \left( k^{1/2} \hat{q}_{ \bm{k}}, k^{-1/2} \hat{p}_{
  \bm{k}}, k^{1/2} \hat{q}_{- \bm{k}}, k^{-1/2} \hat{p}_{-
  \bm{k}}\right)$. We compute the covariance matrix in this partition
\begin{align}
\label{eq:gamma_pmk}
\gamma =
    \begin{pmatrix}
     \gamma_{\bm{k}} & \gamma_{\bm{k},-\bm{k}} \\
    \gamma_{-\bm{k},\bm{k}} & \gamma_{-\bm{k}}
  \end{pmatrix}\, ,
\end{align}
with 
\begin{align}
\gamma_{\bm{k}} = \gamma_{-\bm{k}} = \cosh{(2r_k)} \mathds{I}_{2} =
\left( n_k + \frac{1}{2} \right) \mathds{I}_{2} \, ,
\end{align}
where $\mathds{I}_{2}$ is the 2-dimensional identity matrix and
\begin{align}
\gamma_{\bm{k},-\bm{k}} = \gamma_{-\bm{k},\bm{k}} = -\sinh{(2r_k)} 
\begin{pmatrix}
    \cos{2 \varphi_k} & \sin{2 \varphi_k} \\ \sin{2 \varphi_k} &
    -\cos{2 \varphi_k}
  \end{pmatrix} = \begin{pmatrix}
    \Rez \left( c_k \right)   & \Imz \left( c_k \right)   \\
    \Imz \left( c_k \right)   &  - \Rez \left( c_k \right) 
  \end{pmatrix} \, .
\end{align}
Unlike in the $\textsc{r}/\textsc{i}$ partition, this covariance
matrix is not block-diagonal. It shows that the $\bm{k}$ and $-
\bm{k}$ particles are correlated. The Peres-Horodecki applied to this
covariance matrix reduces to
\cite{campoInflationarySpectraPartially2005a}
\begin{align}
\label{eq:2mss_separability_criterion}
\text{$\hat{\rho}$ separable in $\pm \bm{k}$ partition}
\Longleftrightarrow \left| c_k \right| \leq n_k \, .
\end{align}
This criterion lends itself to a very simple interpretation, the state
will be separable if and only if the correlation of the pairs is
larger than their number. When is this satisfied?  The condition
\eqref{eq:2mss_separability_criterion} is straightforwardly expressed
in terms of the squeezing parameters. We find that the state is
separable if only if $e^{-r_k} \geq 1$, i.e. for the vacuum $r_k =
0$. Therefore, the primordial gravitons pairs $\pm \bm{k}$ are always
entangled. We have found a first quantum feature of their
distribution. Notice that the same analysis could be repeated in the
$\textsc{r}/\textsc{i}$ partition, but since these sectors are not
correlated, it would trivially lead to the conclusion that the state
is always separable in this partition. This illustrates clearly the
dependence of the (non)-separable character of the state on the choice
of subsystems.

The state of the perturbations we have considered so far is pure. It
was shown that, for any entangled pure state, one can build a Bell
inequality that the state
violates~\cite{wernerQuantumStatesEinsteinPodolskyRosen1989}.  The
separability criterion is, in this case, sufficient. How can we find
operators able to violate a Bell inequality for the gravitons? The
considerations of Sec.~\ref{subsec:wkb_classicality} already
demonstrated that, in order to reveal the quantumness of the
distribution, we have to use operators which are non-polynomials in
$\hat{\mu}^{\textsc{s}}_{\bm{k}}$ and
$\hat{\pi}^{\textsc{s}}_{\bm{k}}$. In
\cite{revzenWignerFunctionDistribution2006}, Revzen further introduces
a distinction between what he calls proper and improper operators.

Proper operators are defined as those that cannot be used to violate a
CSH-type \cite{clauserProposedExperimentTest1969} Bell inequality when
the Wigner function of the state is positive. He shows that any
operator $\hat{O}$ whose Weyl transform $\tilde{O}$ takes values in
the set of its eigenvalues is proper. Indeed, the Wigner function then
provides an appropriate local hidden variable theory to describe its
expectation values. Therefore, we have to use operators that do not
fall in this category to build a Bell inequality that can be violated
by primordial gravitational waves. In fact, these operators are not
uncommon. Consider, for example, the number operator
\begin{equation}
    \hat{n}_{k} = \hat{a}^{\dagger}_{\bm{k}} \hat{a}_{\bm{k}} =
    \frac{k}{2 \hbar} \hat{\mu_{\bm{k}}}^2 + \frac{1}{2 \hbar k}
    \hat{\pi_{\bm{k}}}^2 + \frac{1}{2} \, .
\end{equation}
It has a discrete spectrum, while its Weyl transform
$\widetilde{\hat{n}_{k}} = \frac{k}{2 \hbar} \mu_{\bm{k}}^2 +
\frac{1}{2 \hbar k} \pi_{\bm{k}}^2 + \frac{1}{2}$ is a continuous
function of the phase space variables. In
\cite{campoInflationarySpectraPartially2005a}, Campo and Parentani
were the first to exhibit Bell inequalities violated by cosmological
perturbations. They emphasise the necessity to use non-polynomial
operators in the field operator and they use as a building block the
probability of finding the system in a certain 2-mode coherent state
\begin{align}
\begin{split}
Q \left( v , w \right) & = \mathrm{Tr} \left( \hat{\rho} \hat{\Pi}_{
  \bm{k}, -\bm{k} } \right) \, ,\\ & = \frac{1}{ \Delta_k } \exp
\left\{ - \frac{1}{ \Delta_k} \left[ \left( n_k+1 \right) \left(
  \left| v \right|^2 + \left| w\right|^2\right) - 2 \Rez \left( c_k^*
  v w \right) \right] \right\} \, ,
\label{def:husimi_rep}  
\end{split}
\end{align}
where $\Delta_k = \left( n_k +1\right)^2 - \left| c_k \right|^2$ and
$\hat{\Pi}_{ \bm{k}, -\bm{k} } = \left| v , \bm{k} \right \rangle
\left \langle v , \bm{k} \right| \otimes \left| w , -\bm{k} \right
\rangle \left \langle w , -\bm{k} \right|$ projects the subsystem
$\bm{k}$ (respectively $-\bm{k}$) on the coherent state associated to
$v \in \mathbb{C}$ (resp. $w \in \mathbb{C}$). The bounds given on
$n_k$ and $c_k$ in Sec.~\ref{subsec:squeezed_states} ensure that
$\Delta_k$ is a positive quantity. This real and positive function of
$v$ and $w$ is called the Husimi Q-representation of the state
\cite{gardinerQuantumNoiseHandbook2004}\footnote{Like the Wigner
  function, it is a phase-space representation of the state but using
  coherent states as a basis rather than eigenstates of the field
  operators. The authors discuss the quantumness of the perturbation
  using its properties and that of the related Glauber Sudarshan
  P-representation. They argue that the state not admitting a
  P-representation can be considered a non-classical feature. For
  brievity, we will not discuss these aspects here and refer to
  \cite{campoInflationarySpectraPartially2005a,gardinerQuantumNoiseHandbook2004}
  for details.}. For the purpose of building a Bell inequality, it can
be simplified by re-parametrising the arbitrary phase of $v$ to absorb
that of $c_k$. We take $\text{arg}\, v = 2 \text{arg}\, c_k$ so that
$2 \Rez \left( c_k^* v w \right) = 2 \left| c_k \right| \Rez \left(
v^* w \right)$. For a 2-mode squeezed state $\left| c_k \right| =
\sqrt{n_k \left( n_k +1 \right)}$ so that, upon rearranging,
\begin{align}
\begin{split}
Q \left( v , w \right) & = \frac{1}{n_k+1} \exp \left( - \frac{\left|
  v \right|^2}{ n_k+1} \right) \exp \left( - \left| w - v
\sqrt{\frac{n_k }{n_k+1}} \right|^2 \right) \, .
\label{def:husimi_2MSS}  
\end{split}
\end{align}
Since the Husimi representation is also the expectation value of an
operator, it can be used in a Bell inequality.  The authors then use
the Bell inequality demonstrated by
\cite{banaszekNonlocalityEinsteinPodolskyRosenState1998} over $Q
\left( v, w\right)$
\begin{align}
\begin{split}
    C \left( v , w \right) = \left[ Q \left( 0 , 0 \right) + Q \left(
      v , 0 \right) + Q \left( 0 , w \right) - Q \left( v , w \right)
      \right] \left( \frac{n_k + 1}{2} \right) \leq 1 \, .
\label{def:Bell_husimi}
\end{split}
\end{align}
They argue that $C$ is maximal for $w = -v$ in which case it only
depends on $\left|v \right|^2$ and
\begin{equation}
    C_{\text{max}} \left( \left| v \right|^2 \right) = \frac{1}{2}
    \left[1 + 2 e^{-\left| v \right|^2} - e^{-2 \left( 1 +
        \sqrt{\frac{n_k }{n_k +1}} \right)\left| v \right|^2} \right]
    \, .
\end{equation}
One can show that, provided we are not in the vacuum $n_k = 0$,
$C_{\mathrm{max}}$ is always larger than unity in the vicinity of
$v=0$, as illustrated in Fig.~\ref{fig:husimi_bell}; the Bell
inequality is violated. As expected, we have recovered the
separability condition. In a later work
\cite{martinObstructionsBellCMB2017a}, the authors proved that another
inequality, built using operators, also defined in
\cite{banaszekQuantumNonlocalityPhase1999}, that are complementary (in
the sense that their sum is the identity) to the projectors
$\hat{\Pi}_{ \bm{k}, -\bm{k} }$, is violated. They also build other
inequalities using the (GKM and Larsson) pseudo-spin operators in the
same work. They explicitly show that all these operators belong to the
subclass of improper operators identified by Revzen. Since the Weyl
transform of the identity is just the number $1$, we can infer from
their complementary with the projectors $\hat{\Pi}_{ \bm{k}, -\bm{k}
}$ that the operators $\hat{\Pi}_{ \bm{k}, -\bm{k} }$ also belong to
this subclass.

We now introduce a last non-classicality criterion, the quantum
discord. We start by giving the intuition behind its definition and
reviewing some important properties. Technical details in definitions
and proofs are skipped and can be found in
\cite{ollivierIntroducingQuantumDiscord2001,
  hendersonClassicalQuantumTotal2001}. The idea of quantum discord is
also to show that correlations between two subsystems are stronger
than allowed classically. Two measures of the information attached to
these correlations are introduced to that end. These measures are
based on the von Neumann entropy, which, as we have shown, is highly
sensitive to terms that can be neglected when computing field
expectation values. The first measure is the mutual information
\begin{equation}
   \mathcal{I}(A,B) = S(A) + S(B) - S(A, B), 
\label{def:mutual_information_I}
\end{equation}
where $S(A, B)$ is the von-Neumann entropy of the full system while
$S(A)$ and $S(B)$ are the entropies of the subsystems. The latter are
defined by computing the entropy of the reduced density matrices when
one of the subsystems is traced out, e.g. $ \hat{\rho}_{A} =
\mathrm{Tr}_{B} \left( \hat{\rho}\right)$ for the subsystem $A$. They
are also called the entanglement entropy of the state.  The second
measure
\begin{equation}
   \mathcal{J}(A,B) = S(A) - S(A | B), 
\label{def:mutual_information_J}
\end{equation}
where $S(A | B)$ measure the information gained on $A$ by measuring
$B$. Its precise definition in the quantum setting must therefore
include the system state after measuring the system $B$. It is
obtained by minimising the density matrix residual entropy after
having measured a complete set of projections on $B$, i.e. by
maximising the information gain. For a quantum state, we then define
the quantum discord as their difference
\begin{equation}
    \mathcal{D}(A,B) = \mathcal{I}(A,B) - \mathcal{J}(A,B) \, ,
\end{equation}
which is shown to be in general \textit{non-negative}. The key
observation is that, by the Bayes theorem, $\mathcal{I}$ and
$\mathcal{J}$ coincide for a classical system so that the discord
\textit{vanishes}. A non-vanishing discord $\mathcal{D}(A,B) > 0$ is
therefore taken as a non-classical feature. As the other criteria
introduced, the quantum discord depends on the choice of partition
$\hat{R} = \hat{R}_A \bigoplus \hat{R}_B$. However, it does not depend
on the operators chosen to represent them, i.e. it is invariant under
any change of operators within the sectors $A$ and $B$.  We call such
a quantity a \textit{local} symplectic invariant. On the contrary, a
Bell inequality is not necessarily a local symplectic invariant. A
last important property of the discord is that, for a pure state, it
reduces to the entanglement entropy $\mathcal{D}(A,B) = S(A) = S(B)$,
and, for a pure state still, being entangled is equivalent to a
non-vanishing entanglement entropy. Therefore, all criteria introduced
(separability, Bell inequality, quantum discord) are equivalent for
pure states. The cosmological perturbations must therefore have a
non-vanishing quantum discord.

The quantum discord of cosmological perturbations
was computed in
\cite{martinQuantumDiscordCosmic2016b} for the $\pm \bm{k}$
partition\footnote{The quantum discord was already used in a work on
  cosmological perturbations in
  \cite{limQuantumInformationCosmological2015} but the author
  considered correlation of another nature, namely that of the
  perturbations and their environment.}. It reads
\begin{equation}
\label{eq:discord_2MSS}
\mathcal{D}_{\pm \bm{k}} = f\left[ \cosh \left(2 r_k \right) \right],
\end{equation}
where $f$ was defined in Eq.~\eqref{def:function_f}. 
We immediately verify that the discord is non-vanishing provided
that $r_k > 0$, i.e. that we are not in the vacuum.
Taking the de Sitter limit of the above expression, we find 
$\mathcal{D}_{\pm \bm{k}} \approx 2
r_k /\ln 2 \approx 2 N/\ln 2$, the discord grows linearly
with the number of $e$-folds.

The results of this section demonstrate that, as suspected, the
primordial gravitational waves are only classical if we restrict our
attention to field operators $\hat{\mu}$ and $\hat{\pi}$. We showed,
using several criteria, that their state exhibits in principle quantum
features: it is entangled, violates Bell inequalities and has a
non-vanishing quantum discord.  We additionally verified that these
three criteria are equivalent for pure states like the 2-mode squeezed
state considered here.  Still, in any realistic model of the early
Universe, this assumption of purity has to be given up. What has
allowed us so far to simply consider a couple of modes $\pm \bm{k}$ of
the field is that we have neglected all interactions of the
gravitational waves, in particular their intrinsic non-linearities. We
were justified in doing since the latter are weak. Yet, it is well
known that even very weak interactions can lead to an erasure of
non-classical features by inducing \textit{decoherence} of the
system. The most famous example of this is probably that a grain of
dust whose spatial superposition would be turned into a classical
superposition in a fraction of an instant simply by the scattering of
photons from the CMB \cite{joosEmergenceClassicalProperties1985}. The
importance of decoherence in the discussion of quantum features of
cosmological perturbations was quickly realised
\cite{kieferEmergenceClassicalityPrimordial1998,
  kieferQuantumtoclassicalTransitionFluctuations1998a}. We now
investigate how it affects the state, in general, and in particular
the quantum features we have just exhibited.

\subsection{Decoherence of cosmological perturbations}
\label{subsec:decoherence_perturbations}

We start by briefly recalling some basic concepts of decoherence and
refer to \cite{kieferPointerStatesPrimordial2007} for details. The
2-mode squeezed state of a coupled of modes $\pm \bm{k}$ is a pure
state represented by the ket \eqref{eq:squeezed_state_ket}. One can
easily compute its density matrix and express it in the graviton
2-mode number basis
\begin{equation}
    \hat{\rho}_{\textsc{2mss}} = \frac{1}{\cosh^2 \left( 2 r_k \right)
    } \sum_{n , n^{\prime} = 0}^{+\infty} \left[ - \tanh \left( 2 r_k
      \right) \right]^{n+n^{\prime}} e^{2 i \left( n-n^{\prime}
      \right) \varphi_k} \left| n_{\bm{k} } , n_{-\bm{k}} \right
    \rangle \left \langle n^{\prime}_{\bm{k} } , n^{\prime}_{-\bm{k}}
    \right| \, .
\end{equation}
The coefficients on the diagonal $q_n = \tanh^{2 n} \left( 2 r_k
\right) / \cosh^2 \left( 2 r_{\bm{k}} \right)$ give a classical
probability distribution over the 2-mode number states, while the
non-diagonal reflects the quantum interferences between them. If we
discard these terms, the density matrix reads
\begin{equation}
    \hat{\rho}_{\mathrm{th.}} = \frac{1}{\cosh^2 \left( 2 r_k \right)
    } \sum_{n = 0}^{+\infty} \tanh^{2 n} \left( 2 r_k \right) \left|
    n_{\bm{k} } , n_{-\bm{k}} \right \rangle \left \langle n_{\bm{k} }
    , n_{-\bm{k}} \right| \, .
\label{eq:density_matrix_thermal_state}
\end{equation}
The state now represents a classical superposition of different number
states with the same probabilities as $\hat{\rho}_{\textsc{2mss}}$.
Such states are called statistical mixtures and are indeed mixed
states (except if all coefficients but one vanish) since $p_k =
\sum_{n} q_n^2$ and $q_n \leq 1$ . The general idea of decoherence is
that interactions of the system with a large number of unobserved
degrees of freedom, referred to as the environment, precisely
diagonalises the density matrix, driving the state to a statistical
mixture. Equation~\eqref{eq:density_matrix_thermal_state} is actually
the density matrix of a thermal state with, on average, $n_k$
particles in both modes. Since it is fully diagonal, it is considered
the result of a complete decoherence process. A very important point
is that the (non)-diagonal character of the matrix depends on the
basis, e.g. the matrix is originally diagonal in the 2-mode squeezed
state basis. The basis in which decoherence makes the density matrix
diagonal is called the pointer basis.  Once again, we see that the
choice of basis and operators to analyse the state of the system is
crucial.  For cosmological perturbations, several pointer basis were
considered: coherent state basis
\cite{mataczCoherentStateRepresentation1994,
  campoInflationarySpectraPartially2005a}, field amplitude basis
\cite{brandenbergerClassicalPerturbationsDecoherence1991,
  sakagamiEvolutionPureStates1988,
  kieferEmergenceClassicalityPrimordial1998,
  kieferQuantumtoclassicalTransitionFluctuations1998a}, number basis
\cite{brandenbergerClassicalPerturbationsDecoherence1991}, and others
\cite{gasperiniEntropyProductionCosmological1993}\footnote{Notice that
  some of these, \cite{sakagamiEvolutionPureStates1988,
    brandenbergerClassicalPerturbationsDecoherence1991}, predate works
  referred to in the last section. Decoherence was, in fact, already
  investigated in the context of the early Universe before the
  argument of `decoherence without decoherence' was made. It was
  especially used to try to make sense of the solutions of quantum
  cosmology, where both the background and the perturbations are
  treated as quantum fields
  \cite{kieferContinuousMeasurementMinisuperspace1987}.}. Ultimately,
in a realistic model, the pointer basis is given by the eigenstates of
the interaction Hamiltonian selected. The basis thus bears a double
physical sense: it tells us for which type of measurements the system
appears classical, e.g. measures of field amplitude or of number of
particles, and also to which operators of the system is the
environment sensitive. In their follow-up articles
\cite{kieferEmergenceClassicalityPrimordial1998,
  kieferQuantumtoclassicalTransitionFluctuations1998a} to
\cite{polarskiSemiclassicalityDecoherenceCosmological1996b}, the group
of authors (Kiefer, Lesgourgues, Starobinski, Polarski) considered the
effect of decoherence. They argued that the correct pointer basis
should be the field amplitude basis on the ground that
self-interactions of pure gravity are local in the field basis,
i.e. $H_{\mathrm{in}} \propto \hat{\mu}^{n} \left( \bm{x} , \eta
\right) \hat{\pi}^{m} \left( \bm{x} , \eta \right)$. Since these
interactions are contained in the Einstein-Hilbert action, they
constitute a minimal and well-defined source of decoherence.  They
were then taken into account in a more realistic model of decoherence
for the first time in
\cite{burgessDecoherencePrimordialFluctuations2008,
  martineauDecoherencePrimordialFluctuations2007}. There, the system
considered is made up of the observed large wavelengths while the
environment is made-up of the rest of the short unobserved wavelengths
like in stochastic inflation
\cite{starobinskyStochasticSitterInflationary1988}. This approach was
originally performed for scalar perturbations and was later
generalised to tensor perturbations
\cite{gongQuantumNonlinearEvolution2019}.

How is their influence on the state of $\pm \bm{k}$ modes concretely
accounted for? In \cite{burgessDecoherencePrimordialFluctuations2008,
  martineauDecoherencePrimordialFluctuations2007} the process is
followed in time, rather than assumed to have completed
\cite{mataczCoherentStateRepresentation1994,
  brandenbergerEntropyClassicalStochastic1992,
  gasperiniEntropyProductionCosmological1993}, using a master
equation. Earlier papers \cite{sakagamiEvolutionPureStates1988,
  brandenbergerClassicalPerturbationsDecoherence1991} had also used an
equivalent formalism, the Feynman-Vernon influence functional, but
only in solvable toy models with two scalar fields interacting
quadratically. The two formalisms were also used in
\cite{lombardoDecoherenceInflationGeneration2005}, using the
short-long wavelengths splitting and considering a quartic
self-interaction of the scalar field. To derive a master equation, one
starts by postulating that the couple system-environment evolves under
a Hamiltonian
\begin{equation}
  \hat{H}_{\mathrm{tot}}=\hat{H} \otimes \hat{\mathds{I}}_\mathrm{env}
  +\hat{\mathds{I}} \otimes \hat{H}_\mathrm{env} + g
  \hat{H}_\mathrm{int} \, ,
\end{equation}
where the Hamiltonian of interaction is taken to be an integral of a
product of operators acting on the system and the environment
\begin{equation}
    \hat{H}_\mathrm{int} = \int \D^3 {\bm x}\, \hat{A}( \eta
    ,{\bm x}) \otimes \hat{E}( \eta ,{\bm x}) \, .
\end{equation}
Under certain assumptions, essentially perturbative coupling and a
``large'' enough environment unperturbed by the action of the system,
the von Neumann equation over the full density matrix
$\hat{\rho}_{\text{tot}}$ can be reduced to a \textit{master} equation
over the reduced density matrix of the system $\hat{\rho} =
\text{tr}_{\text{env}} \left( \hat{\rho}_{\text{tot}} \right)$. Master
equations became a standard tool to analyse the decoherence of
cosmological perturbations and are very often considered to be of the
Lindblad-type, e.g.
\cite{burgessDecoherencePrimordialFluctuations2008,
  martineauDecoherencePrimordialFluctuations2007,
  kieferPointerStatesPrimordial2007,
  martinObservationalConstraintsQuantum2018},
\begin{equation}
 \frac{\D \hat{\rho}}{\D {\eta}} =-i\left[\hat{H}
   ,\hat{\rho}\right] - g^2 \eta_{\mathrm{c}} \int \D^3{\bm
   x}\, \D^3 {\bm y} \, \langle \hat{E}({\eta},{\bm
   x})\hat{E}({\eta},{\bm y})\rangle \left[\hat{A}({\bm x}) \, ,
   \left[\hat{A}({\bm y}),\hat{\rho}\right]\right],
\label{def:Lindblad_eq}
\end{equation}
where $\eta_{\mathrm{c}}$ is the auto-correlation time of the
environment. This is a Markovian master equation; it assumes that the
environment is effectively stationary with respect to the system,
i.e. $\eta_{\mathrm{c}} \ll \delta \eta$ where $\delta \eta$ is the
typical time-scale of evolution of the system. In addition, the
interaction term is often considered linear in the system field
operators $H_{\mathrm{int}} \propto \left( \alpha \hat{\mu} + \beta
\hat{\pi} \right) \otimes \hat{O}_{\mathrm{env}}$, where
$\hat{O}_{\mathrm{env}}$ acts only on the environment
\cite{polarskiSemiclassicalityDecoherenceCosmological1996b,
  martinObservationalConstraintsQuantum2018}. It is the so-called
Caldeira-Legget model
\cite{caldeiraInfluenceDissipationQuantum1981}. Such interactions can
also be identified as the dominant term when considering pure gravity
\cite{burgessDecoherencePrimordialFluctuations2008,
  gongQuantumNonlinearEvolution2019} and has the great advantage of
preserving gaussianity and homogeneity. The result of the evolution
can therefore be simply analysed by considering a Gaussian decohered
homogeneous density matrix (GHDM). This class of state was introduced
in \cite{campoDecoherenceEntropyPrimordial2008a,
  campoInflationarySpectraPartially2005a} to study decoherence finely,
without having to assume any specific master equation, and still
preserving a ``partially" decohered state rather than assuming from
the on-set the density matrix diagonal. This class also encompasses
the density matrices obtained by the common ansatz that its
non-diagonal terms are suppressed by a Gaussian,
e.g. \cite{kieferEntropyGravitonsProduced2000,
  martinObstructionsBellCMB2017a}. For all these reasons, we will in
this section analyse the effect of decoherence using the GHDM and
follow \cite{campoInflationarySpectraPartially2005a,
  martinDiscordDecoherence2022}.

\begin{figure}
\centering
\includegraphics[width=0.7\textwidth]{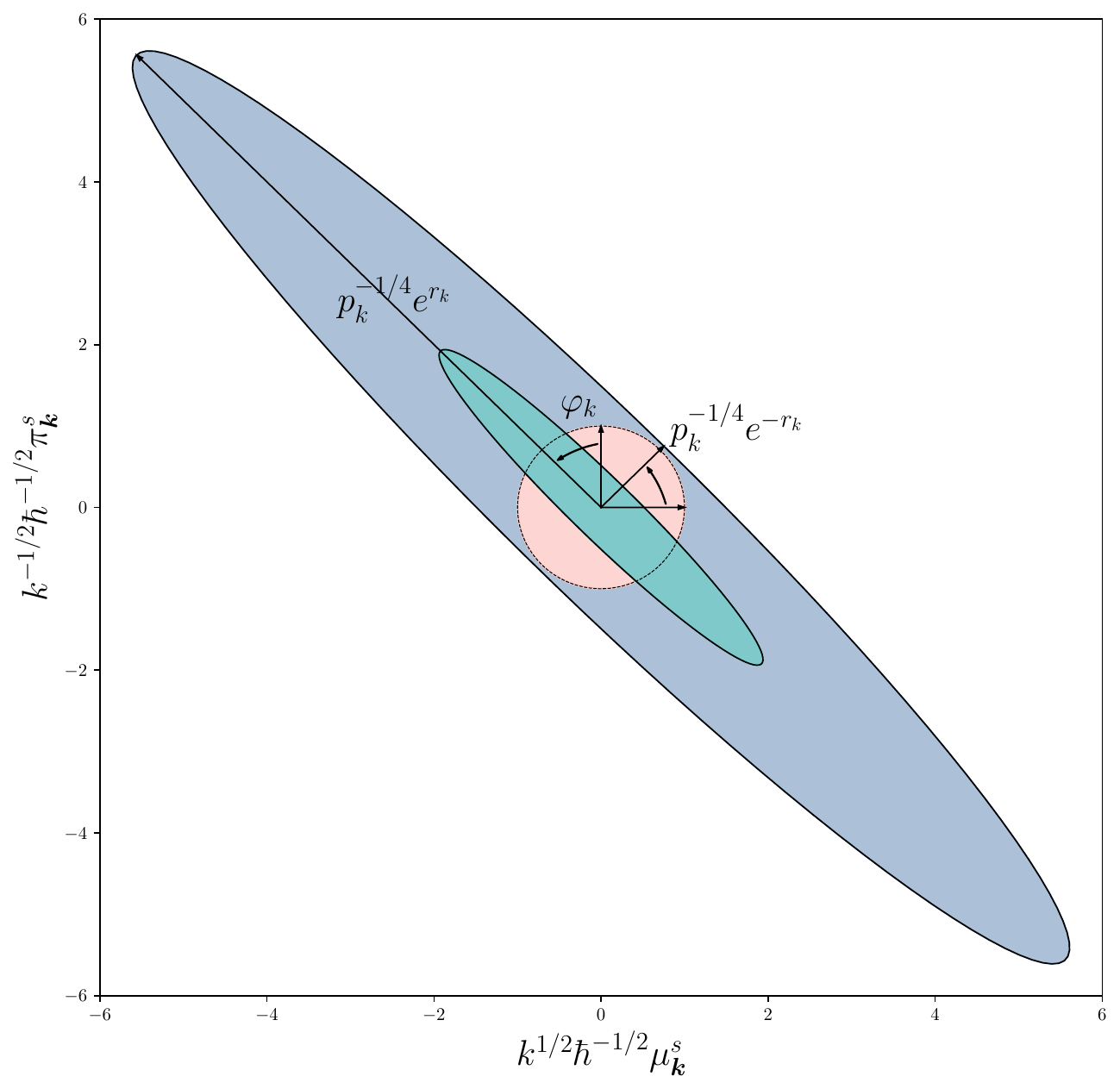}
\caption{$\sqrt{2}$-$\sigma$ contour level of the Wigner function
  $W^{\textsc{s}}$ for $\varphi_k = \pi / 4 $, $r_k = 1$, $p_k = 0.12$
  (blue ellipse) or $p_k=1$ (green ellipse) and the vacuum state $r_k
  = 0$ (pink circle).}
\label{fig:mixed_state_wigner}
\end{figure}

To define the GHDM, we work in Fourier space. First, to avoid a
preferred direction all 1-point correlation functions have to
vanish. The Gaussian state is then completely characterised by its
covariance matrix \eqref{def:covariance_matrix} made of 2-point
correlation functions. By homogeneity, the only non-vanishing 2-point
correlation functions involve $\bm{k}$ and $-\bm{k}$, and we can work
with a single couple of modes $\pm \bm{k}$. \textit{A priori} we have
a $4 \times 4$ matrix, but, as mentioned below
Eq.~\eqref{eq:wavefunction_one_mode_squeezed_state}, homogeneity
further imposes that the matrix is block diagonal in the
$\textsc{r}/\textsc{i}$ partition. We are left with a $2 \times 2$
covariance matrix like that of Eq.~\eqref{def:gamma}. The state is
then fully characterised by the three real covariance matrix elements
$\gamma_{ij}$ in Eq.~\eqref{eq:gammaij}, or alternatively the number
of pairs $n_k$ and their pair correlation $c_k$ (one complex and one
real number) defined in Eq.~\eqref{eq:nk_ck}. The only difference with
the previous analyses is that the constraint imposed by the purity of
the state $p_k = 1$ is now relaxed to $p_k \leq 1$, i.e. $\det \left(
\gamma^{\textsc{s}} \right) = \gamma_{11} \gamma_{22} - \gamma_{12}^2
\geq 1$, or equivalently $\left| c_k \right| \leq \sqrt{ n_{k} \left(
  n_{k} + 1 \right)}$. Notice that these numbers can still not be
arbitrarily chosen in order to keep a \textit{bona fide} quantum state
with purity bounded by one. Finally, to be able to have a simple
geometrical representation, we can use the purity as an effective
extra squeezing parameter and write
\cite{martinDiscordDecoherence2022}
\begin{align}
\gamma_{11} & = p_k^{-1/2} \left[ \cosh \left( 2 r_k \right) - \cos
  \left( 2 \varphi_k \right) \sinh \left( 2 r_k \right) \right] \, ,
\\ \gamma_{12} & = \gamma_{21} = - p_k^{-1/2} \sin \left( 2 \varphi_k
\right) \sinh \left( 2 r_k \right) \, , \\ \gamma_{22} & = p_k^{-1/2}
\left[ \cosh \left( 2 r_k \right) + \cos \left( 2 \varphi_k \right)
  \sinh \left( 2 r_k \right) \right] \, .
\label{def:effective_squeezing}  
\end{align}
One can check that this is a fully general parametrisation of a $2
\times 2$ symmetric matrix, that indeed $\det \left( \gamma \right) =
p_k^{-2}$ and that for $p_k =1$, we recover
Eq.~\eqref{eq:covariance_matrix_squeezed_state}. How is the
geometrical representation affected by this additional parameter? It
is readily seen that the eigenvectors of $\gamma$ are unchanged, and
its eigenvalues simply increased by $p_k^{-1/2} \geq 1$. The effect on
the $\sqrt{2}-\sigma$ contour levels is thus simply a dilation by
$p_k^{-1/4}$. This increased width of the Gaussian was already noticed
as an effect of decoherence in
\cite{kieferEntropyGravitonsProduced2000} and before in a different
context by~\cite{brodierSymplecticEvolutionWigner2004}.  An important
remark is that the existence of a sub-fluctuant mode due to squeezing
is not guaranteed anymore since the semi-minor axis is now of length
$B_k = p_k^{-1/4} e^{-r_k}$ which can always be made larger than one,
the vacuum value, provided that decoherence is strong enough at a
given a value of squeezing $r_k$.  Fig.~\ref{fig:mixed_state_wigner}
illustrates the ellipse corresponding to the state in
Fig.~\ref{fig:wigner_function_squeezing_parameters} after having lost
purity to $p_k = 0.17$; there is no sub-fluctuant direction.  We
mention an alternative parametrisation, used in
\cite{campoInflationarySpectraPartially2005a,
  campoInflationarySpectraViolations2006a}, where the extent of the
breaking of the relation between $n_k$ and $\left| c_k \right|$ is
used to interpolate between a 2-mode squeezed state and a thermal
state at fixed $n_k$. We define $\delta_k$ such that
\begin{equation}
\left| c_k \right| = \left(n_k+1\right)\left(n_k - \delta_k \right).
\label{def:delta_ck}
\end{equation}
$\delta_k = 0$ is a 2-mode squeezed state and $\delta_k = n_k$, the
maximal value, is a thermal state. This parameter is easily related to
the purity and the squeezing via
\begin{equation}
\label{eq:delta_purity}
\delta_k = \frac{1}{2 \sqrt{p_k}} \frac{1 - p_k}{\cosh \left( 2 r_k \right) + \sqrt{p_k} } \, .
\end{equation}

Let us investigate the effect of decoherence using this class of
state.  To start with, how is the level of decoherence of the state
estimated? Several criteria have been used in the literature: the
so-called rate of de-separation
\cite{kieferEmergenceClassicalityPrimordial1998}, evaluating the
suppression of non-diagonal terms
\cite{martineauDecoherencePrimordialFluctuations2007,
  burgessDecoherencePrimordialFluctuations2008,kieferEntropyGravitonsProduced2000},
the positivity time if the initial state is assumed to be non-Gaussian
\cite{kieferPointerStatesPrimordial2007}, $\delta_k$
\cite{campoInflationarySpectraPartially2005a} or simply the purity
$p_k$ \cite{martinObservationalConstraintsQuantum2018}. We will use
the latter since it directly enters our definition of the GHDM
\eqref{def:effective_squeezing}. The purity can also be conveniently
related to the entropy by Eq.~\eqref{def:entropy_neumann}, which still
applies for decohered states. Since the purity has decreased, the
entropy increases and becomes non-vanishing.  For instance, a thermal
state in the 2-mode particle number basis
\eqref{eq:density_matrix_thermal_state} gives $c_k = r_k =0$ and $p_k
= \left( 2 n_k +1 \right)^{-1}$.

Our focus is on how a certain level of decoherence, represented by
$p_k$, can lead to a classical state in the sense of the criteria
discussed in the previous section. As we now show, for mixed states,
the different criteria are, in general, inequivalent and give
different answers \cite{campoDecoherenceEntropyPrimordial2008a}.  The
separability condition Eq.~\eqref{def:separability} is also still
valid for the partially decohered distribution
\cite{campoInflationarySpectraPartially2005a}. It has a very elegant
interpretation when rewritten in terms of the effective squeezing
parameter
\begin{equation}
B_k \hbar^{-1/2} = p_k^{-1/4} e^{-r_k} \leq 1 \, ,
\label{eq:separability_decohered_2MSS}
\end{equation}
i.e. the state becomes separable when there is no sub-fluctuant mode
anymore due to a sufficient level of decoherence $p_k < e^{r_k}$. The
condition can also be written as $\delta_k \geq n_k / \left( n_k +1
\right)$ which, for the very large number of primordial gravitons
expected $n_k \gg 1$, becomes $\delta_k \geq 1$
\cite{campoInflationarySpectraPartially2005a}.

\begin{figure}
\centering
\includegraphics[width=0.7\textwidth]{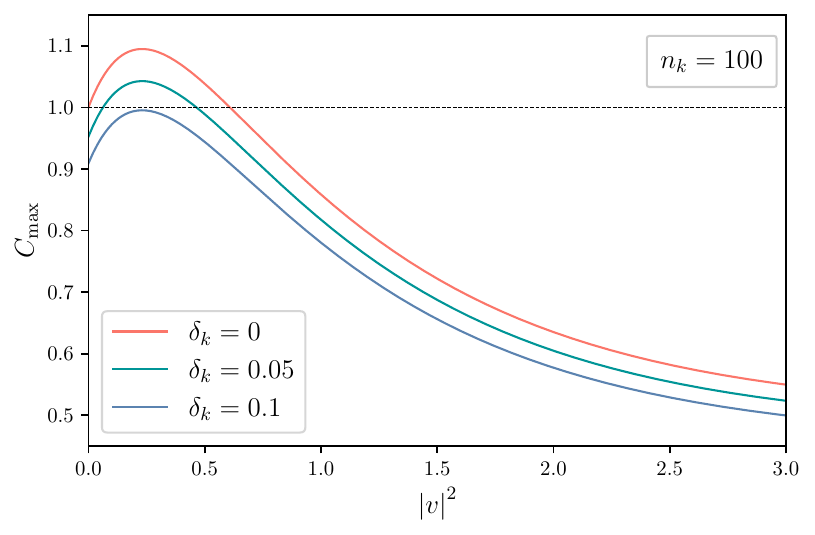}
\caption{$C_{\mathrm{max}}$ as a function of $\left| v \right|^2$ is
  shown in full line for different values of $\delta_k$. The threshold
  of Bell inequality violation $C_{\mathrm{max}}=1$ is shown in dashed
  black line.}
\label{fig:husimi_bell}
\end{figure}

Let us now turn to the Bell inequality of
Eq.~\eqref{def:Bell_husimi}. Its form, its maximisation procedure, and
the formula Eq.~\eqref{def:husimi_rep} are still valid for our
partially decohered state. We plot the value of $C_{\mathrm{max}}$ for
a modest number of gravitons in each polarisation $n_k =100$ and
different values of $\delta_k$ in Fig.~\ref{fig:husimi_bell}. We see
that the maximum of $C_{\mathrm{max}}$ gradually recedes away from
violation as $\delta_k$ increases, and that for $\delta_k = 0.1$, the
inequality is not violated anymore.  In
\cite{campoInflationarySpectraViolations2006a}, the authors give an
approximation in the limit $\delta_k \ll n_k$, which is equivalent to
$\cosh^2 \left( r_k \right) \gg 1$, i.e. in the limit of a very
squeeze state. In this limit, we have
\begin{equation}
    C_{\mathrm{max}} \left( \left| v \right|^2 \right) = \frac{1}{2
      \left(1 + \delta_k \right)} \left[ 1 + \frac{3}{2^{4/3}} + O
      \left( \frac{1 + \delta_k }{n_k} \right) \right] \, .
\end{equation}
so that inequality is violated when
\begin{equation}
    \delta_k < 0.095 \, .
\end{equation}
The threshold is an order of magnitude smaller than that of
separability. This condition is unfortunately not easily expressed in
a comparison between $p_k$ and $r_k$. The perturbations loose their
quantum character in the sense of the Bell inequality
Eq.~\eqref{def:Bell_husimi} faster than in the sense of
separability. This is expected since we recall that separability is a
necessary condition for Bell inequality violation, and here we see
that it is not a sufficient condition anymore; the criteria are
inequivalent for mixed states.

\begin{figure}
\centering
\includegraphics[width=0.7\textwidth]{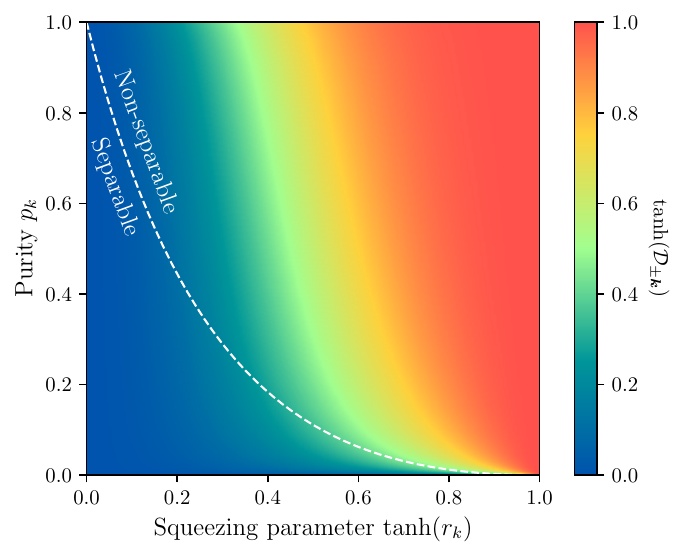}
\caption{ Quantum discord $\mathcal{D}_{\pm \bm{k}}$ of
  Eq.~\eqref{eq:discord_decohered_2MSS} for a partially decohered
  state defined by Eq.~\eqref{def:effective_squeezing} as a function
  of its squeezing $r_k$ and a purity $p_k$.}
\label{fig:discord}
\end{figure}

Finally, let us examine the behaviour of the quantum discord. The
formula \eqref{eq:discord_2MSS} was generalised in
\cite{martinDiscordDecoherence2022} for partially decohered states. A
similar computation in presence of decoherence, although less general,
was previously carried out in
\cite{hollowoodDecoherenceDiscordQuantum2017}. The generalisation
reads
\begin{equation}
\label{eq:discord_decohered_2MSS}
\mathcal{D}_{\pm \bm{k}} =f\left[ p_k^{-1/2} \cosh \left(2 r_k \right)
  \right] -2f\left( p_k^{-1} \right)+ f\left[\frac{p_k^{-1/2} \cosh
    \left(2 r_k \right) + p_k^{-1} } {p_k^{-1/2} \cosh \left(2 r_k
    \right) +1 }\right] \, .
\end{equation}
One notes that the discord does not depend on the squeezing angle
$\varphi_k$. This angle can always be modified by a local symplectic
transformation, and the discord is a local symplectic invariant, so it
must not depend on it.  In Fig.~\ref{fig:discord}, we plot this
formula as a function of $p_k$ and $r_k$, and draw the line delimiting
separable from non-separable states. Its complexity prevents us from
giving a simple threshold for the discord to be, say, larger than $1$
and to compare with separability and Bell
inequality. Figure~\ref{fig:discord} shows clearly that, as for
separability, the value of the discord is dictated by the result of a
competition between the level of squeezing $r_k$ and that of
decoherence $p_k$. These two criteria, along with a Bell inequality of
the type considered in \cite{martinObstructionsBellCMB2017a}, were
recently compared in \cite{Martin:2022kph}.

The overall result of this discussion is that decoherence, if large
enough, does, in the sense of different inequivalent criteria, erase
the quantum features of the state. To be able to complete the
analysis, the only thing necessary is to get a realistic estimation of
the loss of purity in the early universe. Can we get observational
constraints on the interactions generating decoherence and on its
level?  Unfortunately, not for primordial gravitational waves since
they were not detected yet.  However, for scalar perturbations, the
observation of the baryonic acoustic oscillation (BAO) actually
imposes that during inflation, decoherence cannot modify too much the
squeezing parameters~\cite{campoInflationarySpectraPartially2005a}
$r_k \gg 1$ and $\varphi_k \approx - \pi / 2$. In particular, this
implies that complete decoherence during inflation, leading to a
thermal state like Eq.~\eqref{eq:density_matrix_thermal_state}, is
excluded. Indeed, the squeezing parameter $r_k$ would vanish
\cite{kieferEmergenceClassicalityPrimordial1998}. Note that the purity
$p_k$ is not constrained by this argument.  This relation between the
oscillations and strong squeezing had initially led to label the
former a quantum feature
\cite{grishchukSqueezedQuantumStates1990a}. As we have explained, the
squeezing, in its dynamical aspect, can be understood as the presence
of a growing and a decaying mode so that this result can be understood
completely classically as pointed in
\cite{albrechtInflationSqueezedQuantum1994}. This `temporal coherence'
of the perturbations is explained in detail (using a classical point
of view) in \cite{dodelsonCoherentPhaseArgument2003}.  In addition of
this general argument, for precise models of decoherence, other
constraints can be obtained as discussed, for instance, in
Ref.~\cite{martinObservationalConstraintsQuantum2018}.

Let us close this section by coming back to the important question of
the observability of the features. Even in the absence of decoherence,
are the operators that we have used in the discussion measurable?  For
the Bell inequality Eq~\eqref{def:Bell_husimi} they have derived in
\cite{campoInflationarySpectraViolations2006a}, Campo and Parentani
argue that each of the four terms is, in principle,
measurable. However, one needs to measure a difference of order $1$
between these while the intrinsic fluctuations of the factor $n_k$ is
of order $n_k$, which is of order $10^{86}$. The measure is, in
practice, impossible.  The authors of
\cite{martinObstructionsBellCMB2017a} argue that having only access to
the growing mode makes it impossible to measure two of their three
pseudo-spin operators. Verifying their Bell inequality necessitates
measuring at least two, and so is experimentally impossible.  To
address this difficulty, they suggest that one could try to build
Legget-Garg inequalities
\cite{martinLeggettGargInequalitiesSqueezed2016a} that rely on
correlation in time of a single operator and do not require to measure
two non-commuting operators at a given time.
Ref.~\cite{maldacenaModelCosmologicalBell2016} also proposed a
``baroque", to use the term of the author, inflationary model in which
Bell operators are measured during inflation by another field rather
than at later times by observer. The field stores the result in
classical, robust variables that could be read out at later times by
observers.  Finally, the separability and quantum discord, being
directly attached to properties of the density matrix, seem harder to
measure. The possibility of measuring them directly in the
cosmological case has not, to the best of our knowledge, been
analysed. In \cite{martinQuantumDiscordCosmic2016b}, the authors took
another approach and showed that if the perturbations were in a
quantum non-discordant state, and reproduced the power spectrum
measured for scalar perturbations, then they have to be in the thermal
state~\eqref{eq:density_matrix_thermal_state}. As we have just
explained, this is ruled out. Note that this argument \textit{assumes}
that the system is described by a quantum state rather than proves it.

\section{Some perspectives and critics}

To conclude this review, we mention a few perspectives and possible
criticisms of the previously developed issues.

First, the estimation of the minimal level of decoherence of
cosmological perturbations keeps being refined see,
e.g. \cite{burgessEFTHorizonStochastic2015,
  gongQuantumNonlinearEvolution2019, Burgess:2022nwu}. Most authors
conclude that decoherence has completed by the end of inflation, and
the state is classical when the modes become sub-Hubble
again. However, an application of the precise level of decoherence
obtained to a concrete non-classicality criterion is still
missing. Such computation would be essential since we have shown that
the threshold for the emergence of classicality given by the different
criteria depends on both the purity \textit{and} the level of
squeezing.  In addition, some authors have also suggested that the use
of Markovian approximation is not well-justified in the cosmological
context and that a more general master equation is required to achieve
a correct prediction
\cite{colasBenchmarkingCosmologicalMaster2022, Brahma:2022yxu}.

Second, the discussion of Sec.~\ref{sec:quantumness_primordial_GW}
applies to the tensor and scalar perturbations. However, primordial
gravitational waves have the important specificity that they could be
\textit{directly} detected, not only indirectly in the temperature
anisotropies of the CMB, as scalar ones. Direct detection (although
futuristic see \cite{Caprini:2018mtu} ) would bring about exciting
possibilities to search for quantum signatures in gravitational wave
detectors. Several authors,
e.g. \cite{kannoPossibleDetectionNonclassical2019,
  parikhSignaturesQuantizationGravity2021a,
  kannoIndirectDetectionGravitons2021a}, have investigated these. The
squeezed states of gravitons could produce noise in gravitational wave
interferometers, and some of the authors argued that its quantum
character might be revealed by measuring the decoherence it would
induce between two entangled mirrors.

Another possibility that we have not discussed is to use the
interactions of the perturbations, not as a mere source of
decoherence, but as giving new signals in the form of
non-gaussianities that could be used. Focusing on scalar
perturbations, the authors of
Ref.~\cite{greenSignalsQuantumUniverse2020} showed that substituting
the initial quantum vacuum fluctuations by a Gaussian stochastic field
with the same two-point functions would lead to enhanced
non-gaussianities akin to those generated by initial excited
states. Not measuring such an enhancement was then suggested to be a
sign of non-classicality of the initial state (see also
\cite{Ghosh:2022cny}).  With a different approach to
non-gaussianities, the Wigner function of primordial gravitational
waves was calculated in Ref.~\cite{gongQuantumNatureWigner2020},
taking in account the intrinsic non-linearities of gravity. Its
regions of negativity were then explored as a means of exhibiting a
signature of quantumness of the state.  Other works such as
Refs.~\cite{martinNonGaussianitiesQuantum2018, DaddiHammou:2022itk}
took yet another route and provided some constraints on decoherence
based on the level of non-gaussianities.

Finally, some authors criticised the standard approach of analysing
correlations between $\pm \bm{k}$ modes.  The authors of
\cite{agulloDoesInflationSqueeze2022,
  campoDecoherenceEntropyPrimordial2008a} have argued that discussing
correlations between $\pm \bm{k}$ modes is not appropriate as these
two modes do not exist separately outside of Minkowski, in particular
during inflation, and keep being mixed. Just as there is no preferred
choice of vacuum (Sec.~\ref{subsec:graviton_production}), there is no
preferred choice of partition to unambiguously discuss levels of
squeezing and correlations. These critics, we believe, would not apply
to sub-Hubble modes, e.g. in our toy model radiation domination where
$a^{\prime } = 0$. Some recent works
\cite{martinRealspaceEntanglementCosmic2021,
  espinosa-portalesRealspaceBellInequalities2022} do not suffer from
these shortcomings since they perform similar computations for quantum
discord and Bell inequalities, but use real space correlation
functions. Unfortunately, their results tend to show that, even in the
absence of decoherence, no quantum features appear in real space.
Lastly, the formalism presented here does not address the so-called
"quantum measurement problem" in cosmology. In our approach, we used
an ergodicity assumption to justify equating the quantum expectation
values to average values over different patches of the sky. However,
one could argue that we did not discuss how the perturbations
``collapsed" from a homogeneous quantum state to an inhomogeneous
distribution with different values in each patch. For a discussion of
this point, see \cite{sudarskyShortcomingsUnderstandingWhy2011}.

To conclude, it is fair to say that the current status regarding the
quest for quantum features in the primordial gravitational wave
background is not entirely settled. First, on the observational side,
the waves themselves, even in their classical aspects, have yet to be
detected~\cite{Fumagalli:2021dtd}. Experiments in
preparation~\cite{abazajianCMBS4ForecastingConstraints2022,
  campetiMeasuringSpectrumPrimordial2021} might manage to detect
signatures of the waves in the $\bm{B}$-modes of the CMB. However,
direct detection via gravitational wave interferometers seems so far
out of reach~\cite{Caprini:2018mtu}. On the theoretical side, in
recent years, several quantum features of the quantum state for the
primordial gravitational waves predicted in the simplest models have
been exhibited. Unfortunately, no currently available experimental
protocol has yet been designed to detect these features. In addition,
the effect of decoherence has been increasingly more precisely
characterised, and the latest findings tend to show that it might have
erased all the potentially detectable features by the end of
inflation. At this time, most analyses have been restricted to the
simplest inflationary models and at the Gaussian level. More recently,
some promising suggestions and proposals have been made concerning
non-gaussianities, discussing the possible signatures of decoherence,
or other possible hints of a quantum origin of the perturbations.

\section*{Acknowledgements}

We thank Karim Benabed for insights on the possibility of detection of
$\bm{B}$-modes by future missions as well as J\'er\^ome Martin and
Ilya Shapiro for enlightening discussions and remarks on the
manuscript.

\bibliographystyle{spphys}
\bibliography{Refs}

\end{document}